\documentclass[graybox]{svmult}

\usepackage{type1cm}              
\usepackage{makeidx}              
\usepackage{graphicx}              
\usepackage{multicol}               
\usepackage[bottom]{footmisc}
\usepackage{newtxtext}        
\usepackage{newtxmath}       
\usepackage{amsmath,amsfonts}
\usepackage{color,colortbl}
\definecolor{Gray}{gray}{0.9}

\makeindex             

\newcolumntype{g}{>{\columncolor{Gray}}l}

\newcommand{\NF}{N^{\textsc{f}}}
\newcommand{\NL}{N^{\textsc{l}}}
\newcommand{\R}{\mathbb{R}}
\newcommand{\target}{x^\tau}
\newcommand{\xf}{x}
\newcommand{\xl}{y}
\newcommand{\vf}{v}
\newcommand{\vl}{w}
\newcommand{\Ad}{A}
\newcommand{\HFF}{H^{\textsc{f}}}
\newcommand{\HFL}{H^{\textsc{l}}}
\newcommand{\HLF}{K^{\textsc{f}}}
\newcommand{\HLL}{K^{\textsc{l}}}
\newcommand{\barS}{s}
\newcommand{\Crand}{C_z}
\newcommand{\Wz}{z}
\newcommand{\Ct}{C_{\tau}}
\newcommand{\Cchar}{C_{s}}
\newcommand{\Crf}{C^{\textsc{f}}_{r}}
\newcommand{\Crl}{C^{\textsc{l}}_{r}}
\newcommand{\Cat}{C_{at}}
\newcommand{\Caf}{C^{\textsc{f}}_{al}}
\newcommand{\Cal}{C^{\textsc{l}}_{al}}
\newcommand{\rhoF}{m^{\textsc{f}}}
\newcommand{\rhoL}{m^{\textsc{l}}}
\newcommand{\etaF}{\eta^{\textsc{f}}}
\newcommand{\etaL}{\eta^{\textsc{l}}}
\newcommand{\lamF}{\lambda^{\textsc{f}}}
\newcommand{\lamL}{\lambda^{\textsc{l}}}
\newcommand{\noisepar}{\Crand}
\newcommand{\Adcut}{S}

\newcommand{\REV}[1]{{\color{black} #1}}

\newtheorem{alg}{Algorithm}

\begin{document}

\title*{Mathematical models and methods for crowd dynamics control}
\author{Giacomo Albi, Emiliano Cristiani, Lorenzo Pareschi, and Daniele Peri}
\institute{
Giacomo Albi \at Dipartimento di Informatica, Universit\`a di Verona, Verona, Italy, \email{giacomo.albi@univr.it}
\and Emiliano Cristiani \at Istituto per le Applicazioni del Calcolo, Consiglio Nazionale delle Ricerche, Rome, Italy, \email{e.cristiani@iac.cnr.it}
\and Lorenzo Pareschi (corresponding author) \at Dipartimento di Matematica e Informatica, Universit\`a di Ferrara, Ferrara, Italy, \email{lorenzo.pareschi@unife.it}
\and Daniele Peri \at Istituto per le Applicazioni del Calcolo, Consiglio Nazionale delle Ricerche, Rome, Italy, \email{d.peri@iac.cnr.it}
}
\maketitle

\abstract*{In this survey we consider mathematical models and methods recently developed to control crowd dynamics, with particular emphasis on egressing pedestrians. We focus on two control strategies: The first one consists in using special agents, called \emph{leaders}, to steer the crowd towards the desired direction. Leaders can be either hidden in the crowd or recognizable as such. This strategy heavily relies on the power of the \emph{social influence} (herding effect), namely the natural tendency of people to follow group mates in situations of emergency or doubt. 
The second one consists in modify the surrounding environment by adding in the walking area multiple obstacles optimally placed and shaped. The aim of the obstacles is to naturally force people to behave as desired.
Both control strategies discussed in this paper aim at reducing as much as possible the intervention on the crowd. Ideally the natural behavior of people is kept, and people do not even realize they are being led by an external intelligence. 
Mathematical models are discussed at different scales of observation, showing how macroscopic (fluid-dynamic) models can be derived by mesoscopic (kinetic) models which, in turn, can be derived by microscopic (agent-based) models.}

\abstract{In this survey we consider mathematical models and methods recently developed to control crowd dynamics, with particular emphasis on egressing pedestrians. We focus on two control strategies: The first one consists in using special agents, called \emph{leaders}, to steer the crowd towards the desired direction. Leaders can be either hidden in the crowd or recognizable as such. This strategy heavily relies on the power of the \emph{social influence} (herding effect), namely the natural tendency of people to follow group mates in situations of emergency or doubt. 
The second one consists in modify the surrounding environment by adding in the walking area multiple obstacles optimally placed and shaped. The aim of the obstacles is to naturally force people to behave as desired.
Both control strategies discussed in this paper aim at reducing as much as possible the intervention on the crowd. Ideally the natural behavior of people is kept, and people do not even realize they are being led by an external intelligence. 
Mathematical models are discussed at different scales of observation, showing how macroscopic (fluid-dynamic) models can be derived by mesoscopic (kinetic) models which, in turn, can be derived by microscopic (agent-based) models.\footnote{Authors would like to thank the Italian Ministry of Instruction, University and Research (MIUR) to support this research with funds coming from PRIN Project 2017 (No. 2017KKJP4X entitled ``Innovative numerical methods for evolutionary partial differential equations and applications'').}}

\section{Introduction}\label{sec:intro}

This paper aims at presenting a brief survey of some recent developments in the mathematical modeling and control techniques for human crowd dynamics. Here, we define \emph{crowd dynamics control} as the art of steering large masses of people in the desired direction, minimizing verbal directives to individuals and preserving as much as possible their natural behavior. 
In the extreme case, we get methods to steer crowds along predefined paths without the crowd being aware of it, i.e.\ individuals do not even perceive that their (apparently) natural decisions are guided.
Such control techniques are expected to be effective in all situations characterized by the impossibility of directly communicating with the crowd (e.g., in case of very large groups, emergencies, violent crowds reluctant to follow directions indicated by event organizers or police).

Crowd control stems on different research topics and benefits from a multidisciplinary approach. First, physics and psychology are called upon to point out the main behavioral aspects which represent the constitutive ingredients of mathematical models. 
Models will then be used to create a digital twin of the moving crowd. 
After a careful choice of the scale of observation (mainly depending on the size of the crowd and computing resources), numerical analysis is used to solve the equations and get a reproduction of virtual crowds, while real observations and data acquisition are crucial to calibrate the models. 
Calibration is particularly challenging, considering the high variability among persons and the difficulty of measuring some parameters like pushiness, degree of rationality, knowledge of the surrounding environment, etc. 
After that, we have to set up the control problem, defining the \emph{control (design) variables}. Roughly speaking, this means that we need to identify which part of the system is subject to modifications and which part is instead left to the natural state. 
Then,  we must define the \emph{objective function}, i.e.\ the goal of the optimization procedure.
Here optimization techniques come into play to solve the control problem and get the optimal strategy to apply to steer the crowd as desired. Finally, experiments with real crowds are desirable to check the effectiveness of the strategy found in virtual environments.

\REV{Curiously, some metaheuristic optimization techniques like the one adopted in Section \ref{sec:ctrl_obs_GO} are inspired precisely by models for collective behavior: many agents spread in the abstract space of control variables hunting for the optimal strategy of the crowd. This creates a surreal parallelism between the physical space, where people move, and the space of control variables, where people's behaviour lies. In the two spaces, the same abstract mathematical methods can be then used to make agents reach their goals.}

\paragraph{\em Scale of observation, models, and degree of rationality}
Pedestrian dynamics can be observed at different \emph{scales}, and the choice of the point of view drastically changes the modeling framework. A \emph{nanoscopic} approach consists in tracking every single agent, including the position of torso and head \cite{arechavaleta2008IEEETR, chitour2012SICON}. 
A \emph{microscopic} approach consists instead in tracking every single agent assuming s/he is a $0$-dimensional point or a small circle. 
A \emph{mesoscopic} approach is based on the description of average quantities like the density of people, but it keeps the possibility to distinguish one-to-one interactions. Finally, \emph{macroscopic} approach describes only average quantities loosing any kind of granularity. 
\emph{Multiscale} approaches are also possible: One can adopt different scales of observation in different parts of the domain (passing information across an interface) or one can employ two or more scales at the same time and space, to get a fully hybridized description, as in \cite{cristiani2011MMS, cristiani2014book}.

Concerning \emph{pedestrian modeling}, virtually any kind of models have been investigated so far and several reviews and books are available. For a quick introduction, we refer the reader to the surveys \cite{bellomo2011SR, helbing2001RMP} and the books \cite{cristiani2014book, kachroo2008book, NaPaTo10}. 
Some papers deal specifically with egressing/evacuating pedestrians: a very good source of references is the paper \cite{abdelghany2014EJOR}, where evacuation models both with and without optimal planning search are discussed. 
The paper \cite{abdelghany2014EJOR} itself proposes a cellular automata model coupled with a genetic algorithm to find a top-down optimal evacuation plan.  
Evacuation problems were studied by means of lattice models \cite{cirillo2013PhysA, guo2012TRB}, social force models \cite{helbing2000N, parisi2005PhysA}, cellular automata models \cite{abdelghany2014EJOR, wang2015PhysA}, mesoscopic models \cite{agnelli2015M3AS, albi2016SIAP, FTW18}, and macroscopic models \cite{carrillo2016M3AS}. 
Limited visibility issues were considered in \cite{carrillo2016M3AS, cirillo2013PhysA, guo2012TRB}. 
Real experiments involving people can be found in \cite{albi2016SIAP, guo2012TRB}.

It can be useful to recall here that pedestrians can show different \emph{degrees of rationality}, depending on the situation and their knowledge of the surrounding  environment. In an unknown environment with limited visibility we expected people to follow basically a full instinctive behavior, being impossible to make predictions. 
Conversely, an ideal rational pedestrian with a specific target and full knowledge of the environment  can compute her/his path in an optimal manner, is able to forecast the behavior of other pedestrians (even for long time) and is able to understand the impact of the presence of the others at any time along her/his path. In this case dynamics of people are fully coupled in space and time, and a competition among pedestrians naturally arises. Nash equilibria or similar concepts help to find the strategy eventually adopted by the participants. 
For a deep discussion in this direction we refer the reader to the book \cite[Sect.\ 4.4]{cristiani2014book} and papers \cite{cristiani2015NHM, cristiani2015SIAP}.

\paragraph{\em Crowd dynamics control}
The problem of controlling crowds falls in the larger line of research dedicated to \textit{self-organizing agents}. For this, a vast mathematical literature is available and first principles as well as the most important qualitative results are already known. The toy model for such investigations is the Cucker-Smale model \cite{cucker2007IEEE}, introduced in 2007. Controlled versions of the model are widely studied, see, e.g., \cite{AlHePa15, AP18a, BBCK18, BFK15, CFPT13, FS14, HePaSt15}. Let us also mention the seminal paper \cite{couzin2005N}, where authors pointed out that in a group of individuals with tendency to move together, a small percentage of informed individuals is able to steer the whole group in the desired direction.

In this paper we focus on two control strategies: 
\begin{itemize}
\item The first one consists in using special agents, called \emph{leaders}, to steer the crowd towards the desired direction. Leaders can be either \emph{hidden} in the crowd \cite{albi2016SIAP, couzin2005N, duan2014SR, han2006JSSC, han2013PLOSONE} or \emph{recognizable} as such \cite{ABFHKPPS, albi2014Phil, AlPaZa19, borzi2015M3ASa, during2009PRSA, motsch2011JSP}. This strategy heavily relies on the power of the social influence (or herding effect), namely the natural tendency of people to follow group mates in situations of emergency or doubt. 

\emph{En passant}, let us stress that the term ``herding'' is largely ambiguous in the literature, as pointed out in the recent paper \cite{haghani2019JAT}.

\item The second one consists in modify the surrounding environment by adding in the walking area multiple \emph{obstacles} optimally placed and shaped. The aim of the obstacles is to smoothly force people to behave as desired, changing surrounding conditions in such a way that the modified behaviour of people naturally matches the optimal one. This approach can be seen as an inverse application of the Braess's paradox \cite{braess2005TS, hughes2003ARFM}, originally proposed in the context of traffic flow on network. In that case it was noted that adding a new road (i.e.\ a new connection) in the network can lead to a higher degree of congestion. In our framework, an additional constraints lead paradoxically to an improvement of the pedestrian flow.
Several papers investigate numerically the effectiveness of the Braess's paradox by means of both microscopic models (e.g., Helbing's social force model) and macroscopic models. Some papers report the effect of additional obstacles manually placed in the walking area, see, among others, \cite{escobar2003LNCS, frank2011PA, helbing2005TS, hughes2002TRB, matsuoka2015, twarogowska2014AMM}. 
Other papers, instead, employ optimization algorithms, see \cite{cristiani2017AMM, cristiani2019AMM, cristiani2015SIAP, jiang2014PLOS, johansson2007,  shukla2009, zhao2017PhysA}. 
Note that, the resulting optimization problem typically is non convex and high dimensional. 
Efficient optimization algorithms, including Particle Swarm Optimization (PSO), genetic algorithms, differential evolution, and random compass search were used.

\end{itemize}

\paragraph{\em Manuscript organization}
The rest of the manuscript is organized as follows. 
In Sect.\ \ref{sec:models} we introduce the mathematical model for egressing pedestrians which will serve as a guideline for the rest of the paper. Main ingredients are introduced and three scales of observation (microscopic, mesoscopic, macroscopic) are discussed. In addition, we discuss methods to manage obstacles in models, i.e.\ how to prevent pedestrians from entering forbidden zones of the walking area.
In Sect.\ \ref{sec:ctrl_leaders} we present crowd control techniques based on the use of  leaders, either visible (i.e.\ recognizable from the crowd as such) or not.
In Sect.\ \ref{sec:ctrl_obstacles} we analyze crowd control techniques based on the use of  smart obstacles suitably located in the walking area to modify the perception of the environment and suitably modify the optimal paths.
Finally, in Sect.\ \ref{sec:conclusions} we sketch some conclusions and future research directions.
\section{Different levels of description}\label{sec:models}
In this section we describe the model at a general level. In particular we will focus on the different levels of description: microscopic, mesoscopic and macroscopic.

\subsection{Preliminary notions}\label{sec:modelguidelines}
Hereafter, we divide the population between \textit{leaders}, which are the controllers and behave in some optimal way (to be defined), and \textit{followers}, which represent the mass of agents to be controlled. Followers typically cannot distinguish between followers and leaders. 
Our approach consists in describing leaders by a \emph{first order} model (positions are the only state variables), while followers are described by a \emph{second order} model (both positions and velocities are state variables). 
In the latter case, the small inertia typical of pedestrian motion is obtained by means of a fast relaxation towards the target velocity. 

Concerning the way interactions between individuals are modelled, we adopt a mixed approach, assuming short-range interactions to be metrical and long-range ones to be topological.
We recall that, the individual interactions are said to be \textit{metrical} if they involve only mates within a predefined sensory region, regardless of the number of individuals which actually fall in it. Interactions are instead said \textit{topological} if it involves a predefined number of group mates regardless their distance from the considered agent. 

We assume here that pedestrians have a target to reach in minimal time, but the environment is in general unknown. Therefore, since individuals have no idea of the location of their target, we expect that they often look around to explore the environment and see the behavior of the others. This is why we prefer to adhere to isotropic (all-around) interactions.

Next, before introducing the details of the model, let us briefly describe the social forces acting on the agents.

\begin{itemize}
\item\emph{Leaders.}
\begin{itemize}
\item Leaders are subject to an isotropic metrical short-range \textit{repulsion force} directed against all the others, translating the fact that they want to avoid collisions and that a maximal density exists.
\item Leaders are assumed to know the environment and the self-organizing features of the crowd. They respond to an \textit{optimal force} which is the result of an offline optimization procedure, defined as to minimize some cost functional. 
\end{itemize}

\item \emph{Followers.}
\begin{itemize}
\item Similarly to leaders, followers respond to an isotropic metrical short-range \textit{repulsion force} directed against all the others.
\item Followers tend to a \textit{desired velocity} which corresponds to the velocity the would follow if they were alone in the domain. Since they do not know the environment, we assume that followers describe a \textit{random walk} if the exit is not visible (exploration phase) or a sharp motion toward the exit if the exit is visible (evacuation phase). 
%
\item If the exit is not visible, followers are subject to an isotropic topological \textit{alignment force} with all the others, including leaders, i.e., they tend to have the same velocity of the group mates (herding effect). We will distinguish in the sequel between \emph{visible} (recognized by the followers) and \emph{invisible} (not recognized by the followers) leaders.
\end{itemize}
\end{itemize}


\subsection{The microscopic leader-follower model}\label{sec:micromodel}
In this section we introduce the microscopic model for followers and leaders. We denote by $d$ the dimension of the space in which the motion takes place (typically $d=2$), by $\NF$ the number of followers and by $\NL \ll \NF$ the number of leaders. We also denote by $\Omega\equiv\R^d$ the walking area and by $\target\in\Omega$ the target point. To define the target's visibility area, we consider the set $\Sigma$, with $\target\in\Sigma\subset\Omega$, and we assume that the target is completely visible from any point belonging to $\Sigma$ and completely invisible from any point belonging to $\Omega\backslash\Sigma$. 

For every $i=1,\ldots,\NF$, let $(\xf_i(t),\vf_i(t))\in\R^{2d}$ denote position and velocity of the agents belonging to the population of followers at time $t\geq 0$ and, for every  $k=1,\ldots,\NL$, let $(\xl_k(t),\vl_k(t))\in\R^{2d}$  denote position and velocity of the agents among the population of leaders at time $t \geq 0$. Let us also define $\mathbf{\xf}:=(\xf_1,\ldots,\xf_{\NF})$ and $\mathbf{\xl}:=(\xl_1,\ldots,\xl_{\NL})$.

Finally, let us denote by $B_r(x)$ the ball of radius $r>0$ centered at $x\in\Omega$ and by $\mathcal B_\mathcal N(x;\mathbf{\xf},\mathbf{\xl})$ the \emph{minimal} ball centered at $x$ encompassing at least $\mathcal N$ agents, and by $\mathcal N^*$ the actual number of agents in $\mathcal B_\mathcal N(x;\mathbf{\xf},\mathbf{\xl})$. Note that $\mathcal N^*\geq\mathcal N$. 
\begin{remark}
The computation of $\mathcal B_\mathcal N(x;\mathbf{\xf},\mathbf{\xl})$ requires the knowledge of the positions of all the agents, since all the distances $|\xf_i-x|$, $i=1,\ldots,\NF$, and  $|\xl_k-x|$, $k=1,\ldots,\NL$ must be evaluated in order to find the $\mathcal N$ closest agents to $x$.
\end{remark}

\medskip

The microscopic dynamics described by the two populations is given by the following set of ODEs: for $i = 1, \dots, \NF$ and $k = 1, \ldots, \NL$,
\begin{equation}\label{eq:micro}
\left\{
\begin{array}{l}
\dot{x}_i = \vf_i,\\ [1.5mm] 
\dot{v}_i = \Ad(\xf_i,\vf_i) + \sum_{j=1}^{\NF} \HFF(\xf_i,\vf_i,\xf_j,\vf_j;\mathbf{\xf},\mathbf{\xl}) +\sum_{\ell=1}^{\NL} \HFL(\xf_i,\vf_i,\xl_\ell,\vl_\ell;\mathbf{\xf},\mathbf{\xl}),\\ [1.5mm]
\dot{y}_k= w_k = \sum_{j=1}^{\NF} \HLF(\xl_k,\xf_j) +  \sum_{\ell=1}^{\NL} \HLL(\xl_k,\xl_\ell) + u_k.
\end{array}
\right.
\end{equation}
We assume that

\begin{itemize}
\item $\Ad$ is a self-propulsion term, given by the relaxation toward a random direction or the relaxation toward a unit vector pointing to the target (the choice depends on the position), plus a term which translates the tendency to reach a given characteristic speed $\barS \geq 0$ (modulus of the velocity), i.e.,
\begin{align} \label{eq:Ad}
\Ad(x,v) := \theta(x) \Crand(\Wz-v) & + (1-\theta(x))\Ct\left(\frac{\target - x}{|\target - x|} - v\right)+\Cchar(\barS^2-|v|^2)v,
\end{align}
where $\theta:\mathbb{R}^d \rightarrow [0,1]$ is the characteristic function of $\Omega\backslash\Sigma$, $\theta(x) =\chi_{\Omega\backslash\Sigma}(x)$, 
$\Wz$ is a $d$-dimensional random vector with normal distribution $\mathcal N(0,\sigma^2)$, and $\Crand$, $\Ct$, $\Cchar$ are positive constants.

\item
The interactions follower-follower and follower-leader are defined as 
\begin{equation}
\begin{aligned}\label{eq:HFF}
\HFF(x,v,x',v';\mathbf \xf,\mathbf \xl) :=& 
-\Crf R_{\gamma,r}(x,x') + 
\theta(x)\frac{\Caf}{\mathcal N^*}\left(v'-v\right) \chi_{\mathcal B_\mathcal{N}(x;\mathbf \xf,\mathbf \xl)}(x'),\\
\HFL(x,v,y,w;\mathbf \xf,\mathbf \xl) :=& -\Crf R_{\gamma,r}(x,y) + 
\theta(x)\frac{\Cal}{\mathcal N^*}\left(w-v\right) \chi_{\mathcal B_\mathcal{N}(x;\mathbf \xf,\mathbf \xl)}(y)\\
&+\theta(x)\Cat\frac{y-x}{|y-x|},
\end{aligned}
\end{equation}
for given positive constants $\Crf, \Caf, \Cal, \Cat, r$ and $\gamma$. 

In the first equation of \eqref{eq:HFF} the term
\begin{align*}
R_{\gamma,r}(x,x') & = \begin{cases}
e^{-|x'-x|^\gamma}\frac{x'-x}{|x'-x|}, & \text{ if } x'\in B_r(x)\backslash\{x\}, \\
0, & \text{ otherwise,}
\end{cases}
\end{align*}
models a (metrical) repulsive force, while the second term accounts for the (topological) alignment force, which vanishes inside $\Sigma$. Note that, the interaction with the leaders, defined by the second equation in \eqref{eq:HFF}, accounts the previous forces with an additional attraction towards the leaders position.
With the choice $\Cat=0$, $\Caf=\Cal$ we have
$\HFF\equiv \HFL$ and, therefore, the leaders are not recognized by the followers as special. This feature opens a wide range of new applications, including the control of crowds not prone to follow authority's directives.

\item
The interactions leader-follower and leader-leader reduce to a mere (metrical) repulsion, i.e., $\HLF = \HLL = -\Crl R_{\zeta,r}$, where $\Crl>0$ and $\zeta>0$ are in general different from $\Crf$ and $\gamma$, respectively. 

\item
$u_k:\R^+\to\R^{d\NL}$ is the control variable, to be chosen in a set of admissible control functions. Except for the short-range repulsion forces, the behaviour of the leaders is entirely characterized by the control term $u$. More details on the control term will be given in Section 3. 
%
\end{itemize}
\begin{remark}
As a further generalization of the above modeling, the population of leaders can be separated into two populations 
$(y^{\textsc{v}}_k(t),w^{\textsc{v}}_k(t))$, $k=1,\ldots,N^{\textsc{l,v}}$ and  $(y^{\textsc{i}}(t)_k,w^{\textsc{i}}_k(t))$, $k=1,\ldots,N^{\textsc{l,i}}$ with $\NL = N^{\textsc{l,v}} + N^{\textsc{l,i}}$, depending on whether they are recognized by the followers (visible) or not (invisible).  
In the first case, the corresponding interaction function $H^{\textsc{l,v}}\neq \HFF$ since followers will have the tendency to align with greater intensity towards leaders, whereas in the second case we simply have $H^{\textsc{l,i}}= \HFF$. In the sequel, for the sake of simplicity, we present our analysis in the case of system \eqref{eq:micro}, where all leaders are either visible or invisible, leaving to a straightforward generalization the extension of simultaneous coexistence of visible and invisible leaders.   
\end{remark}

\subsection{Boltzmann modelling}\label{sec:mesomodel}

As already mentioned, our main interest in \eqref{eq:micro} lies in the case $\NL \ll \NF$, that is the population of followers exceeds by far the one of leaders. When $\NF$ is very large, a microscopic description of both populations is no longer a viable option. We thus consider the evolution of the distribution of followers at time $t \geq 0$, denoted by $f(t,x,v)$, together with the microscopic equations for the leaders (whose number is still small). To this end, we denote with $\rhoF$ the total mass of followers, i.e.,
\begin{align*}
\rhoF(t) = \int_{\mathbb{R}^{2d}} f(t,x,v) \ dx \ dv,
\end{align*}
which we shall eventually require to be equal to $\NF$. We introduce, for symmetry reasons, the distribution of leaders $g$ and their total mass
\begin{equation}\label{eq:deltaleaders} 
g(t,x,v) = \sum^{\NL}_{k = 1} \delta_{(\xl_k(t),\vl_k(t))}(x,v), 
\qquad 
\rhoL(t) = \int_{\mathbb{R}^{2d}} g(t,x,v) \ dx \ dv = \NL.
\end{equation}

The evolution of $f$ can be then described by a Boltzmann-type dynamics, derived from the above microscopic formulation, which is obtained by analyzing the binary interactions between a follower and another follower and the same follower with a leader. The application of standard methods of binary interactions, see \cite{cercignani1994, PT:13}, shall yield a mesoscopic model for the distribution of followers, to be coupled with the previously presented ODE dynamics for leaders.

To derive the Boltzmann-type dynamics, we assume that, before interacting, each agent has at his/her disposal the values $\mathbf{\xf}$ and $\mathbf{\xl}$ that s/he needs in order to perform its movement: hence, in a binary interaction between two followers with state parameter $(x,v)$ and $(\hat{x},\hat{v})$, the value of $\HFF(x,v,\hat{x},\hat{v};\mathbf{\xf},\mathbf{\xl})$ does not depend on $\mathbf{\xf}$ and $\mathbf{\xl}$. In the case of $\HFF$ of the form \eqref{eq:HFF}, this means that the ball $B_\mathcal{N}(x;\mathbf \xf,\mathbf \xl)$ and the value of $\mathcal{N}^*$ have been already computed before interacting.

Moreover, since we are considering the distributions $f$ and $g$ of followers and leaders, respectively, the vectors $\mathbf{\xf}$ and $\mathbf{\xl}$ are derived from $f$ and $g$ by means of the first moments of $f$ and $g$, $\pi_1 f$ and $\pi_1 g$, respectively, which give the spatial variables of those distribution. Hence, we write $\HFF(x,v,\hat{x},\hat{v};\pi_1 f,\pi_1 g)$ in place of  $\HFF(x,v,\hat{x},\hat{v};\mathbf{\xf},\mathbf{\xl})$ to stress the dependence of this term on $f$ and $g$.

We thus consider two followers with state parameter $(x,v)$ and $(\hat{x},\hat{v})$ respectively, and we describe the evolution of their velocities after the interaction according to
\begin{align}
\begin{cases}
\begin{split}\label{eq:follcoord}
v^* & = v + \etaF \left[\theta(x) \noisepar \xi + \Adcut(x,v) + \rhoF\HFF(x,v,\hat{x},\hat{v};\pi_1 f,\pi_1 g) \right],\\ 
\hat{v}^* & = \hat{v} + \etaF \left[\theta(\hat{x}) \noisepar \xi + \Adcut(\hat{x},\hat{v}) + \rhoF\HFF(\hat{x},\hat{v},x,v;\pi_1 f,\pi_1 g) \right],
\end{split}
\end{cases}
\end{align}
where $\etaF$ is the strength of interaction among followers, $\xi$ is a random variables whose entries are i.i.d.\ following a normal distribution with mean $0$, variance $\varsigma^2$, taking values in a set $\mathcal{B}$, and $\Adcut$ is defined as the deterministic part of the self-propulsion term \eqref{eq:Ad},
\begin{align}\label{eq:Adcut}
\Adcut(x,v)=-\theta(x)\Crand v+(1-\theta(x))\Ct\left(\frac{\target - x}{|\target - x|} - v\right)+\Cchar(\barS^2-|v|^2)v.
\end{align}
We then consider the same follower as before with state parameters $(x,v)$ and a leader agent $(\tilde{x},\tilde{v})$; in this case the modified velocities satisfy
\begin{align}
\begin{cases}
\begin{split}\label{eq:leadcoord}
& v^{**}  = v + \etaL \rhoL \HFL(x,v,\tilde{x},\tilde{v};\pi_1 f,\pi_1 g),\\
& \tilde{v}^*  = \tilde{v}, 
\end{split}
\end{cases}
\end{align}
where $\etaL$ is the strength of the interaction between followers and leaders. Note that \eqref{eq:leadcoord} accounts only the change of the followers' velocities, since leaders are not evolving via binary interactions.


The time evolution of $f$ is then given by a balance between bilinear gain and loss of space and velocity terms according to the two binary interactions \eqref{eq:follcoord} and \eqref{eq:leadcoord}, quantitatively described by the following Boltzmann-type equation
\begin{align}\label{eq:strongBoltz}
\partial_t f+ v \cdot \nabla_x f = \lamF Q(f,f) + \lamL Q(f,g),
\end{align}
where $\lamF$ and $\lamL$ stand for the interaction frequencies among followers and between followers and leaders, respectively. The interaction integrals $Q(f,f)$ and $Q(f,g)$ are defined as
\begin{align*}
Q(f,f)(t) & = \mathbb{E}\left(\int_{\mathbb{R}^{4d}}\left(\frac{1}{J_{\textsc{f}}} f(t,x_*,v_*)f(t,\hat{x}_*,\hat{v}_*) - f(t,x,v)f(t,\hat{x},\hat{v})\right) \ d\hat{x} \ d\hat{v}\right), \\
Q(f,g)(t) & = \mathbb{E}\left(\int_{\mathbb{R}^{4d}}\left(\frac{1}{J_{\textsc{l}}} f(t,x_{**},v_{**})g(t,\tilde{x}_*,\tilde{v}_*) - f(t,x,v)g(t,\tilde{x},\tilde{v})\right)  \ d\tilde{x} \ d\tilde{v}\right),
\end{align*}
where the couples $(x_*,v_*)$ and $(\hat{x}_*,\hat{v}_*)$ are the pre-interaction states that generates $(x,v)$ and $(\hat{x},\hat{v})$ via \eqref{eq:follcoord}, and $J_{\textsc{f}}$ is the Jacobian of the change of variables given by \eqref{eq:follcoord}. Similarly, $(x_{**},v_{**})$ and $(\tilde{x}_*,\tilde{v}_*)$ are the pre-interaction states that generates $(x,v)$ and $(\tilde{x},\tilde{v})$ via \eqref{eq:leadcoord}, and $J_{\textsc{l}}$ is the Jacobian of the change of variables given by \eqref{eq:leadcoord}. Moreover, the expected value $\mathbb{E}$ is computed with respect to $\xi \in \mathcal{B}$.

In what follows, for the sake of compactness, we shall omit the time dependency of $f$ and $g$, and hence of $Q(f,f)$ and $Q(f,g)$ too. In conclusion, we have the following combined ODE-PDE system for the dynamics of microscopic leaders and mesoscopic followers
\begin{align} \label{eq:micromeso}
\left\{
\begin{array}{l}
\partial_t f + v \cdot \nabla_x f = \lamF Q(f,f) + \lamL Q(f,g), \\
\displaystyle\dot{\xl}_k = \vl_k = \int_{\R^{2d}} \HLF(\xl_k,x) f(x,v) \ dx\ dv + \sum_{\ell = 1}^{\NL} \HLL(\xl_k,\xl_{\ell}) + u_k.
\end{array}
\right.
\end{align}

\begin{remark}
If we would have opted for a description of agents as hard-sphere particles, the arising Boltzmann equation \eqref{eq:strongBoltz} would be of Enskog type, see \cite{toscanibellomoenskog}. The relationship between the hard- and soft-sphere descriptions (i.e., where repulsive forces are considered, instead) has been deeply discussed, for instance, in \cite{andersenhard}. In our model, the repulsive force $R_{\gamma,r}$ is not singular at the origin for computational reasons, therefore the parameters $\gamma$ and $r$ have to be chosen properly to avoid arbitrary high density concentrations. 
\end{remark}


\subsection{Mean-field modeling}
A different level of modeling, is obtained by considering directly the limit for large $\NF$ of the dynamic described by \eqref{eq:micro} where all individuals in principle are allowed to interact with all others. This kind of models are typically described by Fokker-Planck equations and can also be obtained directly from \eqref{eq:strongBoltz} in the quasi-invariant limit \cite{PT:13}. 
This technique, analogous to the so-called grazing collision limit in plasma physics, has been thoroughly studied in \cite{Vil02} and allows, as pointed out in \cite{PT:13}, to pass from the binary Boltzmann description introduced in the previous section to the mean-field limit. 

In what follows, we shall assume that our agents densely populate a small region but weakly interact with each other. Formally, we assume that the interaction strengths $\etaF$ and $\etaL$ scale according to a parameter $\varepsilon$, the interaction frequencies $\lamF$ and $\lamL$ scale as $1/\varepsilon$, and we let $\varepsilon \rightarrow 0$. In order to avoid losing the diffusion term in the limit, we also scale the variance of the noise term $\varsigma^2$ as $1/\varepsilon$. More precisely, we set
\begin{equation}\label{eq:scaling}
\etaF = \varepsilon, \qquad \etaL = \varepsilon,  \qquad
\lamF = \frac{1}{\varepsilon \rhoF},  \qquad \lamL  = \frac{1}{\varepsilon \rhoL}, \qquad \varsigma^2 = \frac{\sigma^2}{\varepsilon}. 
\end{equation}
Under the above scaling assumptions, the weak form of the equation \eqref{eq:strongBoltz}, i.e.,
\begin{align}
\frac{\partial}{\partial t} \left \langle f, \varphi \right\rangle + \left \langle f, v \cdot \nabla_x \varphi \right\rangle =  \lamF\left\langle Q(f,f),\varphi \right\rangle + \lamL \left\langle Q(f,g),\varphi \right\rangle,
\label{eq:Bolweek}
\end{align}
for a compactly supported test function $\varphi$, where
\begin{align}
\left\langle Q(f,f),\varphi \right\rangle & = \mathbb{E}\left(\int_{\mathbb{R}^{4d}} \left(\varphi(x,v^{*}) - \varphi(x,v) \right)f(x,v)f(\hat{x},\hat{v}) \ dx \ dv \ d\hat{x} \ d\hat{v}\right), \label{eq:collopF} \\
\left\langle Q(f,g),\varphi \right\rangle & = \mathbb{E}\left(\int_{\mathbb{R}^{4d}} \left(\varphi(x,v^{**}) - \varphi(x,v) \right)f(x,v)g(\tilde{x},\tilde{v}) \ dx \ dv \ d\tilde{x} \ d\tilde{v} \right), \label{eq:collopL}
\end{align}
 reduces to the following Fokker-Planck equation (see \cite{PT:13} for more details) 

\begin{align} \label{eq:FokkerPlanck}
\frac{\partial}{\partial t} \left \langle f, \varphi \right\rangle + \left \langle f, v \cdot \nabla_x \varphi \right\rangle =  \left\langle f, \nabla_v \varphi \cdot \mathcal{G}\left[f,g\right] + \frac{1}{2}\sigma^2 (\theta \noisepar)^2 \Delta_v \varphi \right\rangle,
\end{align}
where
\begin{align*}
\mathcal{G}\left[f,g\right] & = \Adcut + \mathcal{H}^{\textsc{f}}[f] + \mathcal{H}^{\textsc{l}}[g].
\end{align*}
with 
\begin{eqnarray*}
\mathcal{H}^{\textsc{f}}[f](x,v) & =& \int_{\mathbb{R}^{2d}} \HFF(x,v,\hat{x},\hat{v};\pi_1 f,\pi_1 g) f(\hat{x},\hat{v}) \ d\hat{x} \ d\hat{v},\\
\mathcal{H}^{\textsc{l}}[g](x,v) &=& \int_{\R^{2d}} \HFL(x,v,\tilde{x},\tilde{v};\pi_1 f, \pi_1 g) g(\tilde{x},\tilde{v}) \ d\tilde{x} \ d\tilde{v}.
\end{eqnarray*}

%
%
Since $\varphi$ has compact support, equation \eqref{eq:FokkerPlanck} can be recast in strong form by means of integration by parts. Coupling the resulting PDE with the microscopic ODEs for the leaders  $k=1,\ldots,\NL$, we eventually obtain the system
\begin{equation}\label{eq:FokkerPlanckStrong}
\left\{ 
\begin{array}{l}
\partial_t f + v \cdot \nabla_x f = - \nabla_v \cdot \left(\mathcal{G}\left[f,g\right]f\right) + \frac{1}{2} \sigma^2 (\theta \noisepar)^2 \Delta_v f,\\ [1mm]
\displaystyle
\dot{\xl}_k = \vl_k = \int_{\R^{2d}} \HLF(\xl_k,x) f(x,v) \ dx\ dv + \sum_{\ell = 1}^{\NL} \HLL(\xl_k,\xl_{\ell}) + u_k.
\end{array}
\right.
\end{equation}


\subsection{Macroscopic modelling}
In terms of model hierarchy one could imaging to compute the moments of \eqref{eq:FokkerPlanckStrong} in order to further reduce the complexity. Let us stress that deriving a consistent macroscopic system from the kinetic equation is in general a difficult task, since equilibrium states are difficult to obtain, therefore no closure of the moments equations is possible. For self-organizing models similar to \eqref{eq:FokkerPlanckStrong}, in the noiseless case (i.e.\ $\sigma\equiv0$), a standard way to obtain a closed hydrodynamic system is to assume the velocity distribution to be mono-kinetic, i.e.\ $f(t,x,v)=\rho(t,x)\delta(v-V(t,x))$, and the fluctuations to be negligible, thus computing the moments of \eqref{eq:FokkerPlanckStrong} leads to the following macroscopic system for the density $\rho$ and the bulk velocity $V$,
\begin{equation}
\left\{ 
\begin{array}{l}
\partial_t \rho + \nabla_x\cdot (\rho V) = 0,\\ [2mm]
\partial_t (\rho V)+ \nabla_x\cdot (\rho V\otimes V) = \mathcal{G}_m\left[\rho,\rho^{\textsc{l}},V,V^{\textsc{l}}\right]\rho,\\ [1mm]
\displaystyle
\dot{\xl}_k = \vl_k = \int_{\R^{d}} \HLF(\xl_k,x) \rho(t,x) \ dx+ \sum_{\ell = 1}^{\NL} \HLL(\xl_k,\xl_{\ell}) + u_k,
\end{array}
\right.
\end{equation}
where $\rho^{\textsc{l}}(x,t), V^{\textsc{l}}(x,t)$ represent the leaders' macroscopic density and bulk velocity, respectively, and $\mathcal{G}_m$ the macroscopic interaction operator, see \cite{albi2013AML,carrillo2010particle} for further details. 
For $\sigma>0$, the derivation of a macroscopic system depends highly on the scaling regime between the noise and the interaction terms, see for example \cite{carrillo2009double,carrillo2010particle,karper2015hydrodynamic}. Furthermore, the presence of diffusion operator in model \eqref{eq:FokkerPlanckStrong} depends on the spatial domain, therefore the derivation of a reasonable macroscopic model is not trivial and it is left for further studies.

\subsection{Interaction with obstacles}\label{sec:interactionwithobstacles}
So far we have considered pedestrians influenced by the leaders action but free to move in any direction of the space. In practical applications, however, dynamics are often constrained by walls or other kind of obstacles. Including obstacles in mathematical models is not as trivial as one can imagine. 
We refer the reader to \cite[Sect.\ 2]{cristiani2017AMM} for a review of obstacles' handling techniques proposed in the literature. Here we recall just the three most common procedures, the first is the one we use in the numerical tests presented in this paper.

\smallskip

\noindent {\bf Cut off of the velocity field.} An easy method to deal with obstacles is obtained by computing the velocity field first neglecting the presence of the obstacles, then nullifying the component of the velocity vector which points inside the obstacle. This method is used in, e.g., \cite{albi2016SIAP, cristiani2011MMS, cristiani2015SIAP}. The method requires to pay attention that pedestrians do not stop walking completely because both components of the velocity vector vanish. This can happen around corners, stair-shaped obstacles and when obstacles are very close to each other (i.e.\ the distance is comparable with the spatial resolution of the numerical grid). A similar but more sophisticated approach can be found in \cite[Sect.\ 3]{cristiani2017AMM}.

\smallskip

\noindent {\bf Repulsive obstacles.} 
Another easy method used to manage obstacles is obtained assuming that they generate a repulsive (social) force, exactly as pedestrians themselves do. In other words, obstacles are treated as frozen pedestrians. In this way one can use a repulsion function of the same kind to model both the interactions with group mates and with obstacles.
This method is extensively used in microscopic models, see, e.g., \cite{colombi2016CAM, frank2011PA, helbing1995PRE, loehner2010AMM, okazaki1979TAIJa, okazaki1979TAIJb, okazaki1979TAIJc} and also in macroscopic and multiscale models, see, e.g., \cite{colombo2012M3AS, coscia2008M3AS, etikyala2014M3AS, piccoli2009CMT}, with or without the pre-evaluation of the distance-to-obstacle function. The main drawback of this approach is that it is quite difficult to tune the strength of the repulsion force in such a way that the resulting behaviour is both admissible and realistic. Indeed, if the force is too small there is the risk that pedestrians enter the obstacles, while if it is too large pedestrians bypass the obstacles excessively far away. The paper \cite{colombo2012M3AS} proposes a method to tune automatically the strength of the repulsion. 

\smallskip

\noindent {\bf Rational turnaround.}
In more sophisticated models which take into account the rationality and predictive ability of pedestrians, obstacles can be managed including them into the decision-making process. 
For example, in the Hughes's model \cite{hughes2002TRB} pedestrians move, at each given time, along the fastest path toward the target, considering that crowded regions slow down the walking speed. In this framework, obstacles are easily included assuming that inside them the speed is null,
so that the computation of the fastest path will circumvent them automatically. 
\section{Crowd controls through leaders}\label{sec:ctrl_leaders}
As discussed in the previous Section, in order to steer the crowd towards a desired direction or target position, we want to exploit the tendency of people to follow group mates in situations of emergency or doubt (social influence or herding effect). 
In particular by controlling few leaders and their trajectories
we want to drive the whole system of followers. In this Section, we will formulate this problem in the context of optimal control theory and discuss its numerical solution. The main challenge is represented by the complexity induced by the non-linearities, and the high-dimensionality of models \eqref{eq:micro} and \eqref{eq:micromeso}. Hence, we are interested in efficient methods to solve this optimization problem, synthesizing strategies scalable at various levels: from micro to macro.

\subsection{Optimal control framework}
The functional to be minimized can be chosen in several ways, the effectiveness mostly depends on the optimization method which is used afterwards. The most natural functional to be minimized for a crowd of egressing pedestrian, is the \textit{evacuation time}, which can be defined as follows
\begin{equation}\label{defJ_evactime}
{\min}\{t>0~|~x_i(t)\notin\Omega \quad \forall i=1,\ldots,\NF\},
\end{equation}
subject to (\ref{eq:micro}) or (\ref{eq:micromeso}) and with $u(\cdot)\in U_{\textup{adm}}$, where $U_{\textup{adm}}$ is the set of admissible controls (including for instance box constraints to avoid excessive velocities). 
Such functional can be extremely nonregular, therefore the search of local minima is particularly difficult. Moreover the evacuation of the total mass in many situations can not completely be reached, in particular for the mesoscopic model where we account for a diffusion term.

In the sequel we propose two alternative optimal control problems, both designed to improve the {evacuation time}, and associated to different optimization methods for their solution.

\paragraph{\em Quadratic cost functional and Model Predictive Control } 
A first approximation of \eqref{defJ_evactime} can be designed introducing a quadratic cost  as follows
\begin{equation}\label{defJ}
\begin{split}
\ell(\mathbf{\xf},\mathbf{\xl},u)=C_1\sum_{i=1}^{\NF}\|x_i-\target\|^2+
C_2\sum_{i=1}^{\NF}\sum_{k=1}^{\NL}\|x_i-y_k\|^2+
C_3\sum_{k=1}^{\NL}\|u_k\|^2,
\end{split}
\end{equation}
for some positive constants $C_1,C_2$ and $C_3$. The first term promotes the fact that followers have to reach the exit, the second forces leaders to keep contact with the crowd, and the last term penalizes excessive velocities. This minimization is performed along a fixed time frame $[0,T]$
\begin{equation}\label{eq:finhor}
\underset{u(\cdot)\in U_{\textup{adm}}}{\min} \int\limits_{0}^{T}\ell(\mathbf{\xf}(t),\mathbf{\xl}(t),u(t))\,dt,\qquad \text{subject to (\ref{eq:micro}) or (\ref{eq:micromeso})}.
\end{equation}
For this type of problem optimal solutions are typically out of reach, therefore we have to rely on sub-optimal strategies. A computationally efficient way to address the optimal control problem \eqref{eq:finhor} is by Model Predictive Control (MPC) \cite{mayne2000A}, the method works as follows.
\begin{alg}[MPC]~
\begin{enumerate}
\item 
Set the time step  $\Delta t$ with $\bar n=0,\ldots,N_T$ such that $T=N_T\Delta t$, and the predictive parameter $N_{\texttt{mpc}}$, where $N_{\texttt{mpc}}\ll N_T$.
\item
\texttt{while} $\bar n<N_T$
\begin{enumerate}
\item Solve the reduced minimization problem
\begin{equation}
\underset{u(\cdot)\in U_{\textup{adm}}}{\min} 
\sum_{n=\bar n}^ {\bar n+N_{\texttt{mpc}}-1} 
\ell(\mathbf{x}(n\Delta t),\mathbf{y}(n\Delta t),u(n\Delta t))
\end{equation}
subject to a discretization of the dynamics \eqref{eq:micro}.
\item Generate an optimal sequence of controls $\{u(\bar n\Delta t),\ldots,u((\bar n+N_{\texttt{mpc}}-1)\Delta t)\}$.
\item Evolve the dynamics of \eqref{eq:micro} for a time step $\Delta t$ with $u(\bar n\Delta t)$.
\item Update $\bar n\leftarrow\bar n+1$.
\end{enumerate}
\texttt{repeat}
 \end{enumerate}
\end{alg}
 Note that for $N_{\texttt{mpc}}=T/\Delta t$ the MPC approach solves the full time frame problem $\eqref{eq:finhor}$, whereas for $N_{\texttt{mpc}}=2$, it recovers an instantaneous controller. Such flexibility is complemented with a robust behavior, as the optimization is re-initialized every time step, allowing to address perturbations along the optimal trajectory.

\paragraph{\em Evacuated mass functional and Compass Search}
Complete evacuation of the crowd is not always feasible, therefore we consider as milder request to maximize the evacuated mass at final time $T$, minimizing the total mass inside the domain $\Omega$ as follows
\begin{equation}\label{defJ_evacmass}
\underset{u(\cdot)\in U_{\textup{adm}}}{\min}\left\{\rhoF(T|u)=\int_{\mathbb{R}^{d}}\int_{\Omega}f(T,x,v)\,dx\,dv\right\},\qquad \text{subject to (\ref{eq:micromeso})},
\end{equation}
where in the microscopic case the integral over the density $f(T,x,v)$ has to be interpreted in the empirical sense as sum of Dirac masses namely 
\[
\rhoF(T|u)=\frac{1}{\NF}\sum_{i=1}^{\NF}\delta(x_i(t),v_i(t))\chi_\Omega(x_i(t)).
\]
In order to minimize such functional we move towards random methods as compass search (see \cite{audet2014MPC} and references therein), alternative methods are genetic algorithms, or particle swarm optimization which will be discussed in more details in Section \ref{sec:ctrl_obstacles}. 

First of all, we consider only piecewise constant trajectories, introducing suitable \emph{switching times} for the leaders' controls. More precisely, we assume that leaders move at constant velocity for a given fixed time interval and when the switching time is reached, a new velocity vector is chosen. Therefore, the control variables are the velocities at the switching times for each leader. In order to optimize such strategy we define the following Compass Search (CS) algorithm.

\begin{alg}[CS]~
\begin{enumerate}
\item Select a discrete set of sample times $S_M=\{t_1,t_2,\ldots,t_M\}$, the parameters $k =0$, $k_{\texttt {max}}$ and $m_E$.
\item Select an initial strategy $u^*$  piecewise constant over the set $S_M,$
 e.g. constant direction and velocity speed towards the target $\target$ ({\em go-to-target})
\[u^{*}_j(t) = -\frac{y_j(0)-\target}{\|y_j(0)-\target\|},\quad j=1,\ldots,\NL,\]
compute the functional $m(T |u^{*})$.
\item Perform a perturbation of the the piecewise constant $u^{*}(t)$ with small random variations over the time-set $S_M$
\begin{equation}\label{eq:randvar}
u^{(k)}(t_m) = u^*(t_m)+B_m,\quad m=1,\ldots,M \tag{$\mathcal P$}
\end{equation}
where $B_m$ is a random perturbation of the velocity at time $t_m$. Finally compute $m(T |u^{(k)})$.
\item \texttt{while} $k<k_{\texttt{max}}$ AND $m(T |u^{*})<m_E$
\begin{enumerate}
\item Update $ k\leftarrow k+1$.
\item Perform the perturbation \eqref{eq:randvar} and compute $m(T |u^{(k)})$
\item \texttt{If} $m(T |u^{(k)})\leq  m(T|u^{*})$
\\ \qquad set $u^{*}\leftarrow u^{(k)}$ and $m(T |u^{*})\leftarrow m(T |u^{(k)})$.

\end{enumerate}
\texttt{repeat}
\end{enumerate}
\end{alg}

\begin{remark}
In the following some remarks concerning the above control settings.
\begin{itemize}
\item  Both MPC and CS approaches produce suboptimal controls, but they offer a good compromise in terms of computational efficiency.
\item  Controlling directly the velocities rather than the accelerations makes the optimization problem much simpler because minimal control variations have an immediate impact on the dynamics.
\item In the mesoscopic scale, both functionals \eqref{defJ_evactime} and \eqref{defJ} can be considered, however the major difficulty to reach a complete evacuation of the continuous density is mainly due to the presence of the diffusion term and to the invisible interaction with respect to the leaders. 
\REV{ Indeed the action of the Laplacian outside the visibility area causes the followers' density to spread overall the domain. Hence without any further assumptions  such non-linear diffusion, boundary conditions, or stronger interaction terms among followers and followers-leaders. Thus, in order to deal with the optimal control of the mean-field model, we will consider functional \eqref{defJ_evacmass}, namely the mass evacuated at the final time $T$.}
\end{itemize}
\end{remark}

\subsection{Numerical experiments}
In this section, we present some numerical tests to validate our modeling framework at the microscopic and mesoscopic level. We explore three different scenarios for pedestrians: in Setting 1 (S\#1) we discuss the difference between visible and invisible leaders; in Setting 2 (S\#2) and 3 (S\#3) we explore situation without and with obstacles respectively.

The dynamics at microscopic level \eqref{eq:micro} is discretized by means of the explicit Euler method with a time step $\Delta t=0.1$. The evolution of the kinetic density in \eqref{eq:FokkerPlanckStrong} is approximated by means of binary interaction algorithms, which approximates the mesoscopic model \eqref{eq:micromeso} simulating the Boltzmann dynamics  \eqref{eq:strongBoltz} with a Monte-Carlo method for small values of the parameter $\varepsilon$, as presented in \cite{albi2013MMS}.  We choose $\varepsilon=0.02$, $\Delta t = 0.01$ and a sample of $N_s=O(10^4)$ particles to reconstruct the kinetic density for Setting 0 and 1, and $N_s=O(4\times10^3)$ for S\#2. This type of approach is inspired by numerical methods for plasma physics and it allows to solve the interaction dynamics with a reduced computational cost compared with mesh-based methods, and an accuracy of $O(N_s^{-1/2})$. For further details on this class of binary interaction algorithms see \cite{albi2013MMS, PT:13}. 

Concerning optimization, in the microscopic case we adopt either the compass search with functional \eqref{defJ_evactime} or MPC with functional \eqref{eq:finhor}. In the mesoscopic case we adopt the compass search with functional \eqref{defJ_evacmass}. 

In S\#2 and S\#3 we set the compass search switching times every $20$ time steps, and in S\#3 every $50$, having fixed the maximal random variation to $1$ for each component of the velocity. In S\#1, the inner optimization block of the MPC procedure is performed via a direct formulation, by means of the \texttt{fmincon} routine in MatLab, which solves the optimization problem via an SQP method.

In Table \eqref{tab:all_parameters} we  report the various parameters used for the different settings.

\begin{table}[!h]
\label{tab:all_parameters}
\begin{center}
\caption{Model parameters for the different scenarios.}
\begin{tabular}{|c|c|c|c|c|c|c|c|c|c|c|c|c|c|c|}
\hline
Setting & $\NL$ & $\NF$ & $\mathcal N$ & $\Crf$ & $\Crl$ & $\Cal$ &$\Caf$ & $\Cat$ & $\Crand$ & $\Ct$ & $\Cchar$ & $\barS^2$ & $r=\zeta$ & $\gamma$\\
 &  & &  &  & &  & &  &  &  &  &  & & \\
\hline\hline
\#1 & 3 & 150   & 10 & 2 & 1.5 & 3 & 3 & 0.01/0 & -- & 1 & 0.5 & 0.4 & 1 &1 \\
\hline
\#2 & 0-3 & 150   & 10 & 2 & 1.5 & 3 &3& 0  & 1  & 1 & 0.5 & 0.4 & 1 &1\\
\hline
\#3 & 0-2 & 150   & 10 & 2 & 1.5 & 3 &3& 0  & 1  & 1 & 0.5 & 0.4 & 1 &1\\
\hline
\end{tabular}
\end{center}
\end{table}

\subsubsection{S\#1: Visible vs invisible leaders}
We investigate numerically the difference among invisible and visible leaders, namely we distinguish situations where leaders are undercover with respect to cases where leaders act as an attractor of the crowd. 

In this first setting the crowd of followers is distributed uniformly in the space domain with initial velocity randomly distributed with zero average, the leaders are positioned on the far right side of the crowd moving with fixed velocity $\vl_\ell = (|\texttt{v}|,0)^T$ for every $\ell=1,2,3$, and no target is visible.%

\begin{figure}[h!]
\begin{center}
\includegraphics[width=3.85cm]{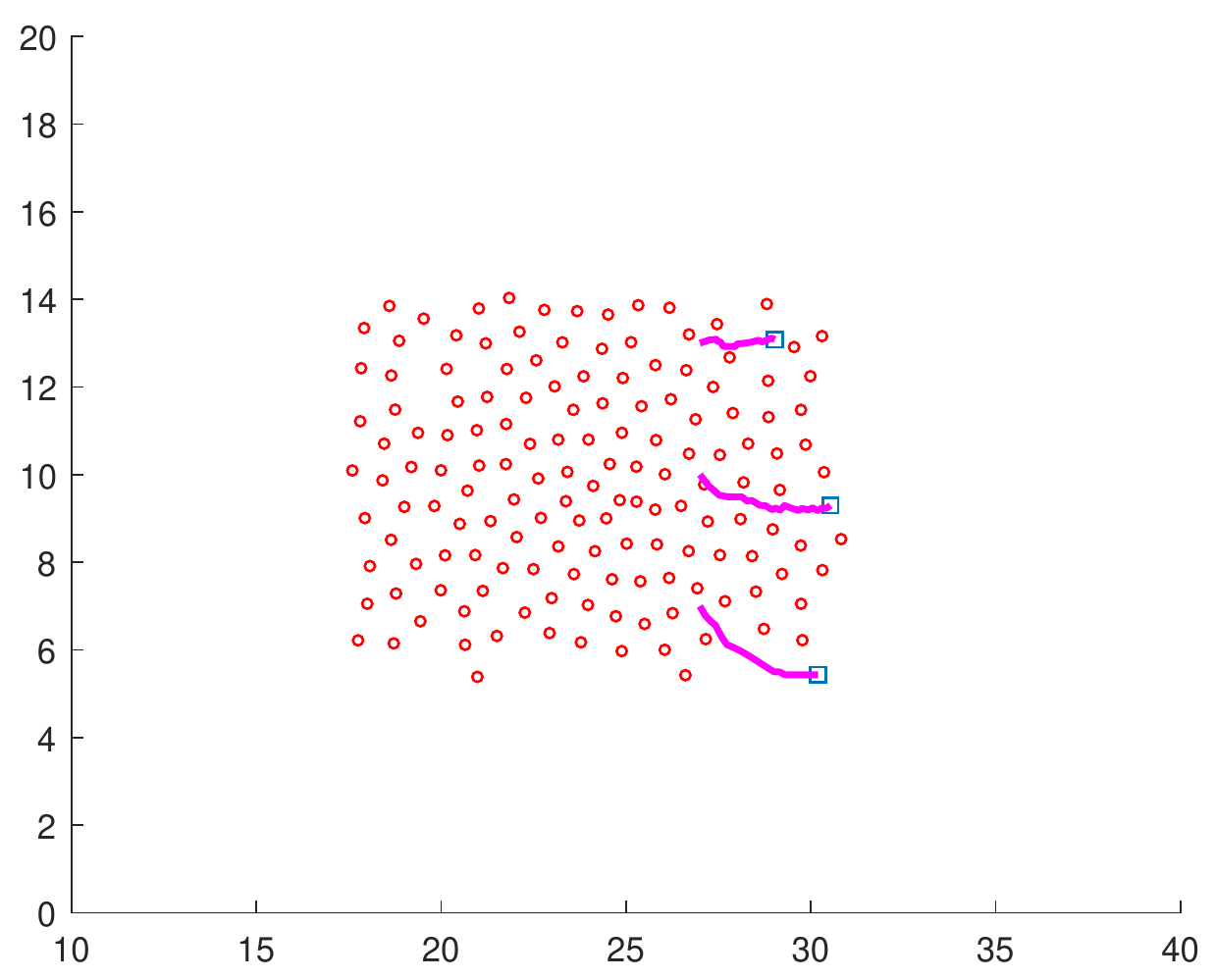}
\includegraphics[width=3.85cm]{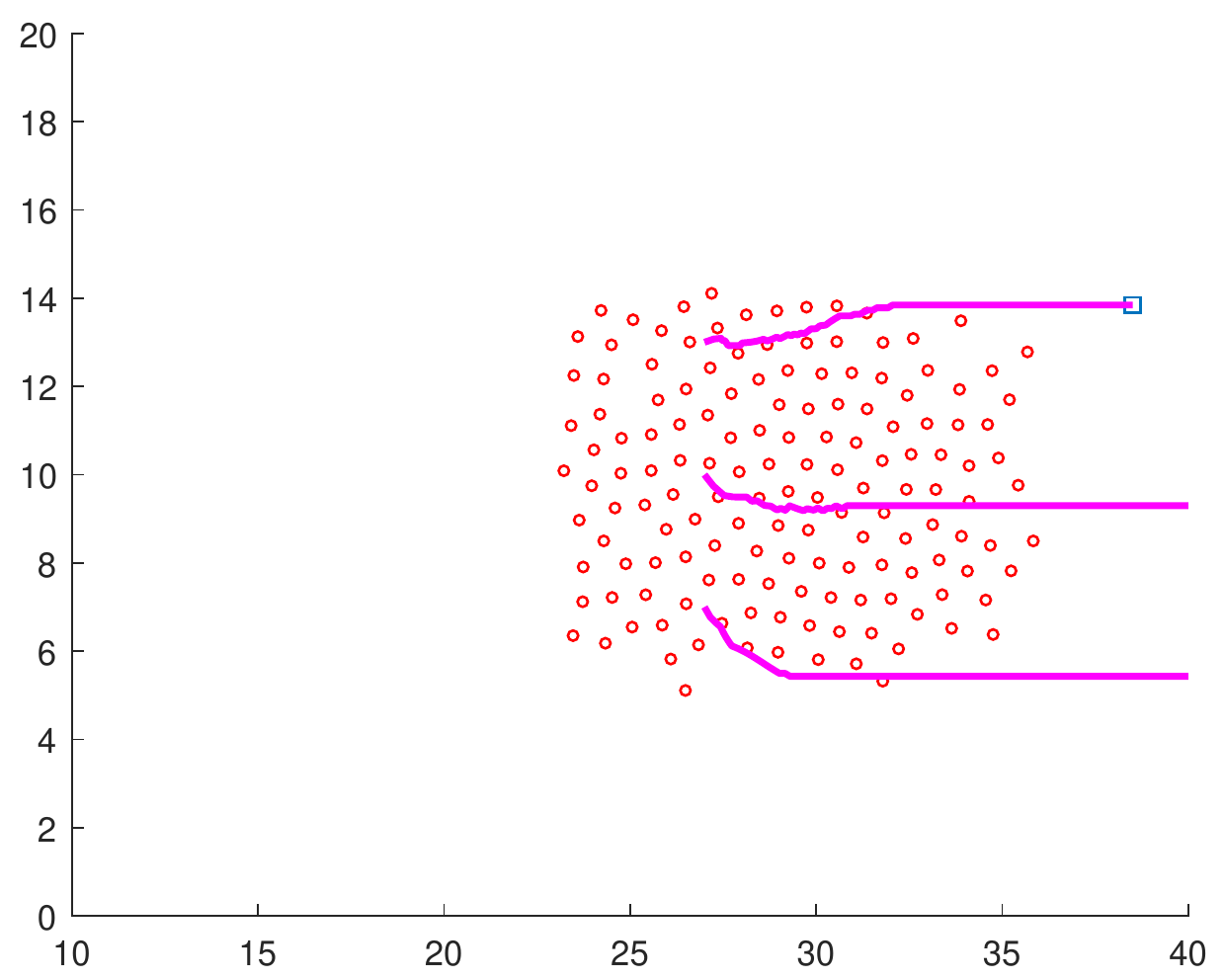}
\includegraphics[width=3.85cm]{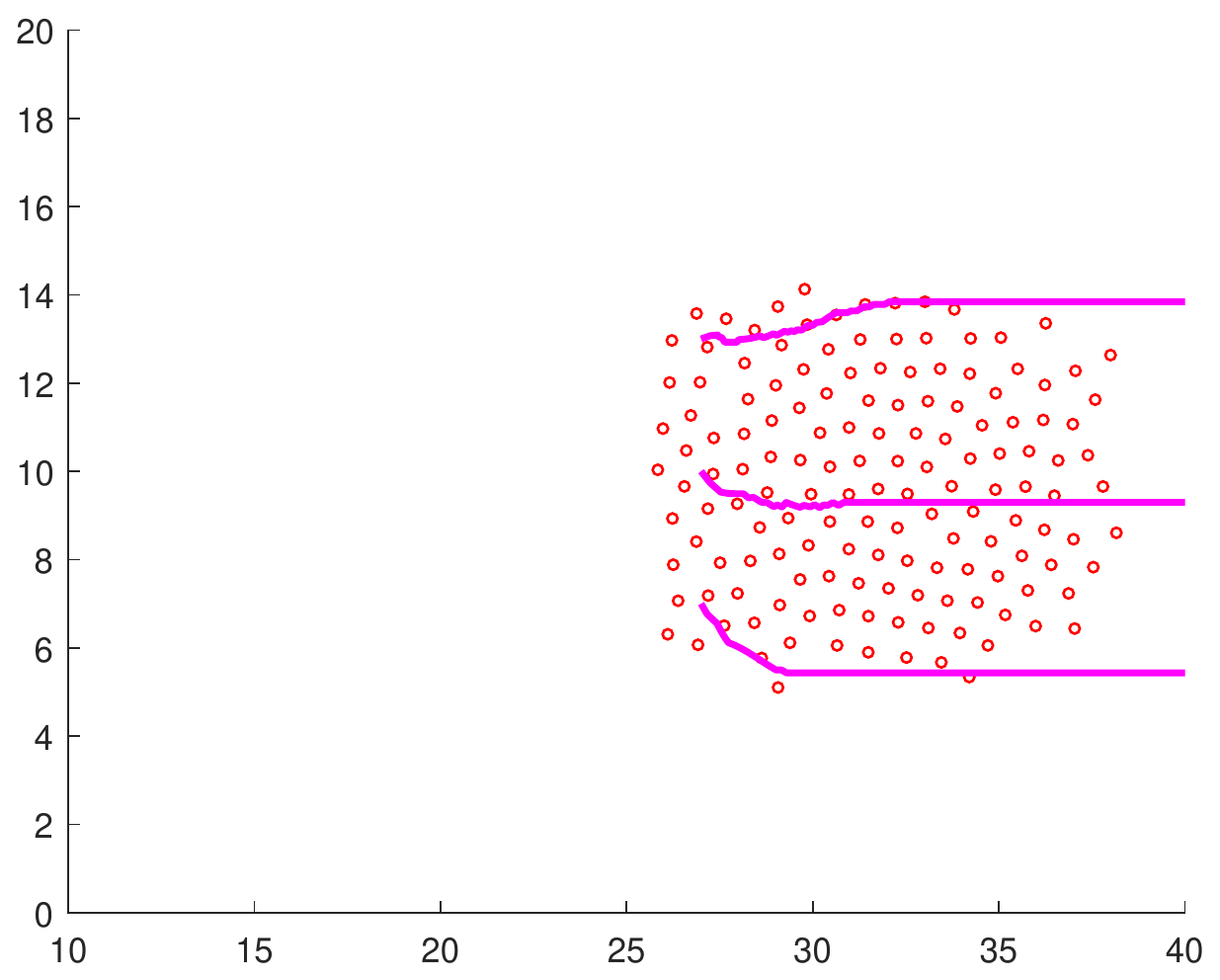}
\\
\includegraphics[width=3.85cm]{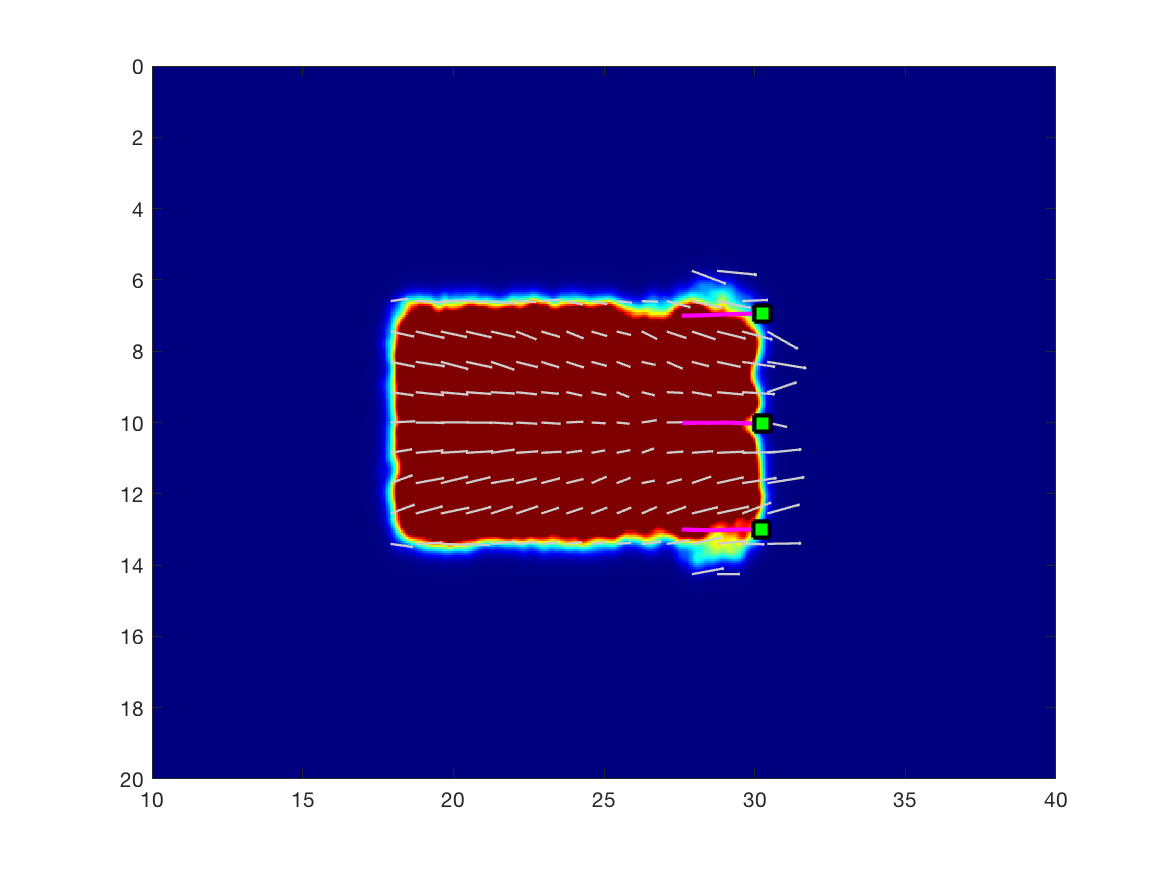}
\includegraphics[width=3.85cm]{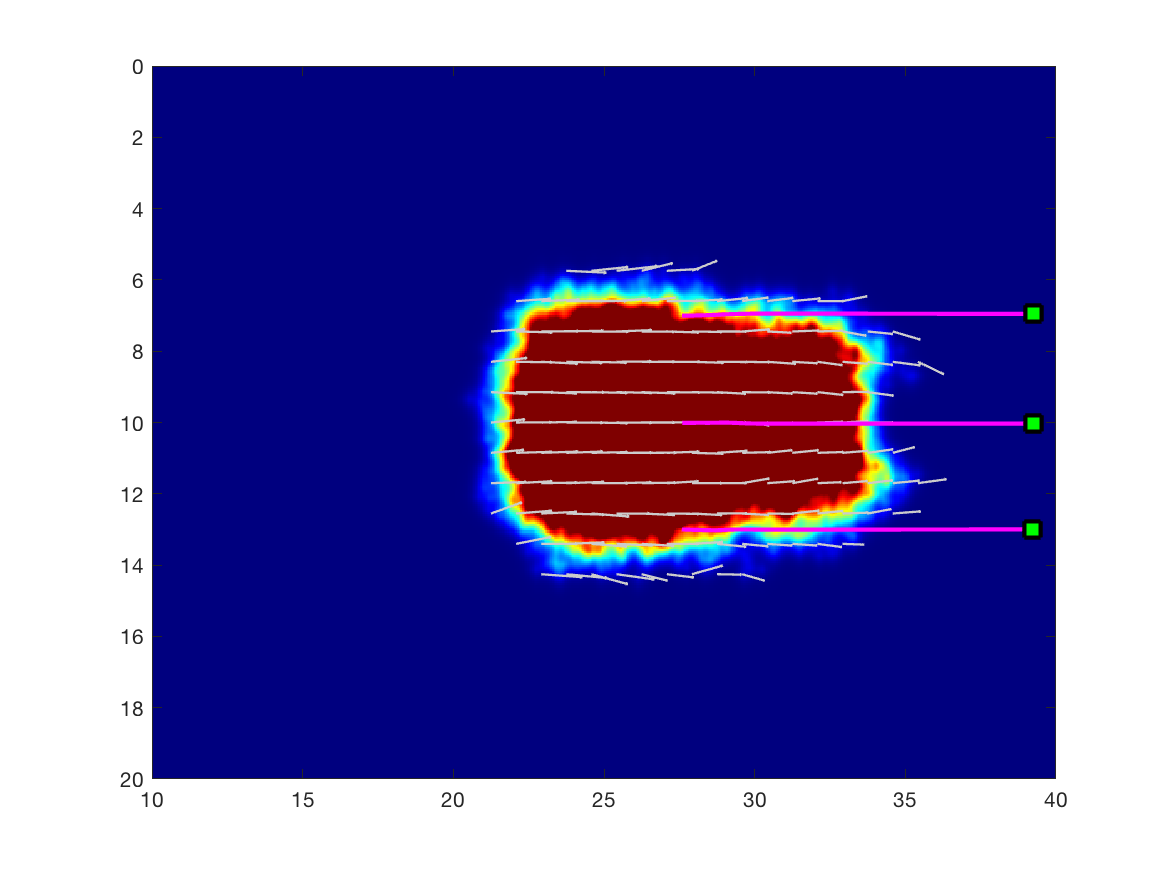}
\includegraphics[width=3.85cm]{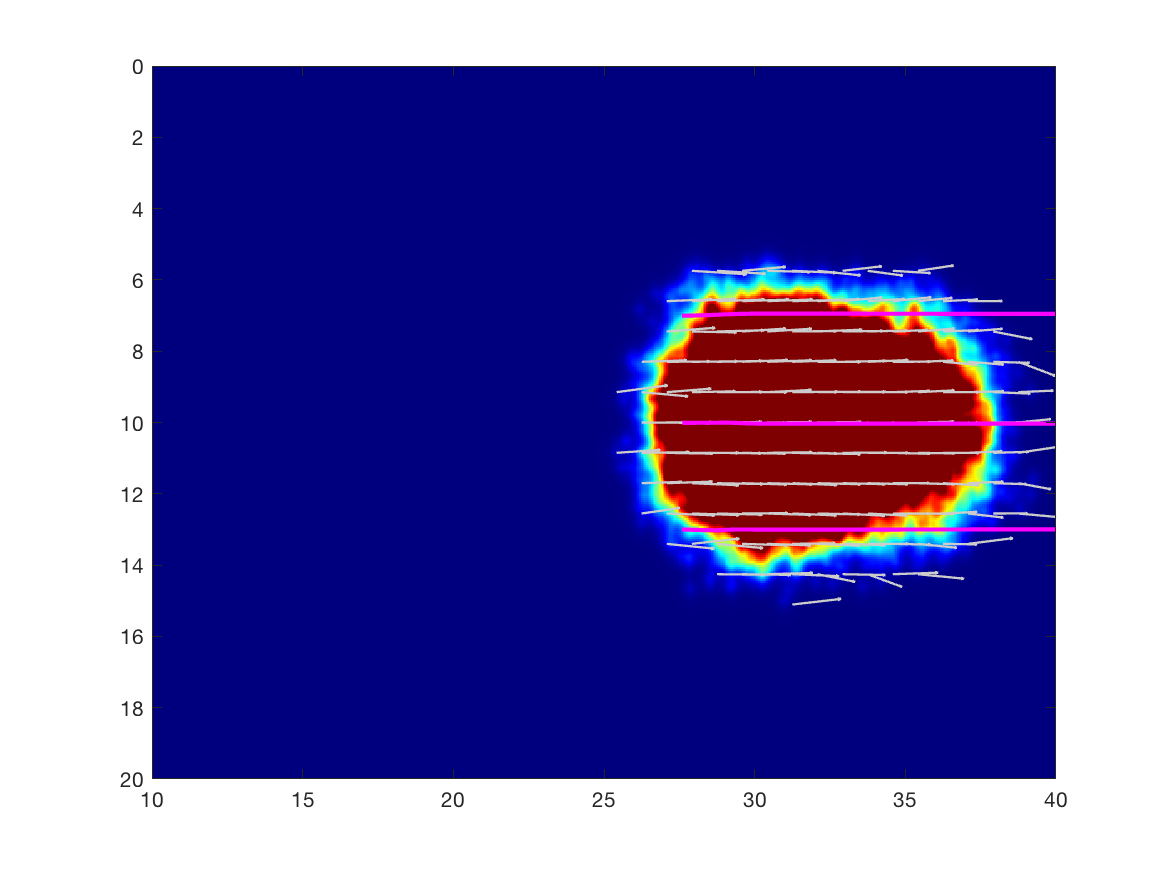}
\caption{{\em S\#1- Visible leaders}. Evolution of microscopic density guided by three visible leaders. Top row shows the microscopic system, bottom row the mesoscopic system.}\label{visible_vs_invisible1}
\end{center}
\end{figure}

In Figure \ref{visible_vs_invisible1} we observe the evolution of microscopic and mesoscopic models density for visible leaders and speed ${\texttt v} = 1.5$. In both cases is evident that the visibility plays a central role in attracting the whole crowd in the direction of leaders movement.

\begin{figure}[h!]
\begin{center}
\includegraphics[width=3.85cm]{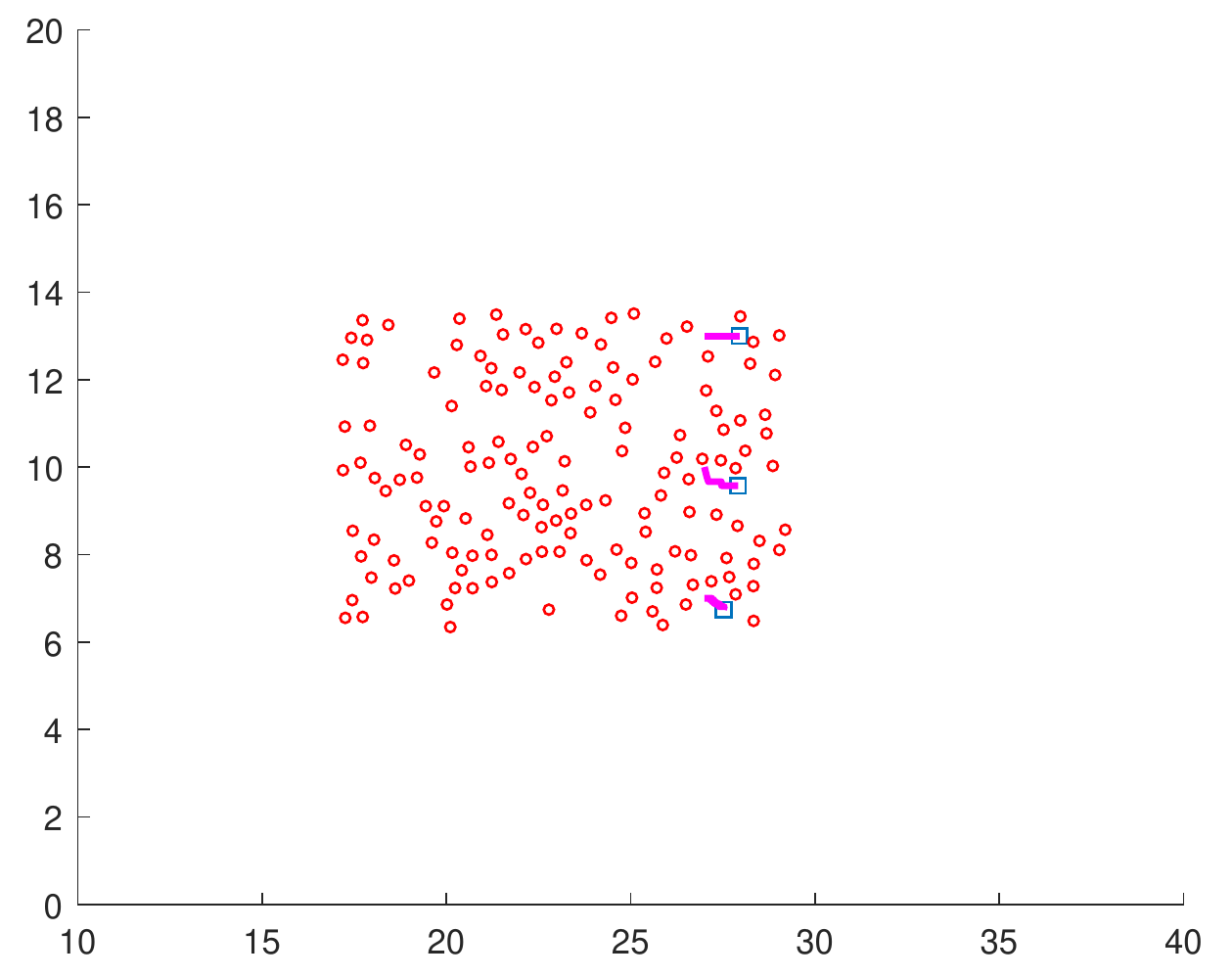}
\includegraphics[width=3.85cm]{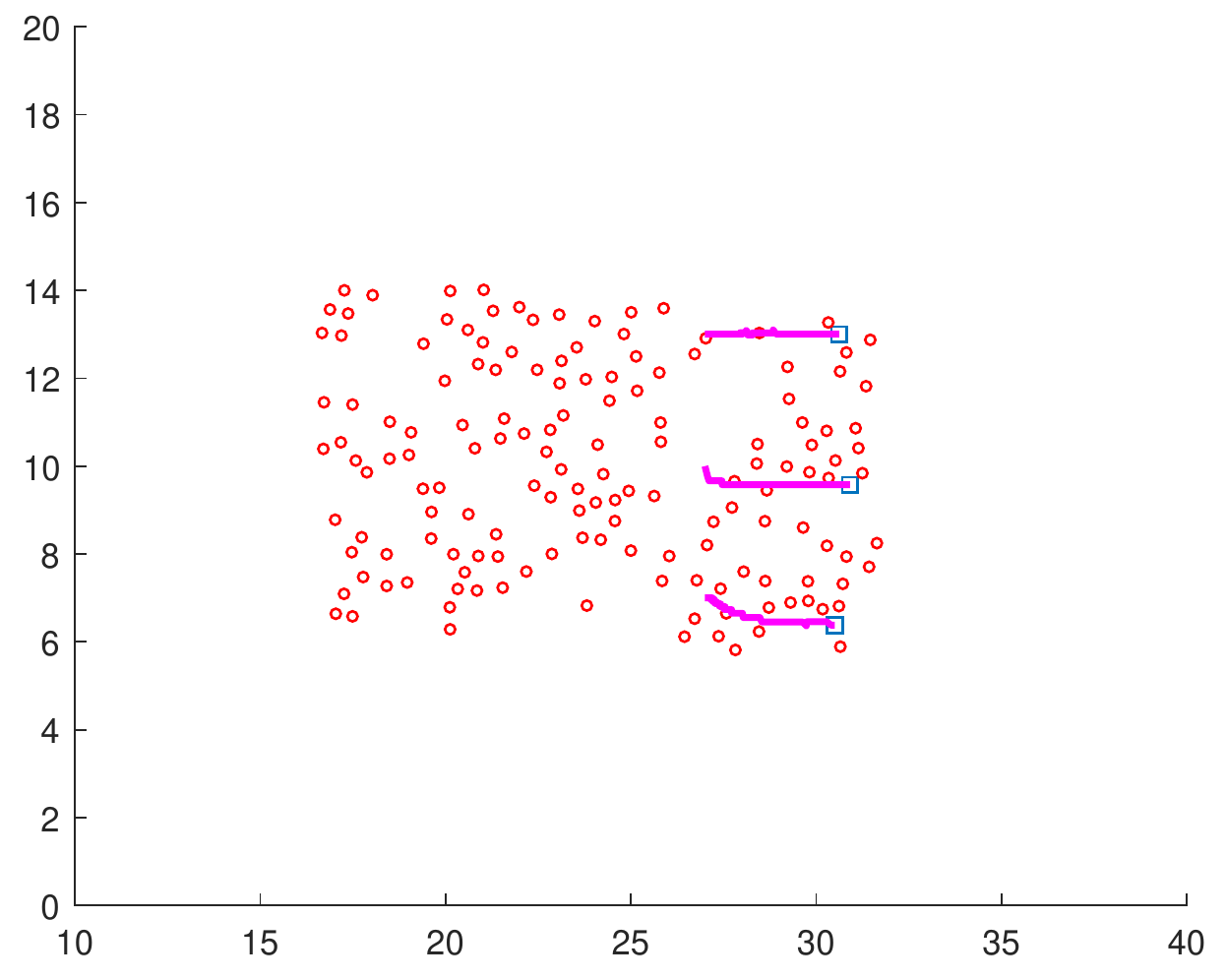}
\includegraphics[width=3.85cm]{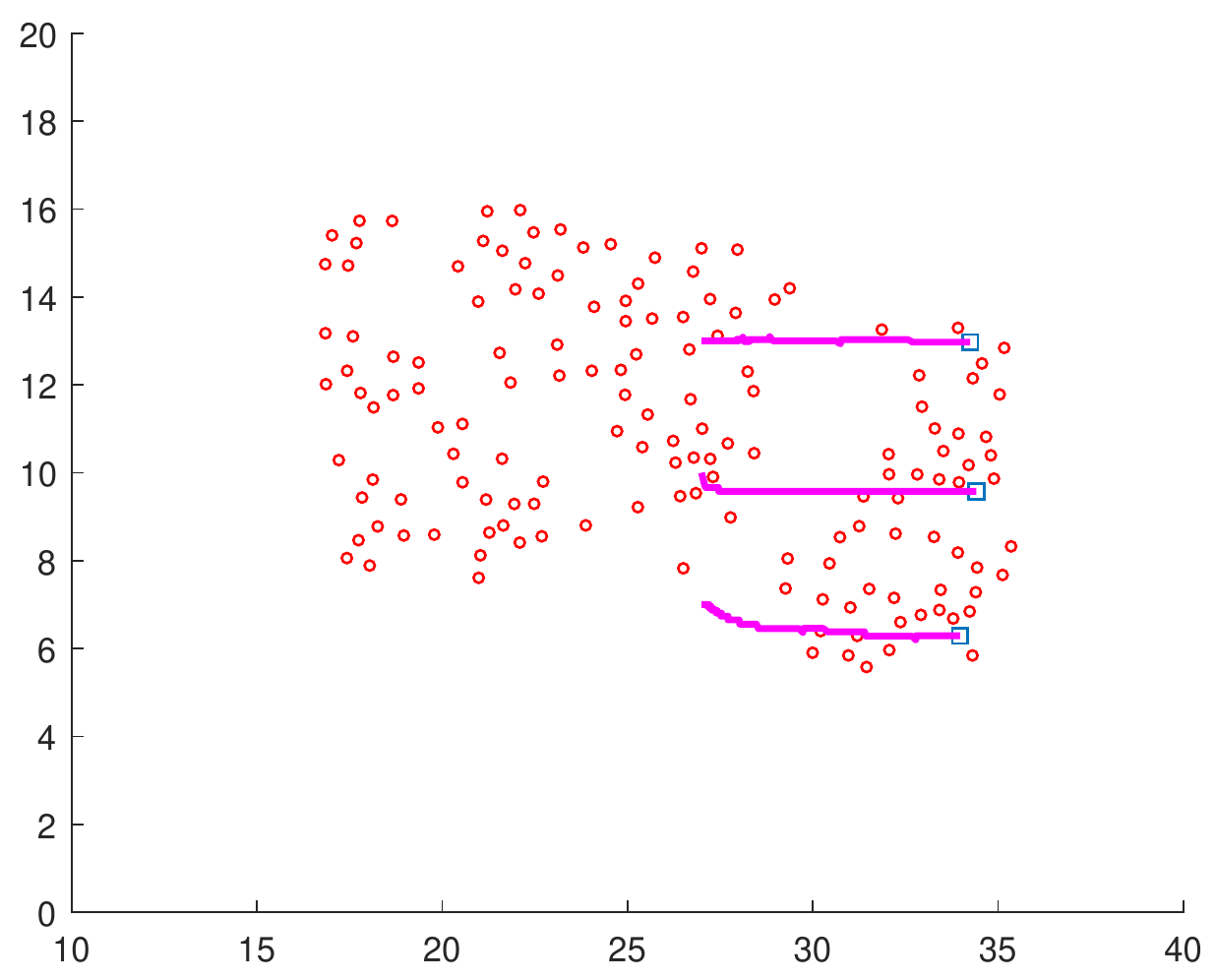}
\\
\includegraphics[width=3.85cm]{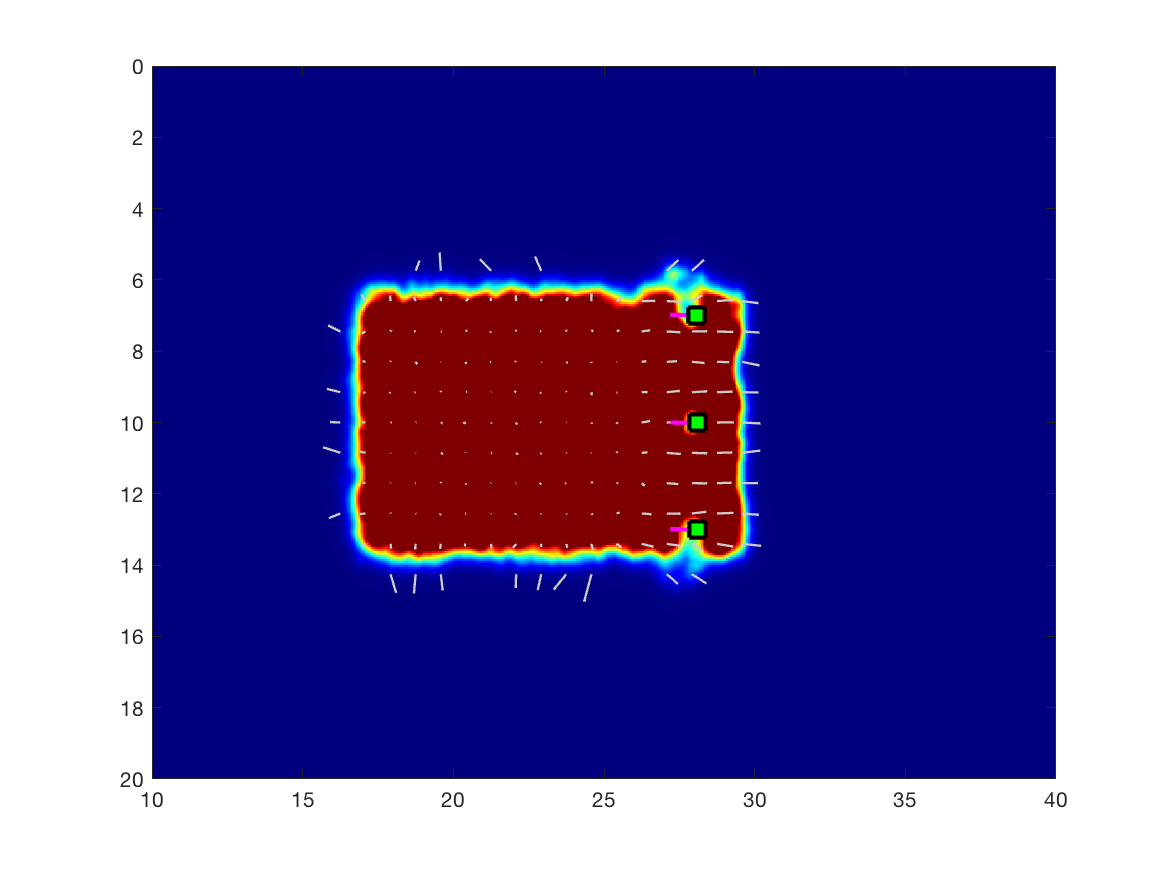}
\includegraphics[width=3.85cm]{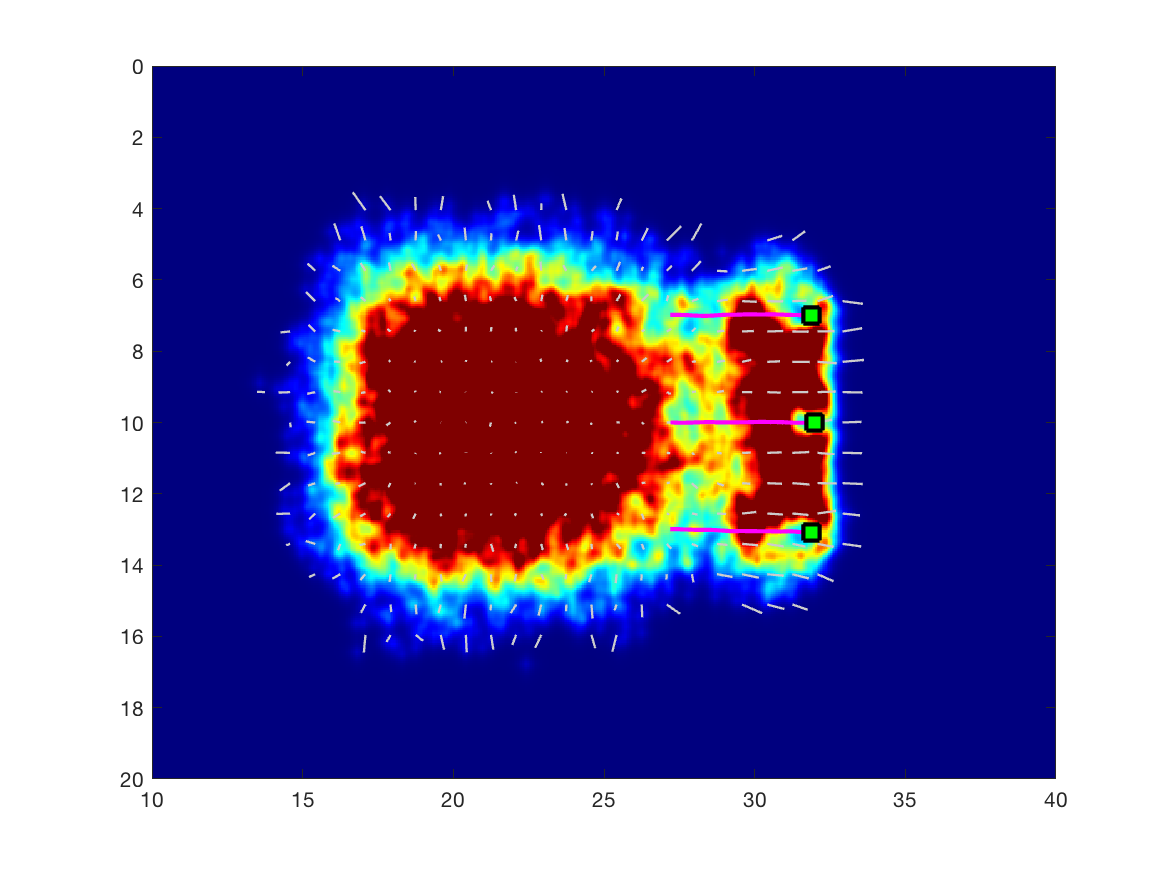}
\includegraphics[width=3.85cm]{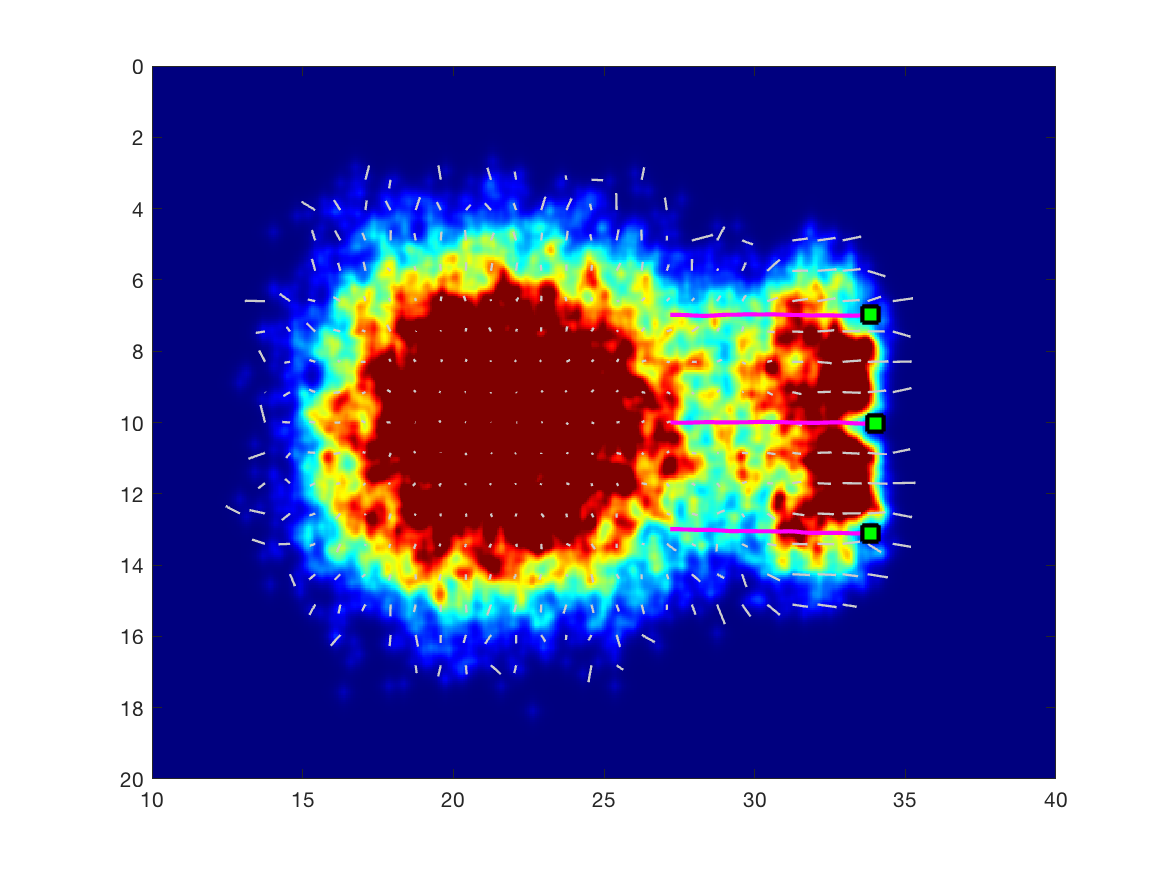}
\caption{{\em S\#1- Invisible leaders}. Evolution of microscopic density guided by three invisible leaders. Top row shows the microscopic system, bottom row the mesoscopic system.}\label{visible_vs_invisible2}
\end{center}
\end{figure}
In Figure \ref{visible_vs_invisible2} we observe the same situation with invisible leaders reducing the speed to ${\texttt v} = 0.5$, in order to let the leaders interact for longer time with the followers.  Indeed, in this case the followers are only partially influenced by the leaders, and only the mass close to them is driven towards the right direction, the remaining part spreads in the domain.

These experiments confirm that the action of invisible leaders is in general more subtle on the crowd influence, and determining effective strategies poses an additional challenge to the crowd control.

Moreover we can also infer that for the invisible case the initial positioning of leaders is of paramount importance to maximize their impact on the crowd dynamics.

\subsubsection{S\#2: Invisible leaders guiding a crowd}
We consider now the case of invisible leaders. We compare microscopic and mesoscopic framework and the evolution of the followers according to three strategies of the leaders: 
	\begin{itemize}
		\item \REV{no action (leaders behave as normal followers);}
		\item \REV{go-to-target (leaders point straight at the target);}
		\item \REV{optimized strategy (an optimization algorithm is used to find the optimal strategy with respect to some criterion).}
	\end{itemize}

Differently from S\#1 the crowd is now placed between the leaders and the exit. In this way leaders, moving to the exit, break more easily the initial uncertainty and triggers the crowd of followers toward the correct direction.
 
\begin{figure}[b!]
\centering
\includegraphics[width=0.3\textwidth]{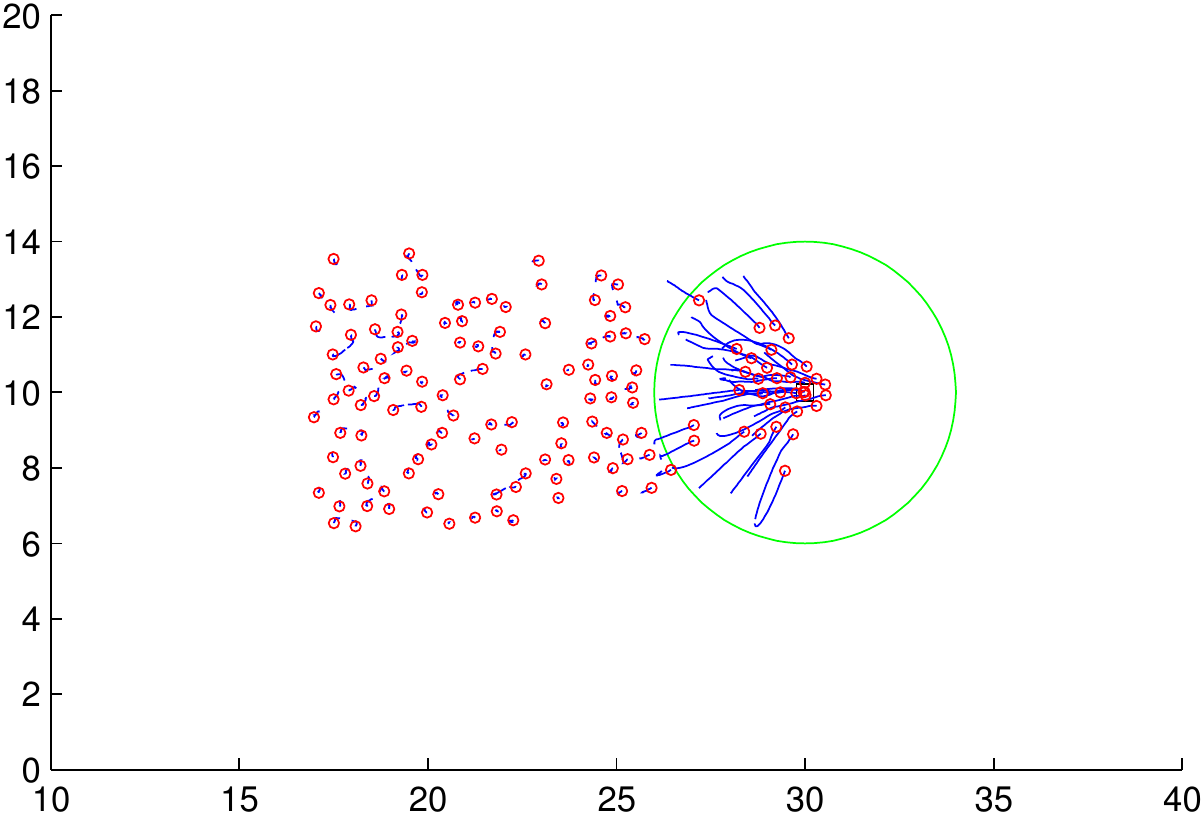}\quad
\includegraphics[width=0.3\textwidth]{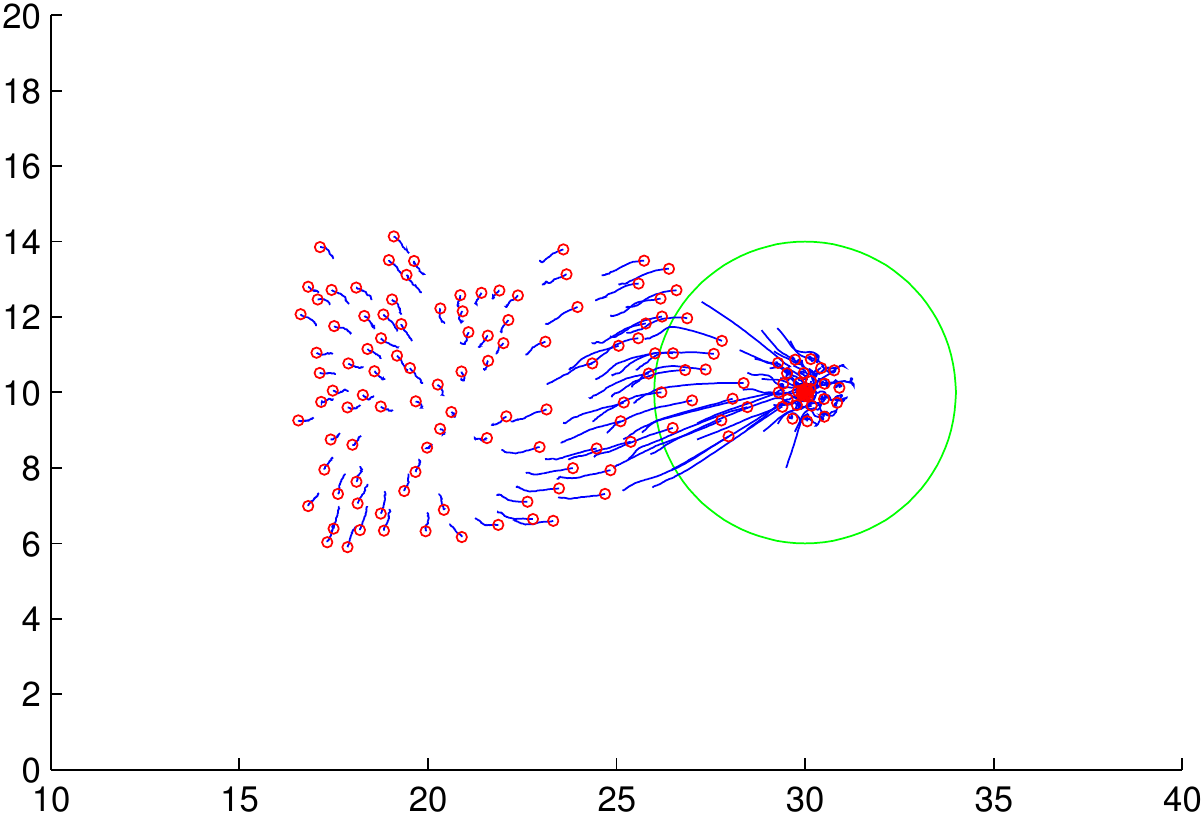}\quad
\includegraphics[width=0.3\textwidth]{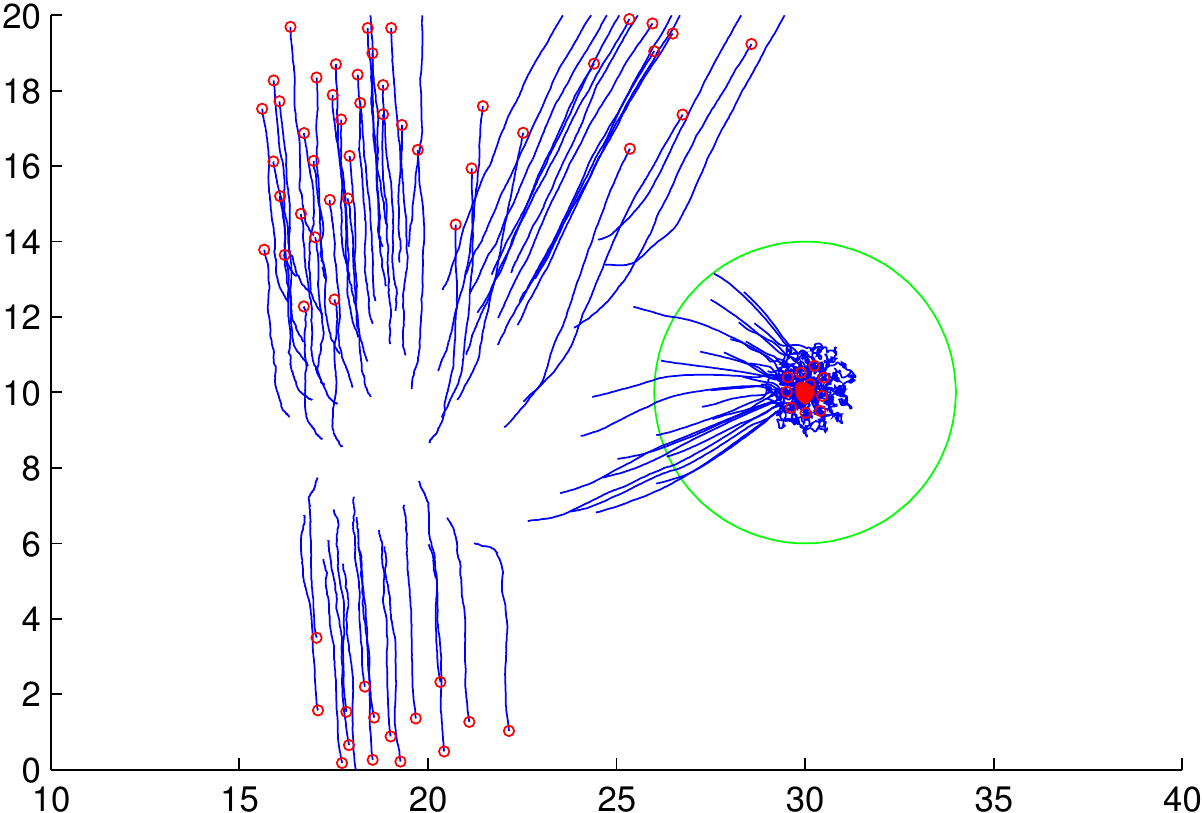}\\ [5mm]
\includegraphics[width=0.3\textwidth]{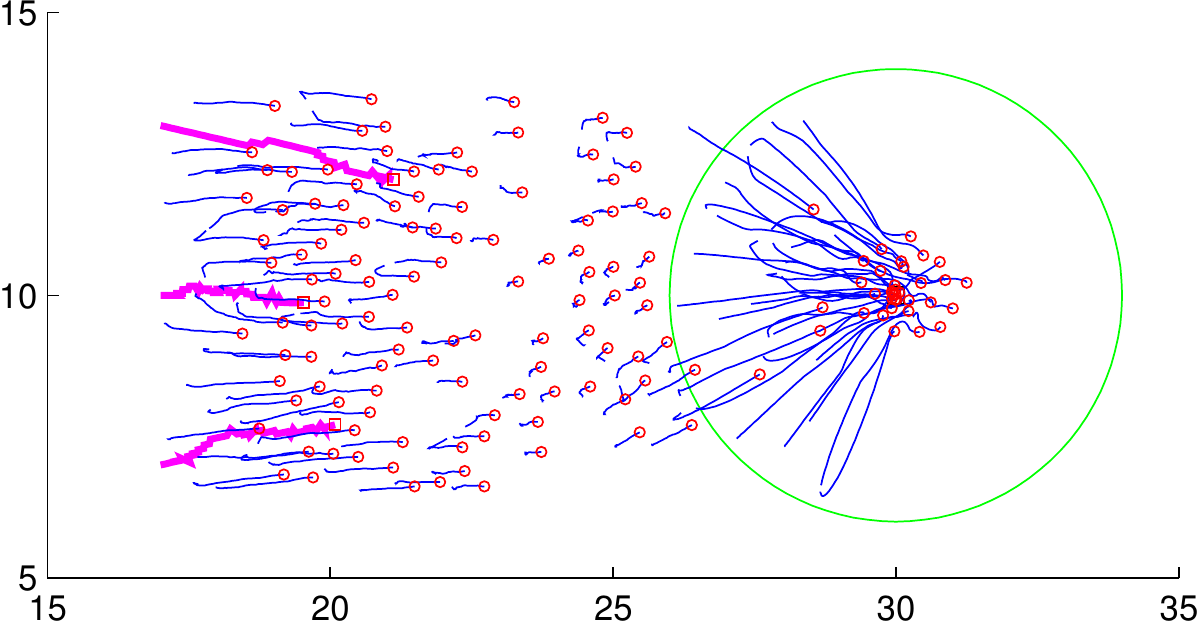}\quad
\includegraphics[width=0.3\textwidth]{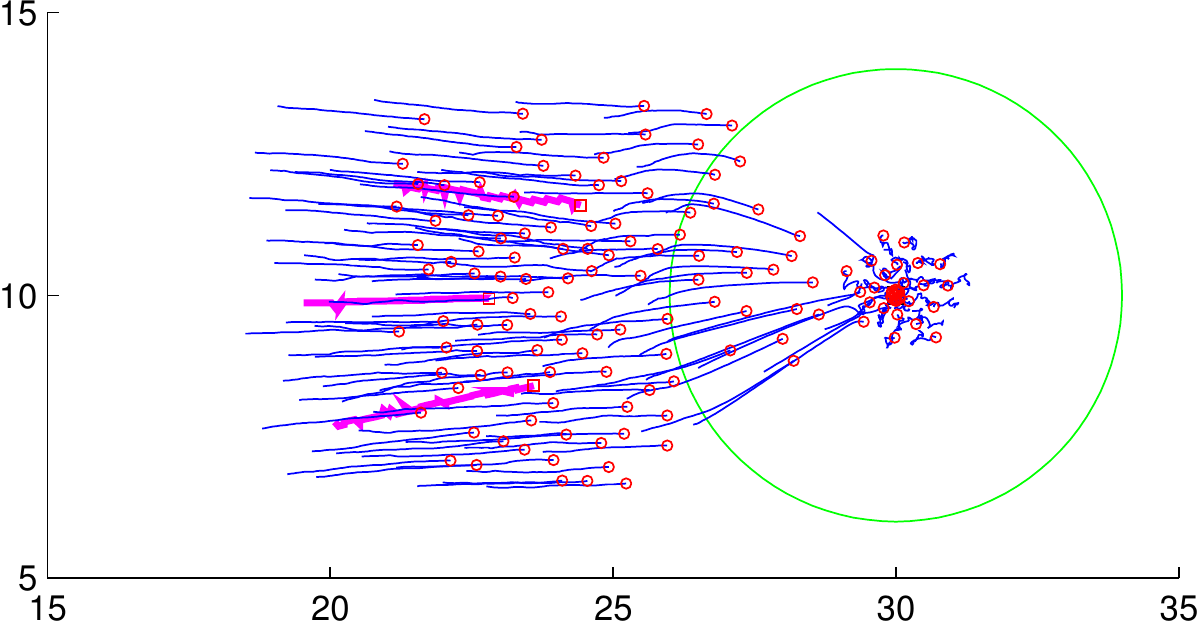}\quad
\includegraphics[width=0.3\textwidth]{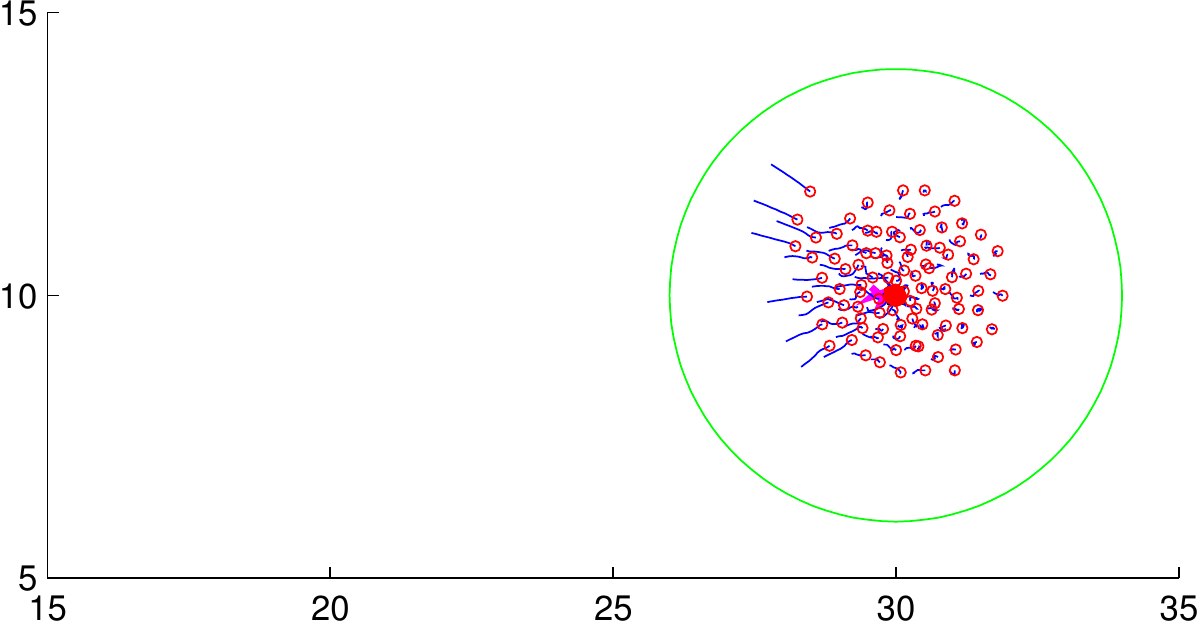}\\ [4mm]
\includegraphics[width=0.3\textwidth]{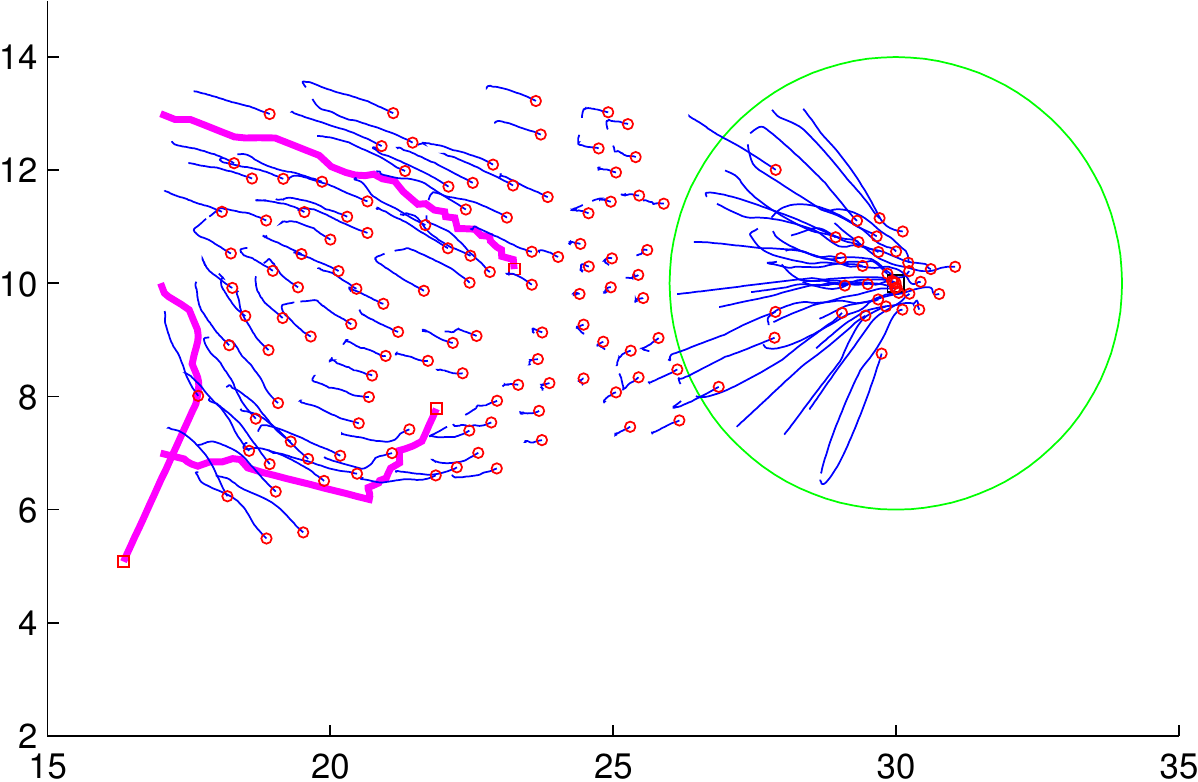}\quad
\includegraphics[width=0.3\textwidth]{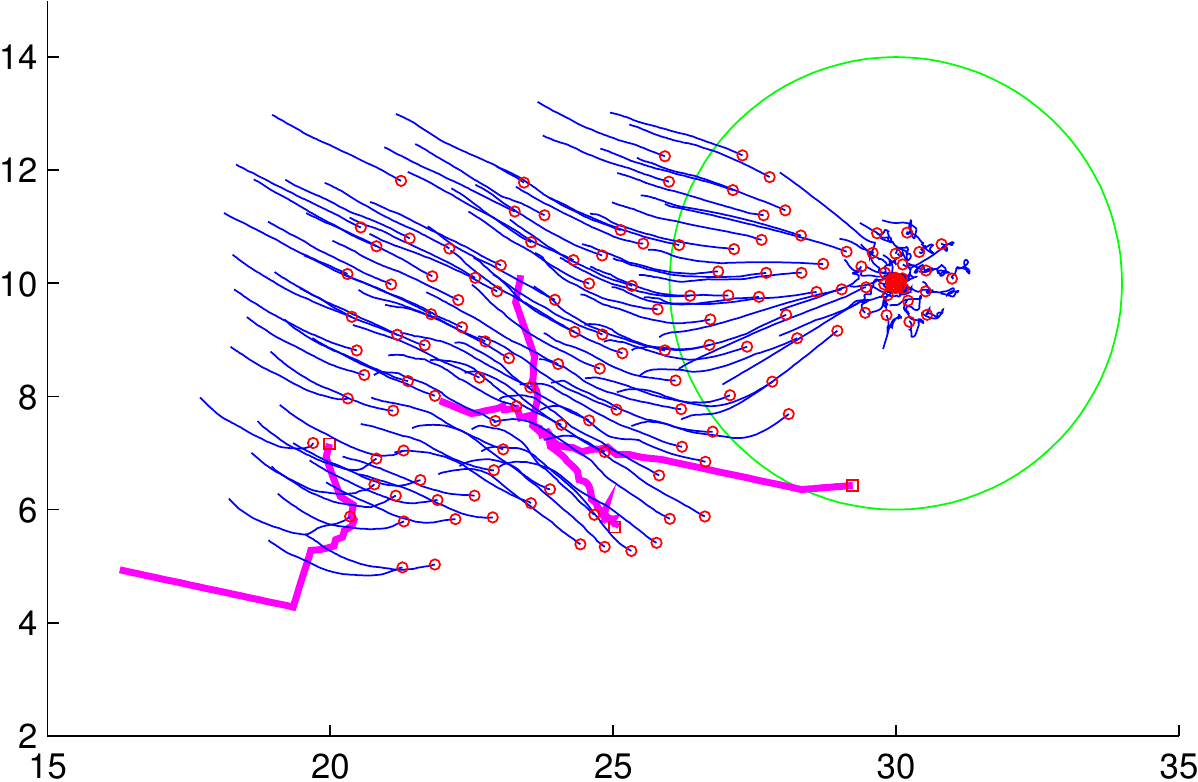}\quad
\includegraphics[width=0.3\textwidth]{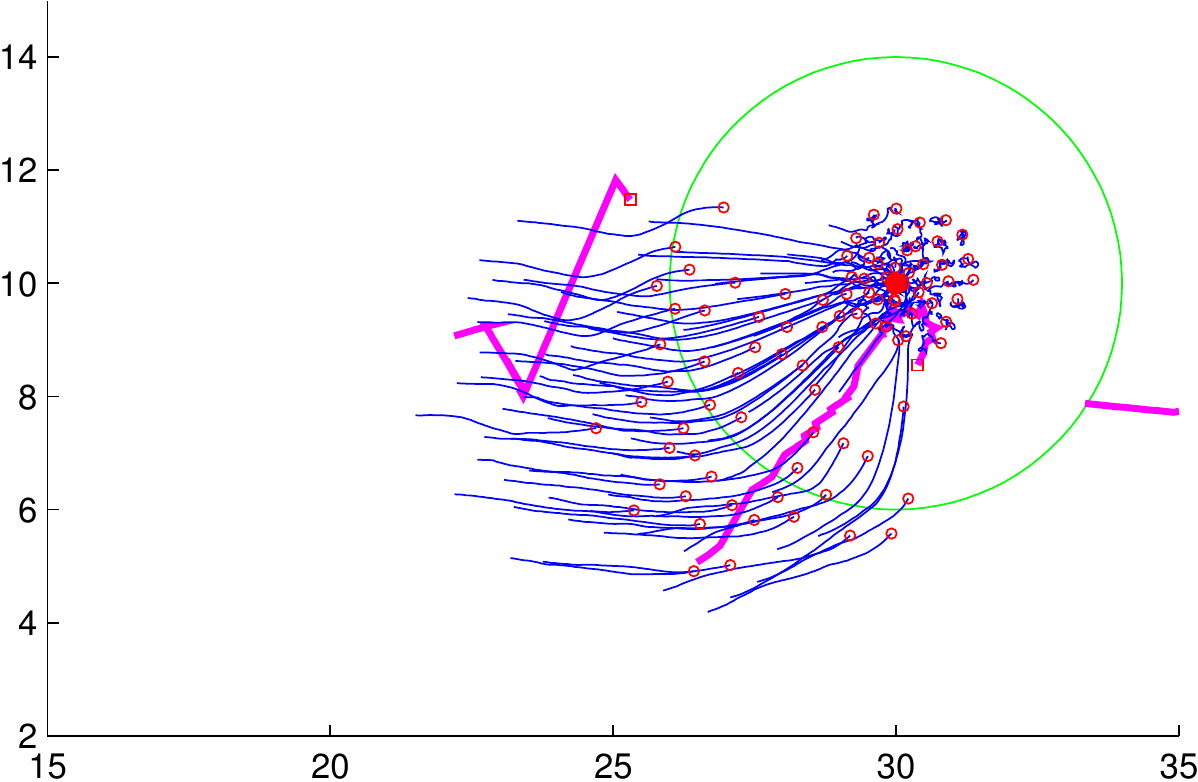}
\caption{{\em S\#2- Microscopic dynamics}. First row: no leaders. Second row: three leaders, go-to-target strategy. Third row: three leaders, optimal strategy (compass search).}
\label{fig:S1_micro}
\end{figure}

\paragraph{\em {Microscopic model}.}
Figure \ref{fig:S1_micro}(first row) shows the evolution of the agents computed by the microscopic model, without leaders. Followers having a direct view of the exit immediately point towards it, and some group mates close to them follow thanks to the alignment force. On the contrary, farthest people split in several but cohesive groups with random direction and never reach the exit. 

Figure \ref{fig:S1_micro}(second row) shows the evolution of the agents with three leaders. The leaders' strategy is defined manually. More precisely, at any time the control is equal to the unit vector pointing towards the exit from the current position (go-to-target strategy). 

The initial position of the leaders play a central role. Indeed having placed them on the left-side of the crowd, their motion generates a larger influence into the followers dynamics. 
This is in contrast with the behavior observed in Figure \ref{visible_vs_invisible1}: where only a small portion of the pedestrians was triggered by the leaders positioned on the right-hand side of the crowds.

Note also that the final leaders' trajectories are not straight lines because of the additional repulsion force. As it can be seen, the crowd behavior changes completely since, this time, the whole crowd reaches the exit. However followers form a heavy congestion around the exit. It is interesting to note that the shape of the congestion is circular: this is in line with the results of other social force models as well as physical observation, which report the formation of an ``arch'' near the exits. The arch is correctly substituted here by a full circle due to the absence of walls. Note that the congestion notably delays the evacuation. This suggests that the strategy of the leaders is not optimal and can be improved by an optimization method.

Fig.\ \ref{fig:S1_micro}(third row) shows the evolution of the agents with three leaders and the optimal strategy obtained by the compass search algorithm. Surprisingly enough, the optimizator prescribes that leaders \emph{divert} some pedestrians from the right direction, so as not to steer the whole crowd to the exit at the same time. In this way congestion is avoided and pedestrian flow through the exit is increased.

In this test we have also run the MPC optimization, including a box constraint $u_k(t)\in[-1,1]$. 
We choose $C_1=1$, and $C_2=C_3=10^{-5}$.  MPC results are consistent in the sense that for $N_{\texttt{mpc}}=2$, the algorithm recovers a controlled behavior similar to the application of the instantaneous controller (or go-to-target strategy). Increasing the time frame up to $N_{\texttt{mpc}}=6$ improves both congestion and evacuation times, but results still remain non competitive if compared to the whole time frame optimization performed with a compass search. 

\begin{figure}[h!]
\centering
\ \includegraphics[scale=0.4]{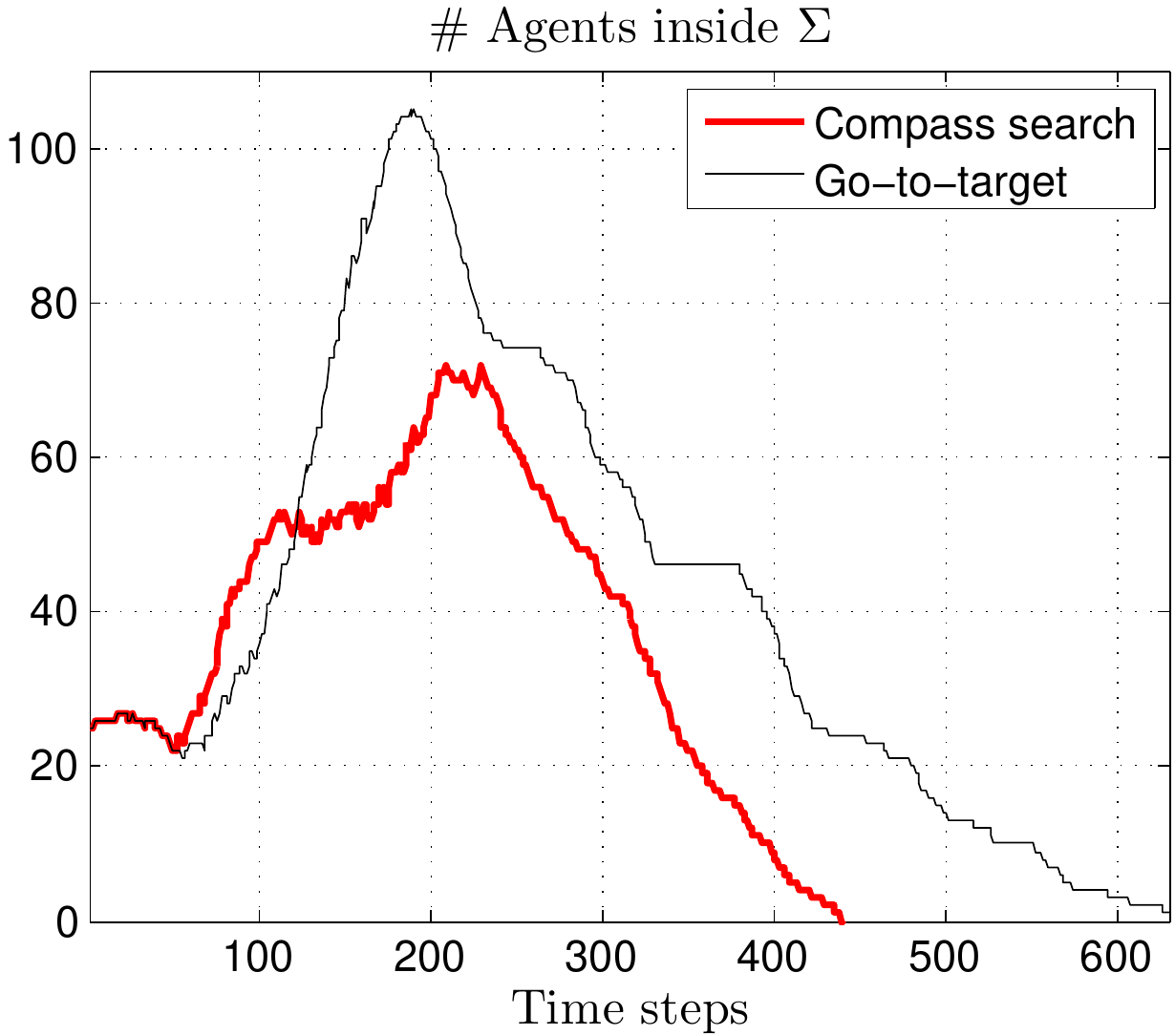}\qquad \ \
\includegraphics[scale=0.4]{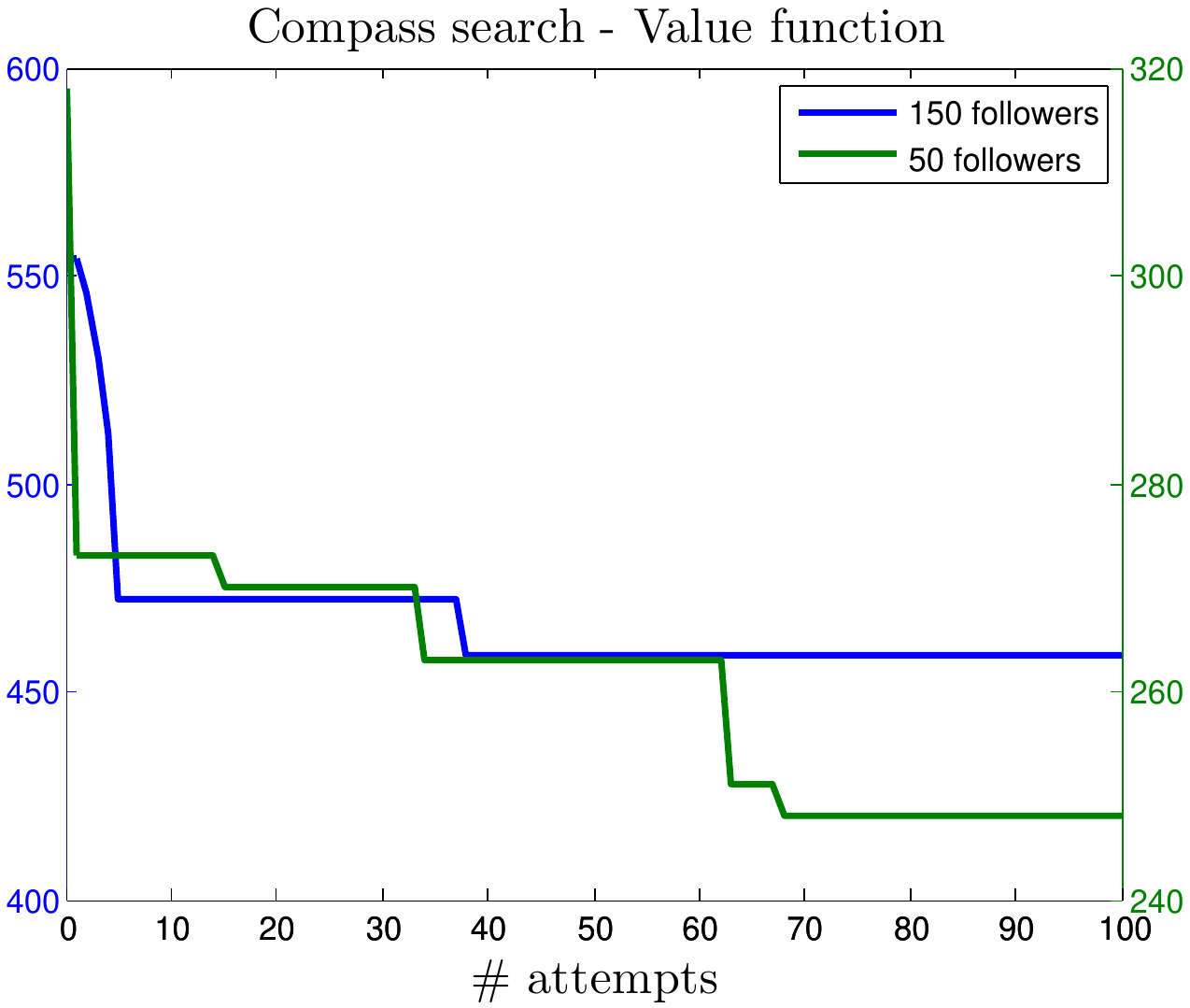}\vskip0.2cm
\includegraphics[scale=0.39]{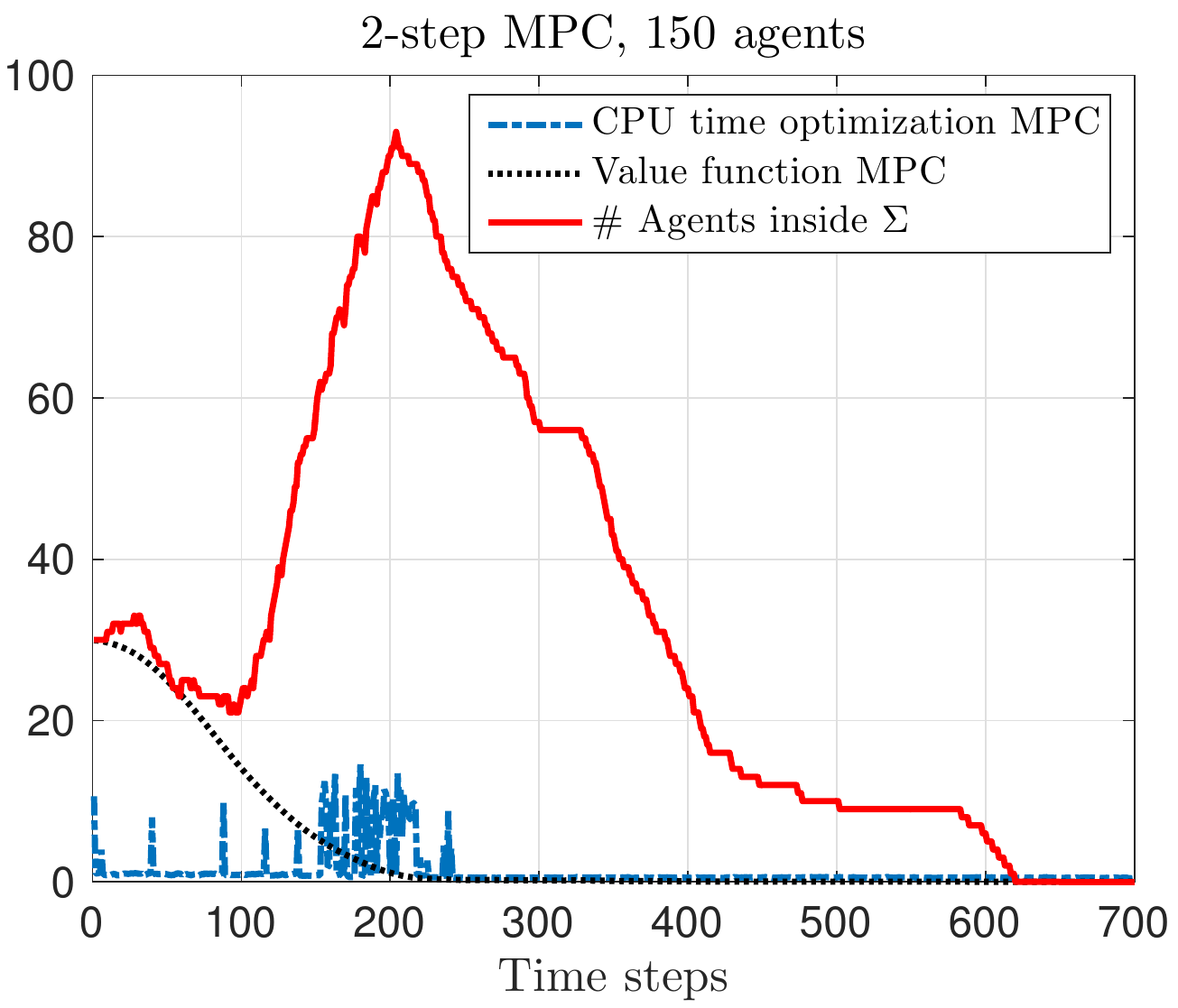}\qquad
\includegraphics[scale=0.39]{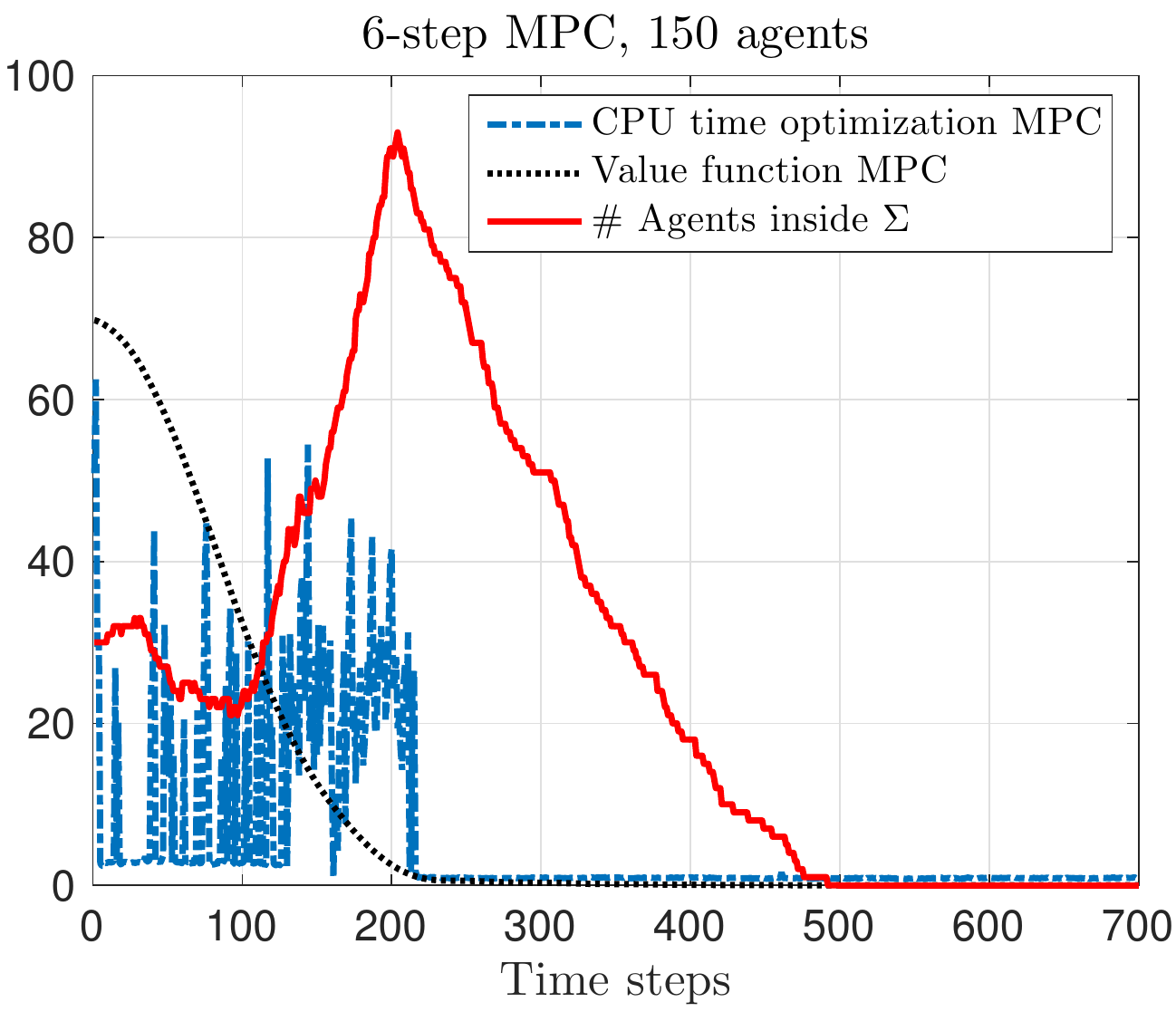}
\caption{{\em S\#2-Microscopic dynamics}. Optimization of the microscopic dynamics. Top-left: occupancy of the exit's visibility zone $\Sigma$ as a function of time for optimal strategy (compass search) and go-to-target strategy. Top-right: decrease of the value function \eqref{defJ_evactime} as a function of the iterations of the compass search (for 50 and 150 followers). Bottom: MPC optimization. occupancy of the exit's visibility zone $\Sigma$ as a function of time, CPU time of the optimization call embedded in the MPC solver, and the evolution of the corresponding value (2-step and 6-step).}
\label{fig:S1_hystogram}
\end{figure}

In Fig.\ \ref{fig:S1_hystogram} we compare the occupancy of the exit's visibility zone as a function of time for go-to-target strategy and optimal strategies (compass search, 2-step, and 6-step MPC). We also show the decrease of the value function as a function of attempts (compass search) and time (MPC). Evacuation times are compared in Table \ref{tab:S1_evactimes}. It can be seen that only the long-term optimization strategies are efficient, being able to moderate congestion and clogging around the exit.

\begin{table}[!h]
\caption{S\#1. Evacuation times (time steps). CS=compass search, IG=initial guess.}
\label{tab:S1_evactimes}
\begin{center}
\begin{tabular}{|c|c|c|c|c|c|}
\hline
 & no leaders & go-to-target & 2-MPC & 6-MPC & CS (IG)\\
\hline
$\NF=50$ & 335 & 297 & 342 & 278 & 248 (318)\\ \hline
$\NF=150$ & $\infty$ & 629 & 619 & 491 & 459 (554)\\
\hline
\end{tabular}
\end{center}
\end{table}
This suggests a quite unethical but effective evacuation procedure, namely misleading some people to a false target and then leading them back to the right one, when exit conditions are safer. Note that in real-life situations, most of the injuries are actually caused by overcompression and suffocation rather than urgency.  

\paragraph{\em Mesoscopic model}

\begin{figure}[b!]
\centering
\includegraphics[width=0.3\textwidth]{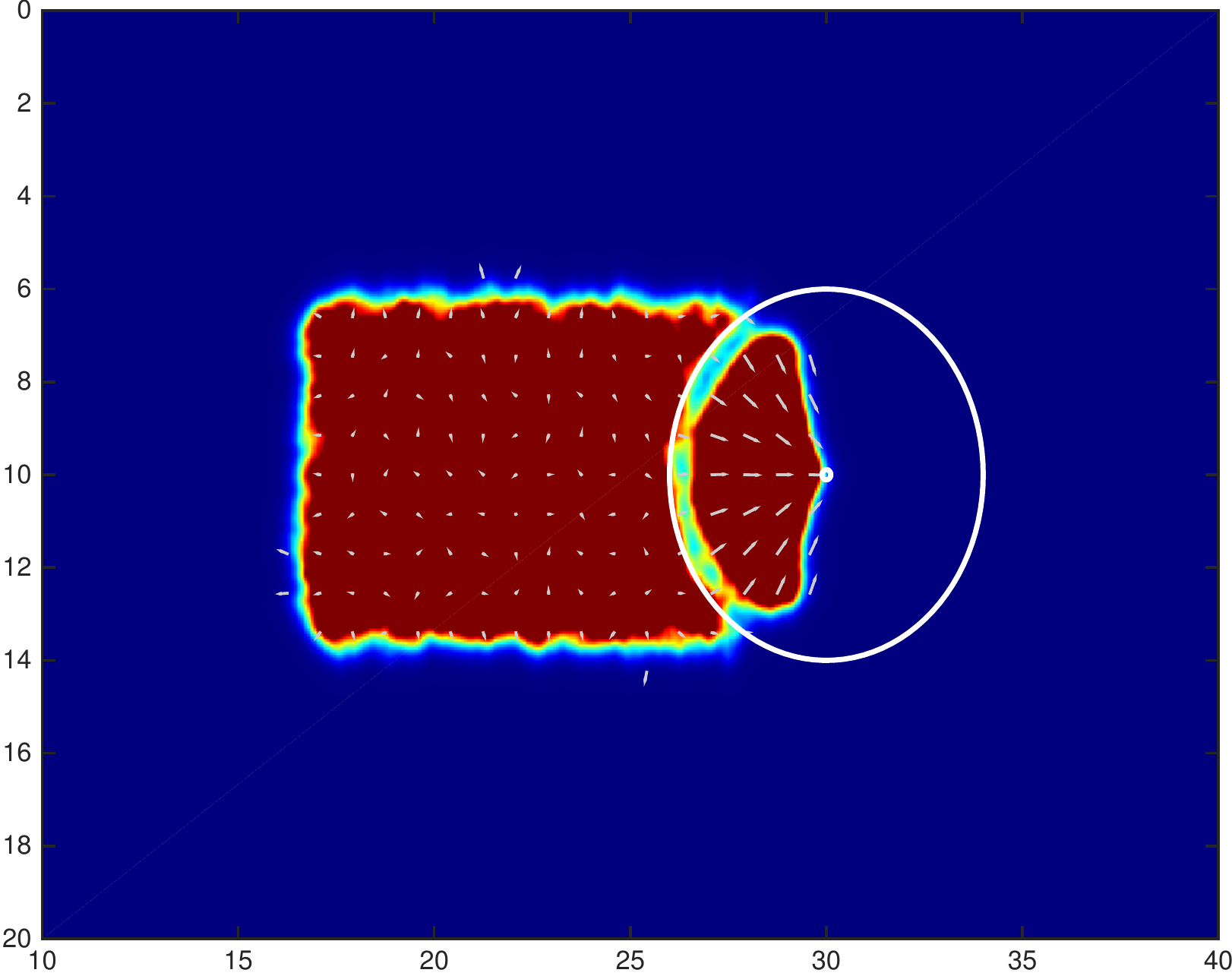}\quad
\includegraphics[width=0.3\textwidth]{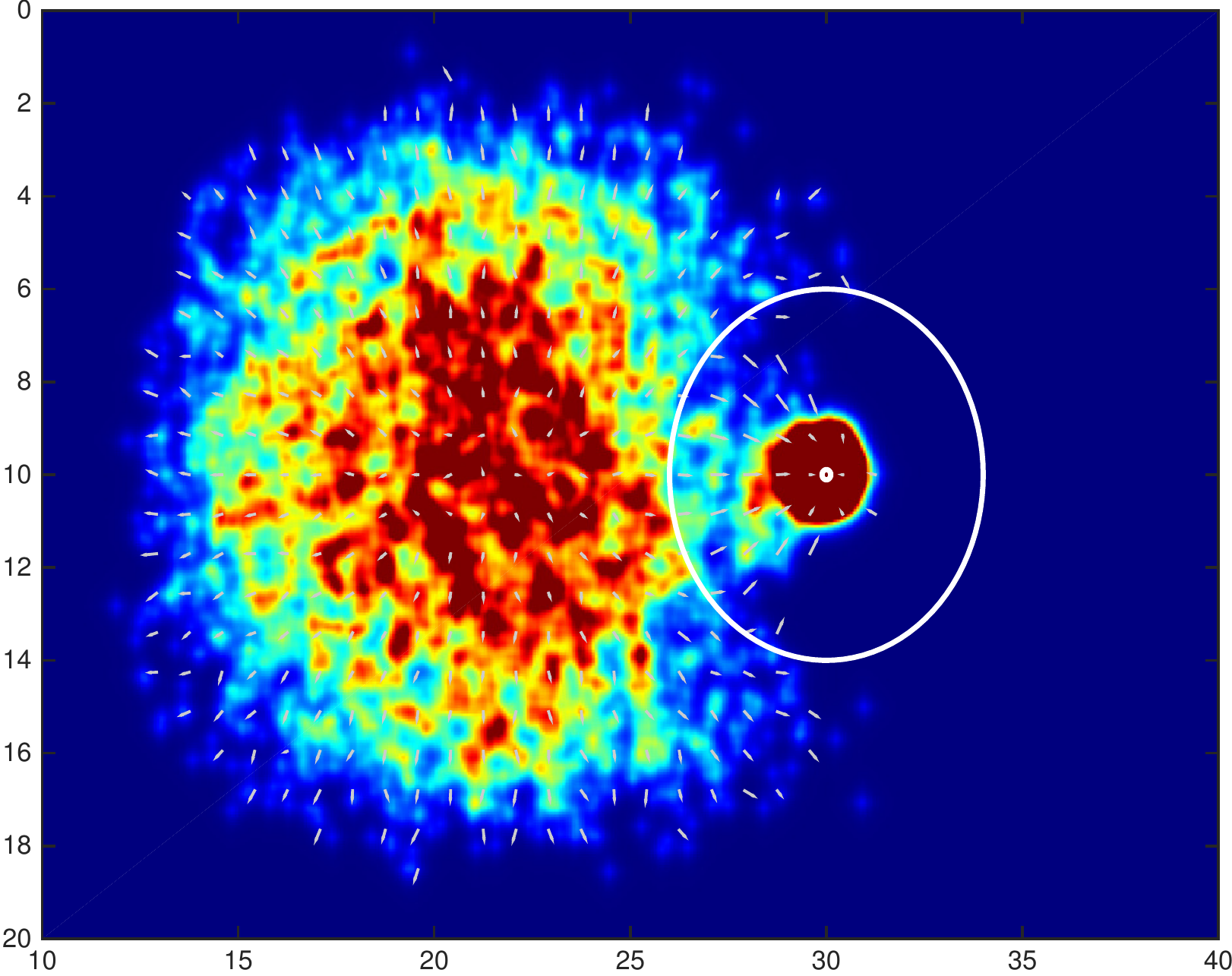}\quad
\includegraphics[width=0.3\textwidth]{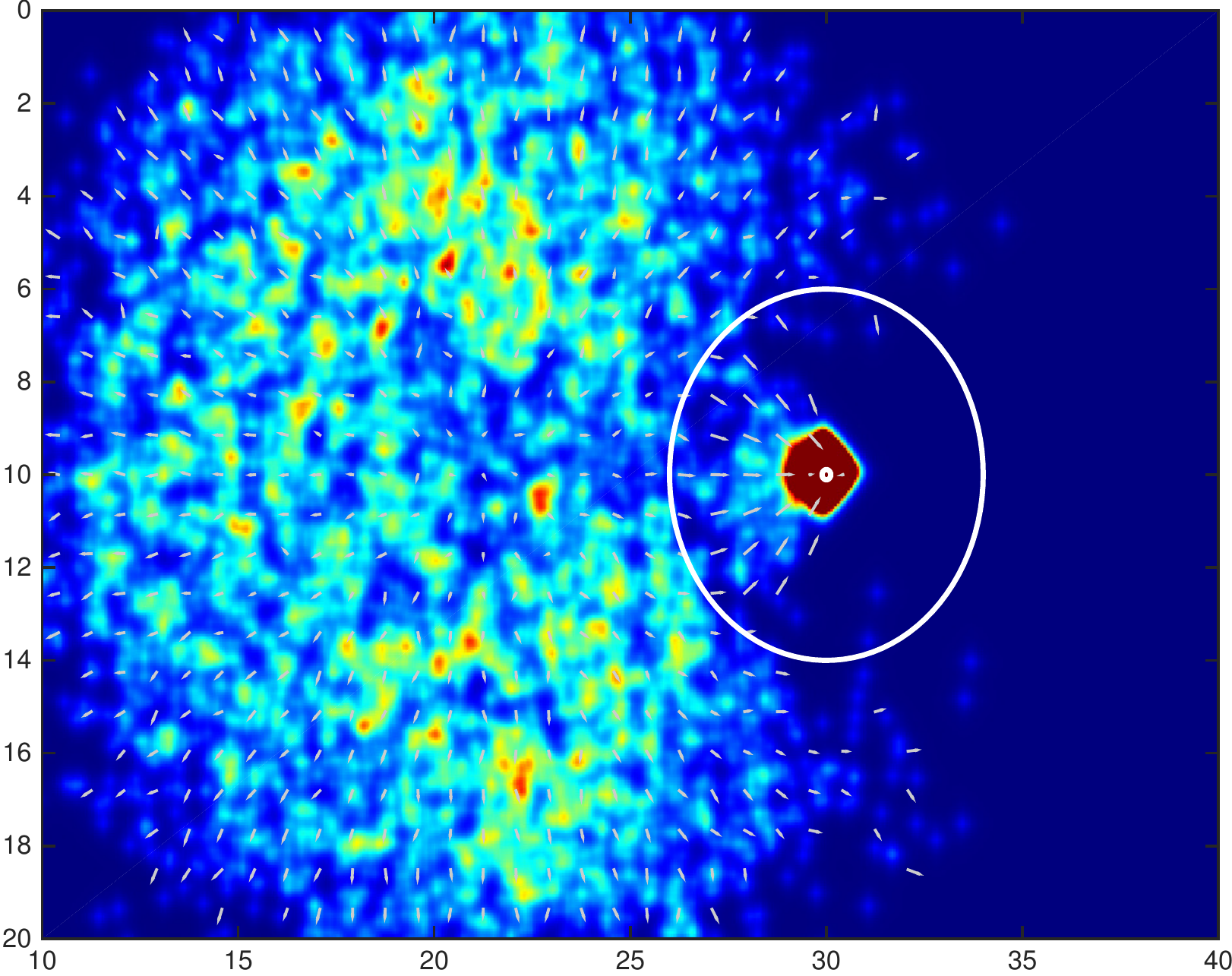}\\
\includegraphics[width=0.3\textwidth]{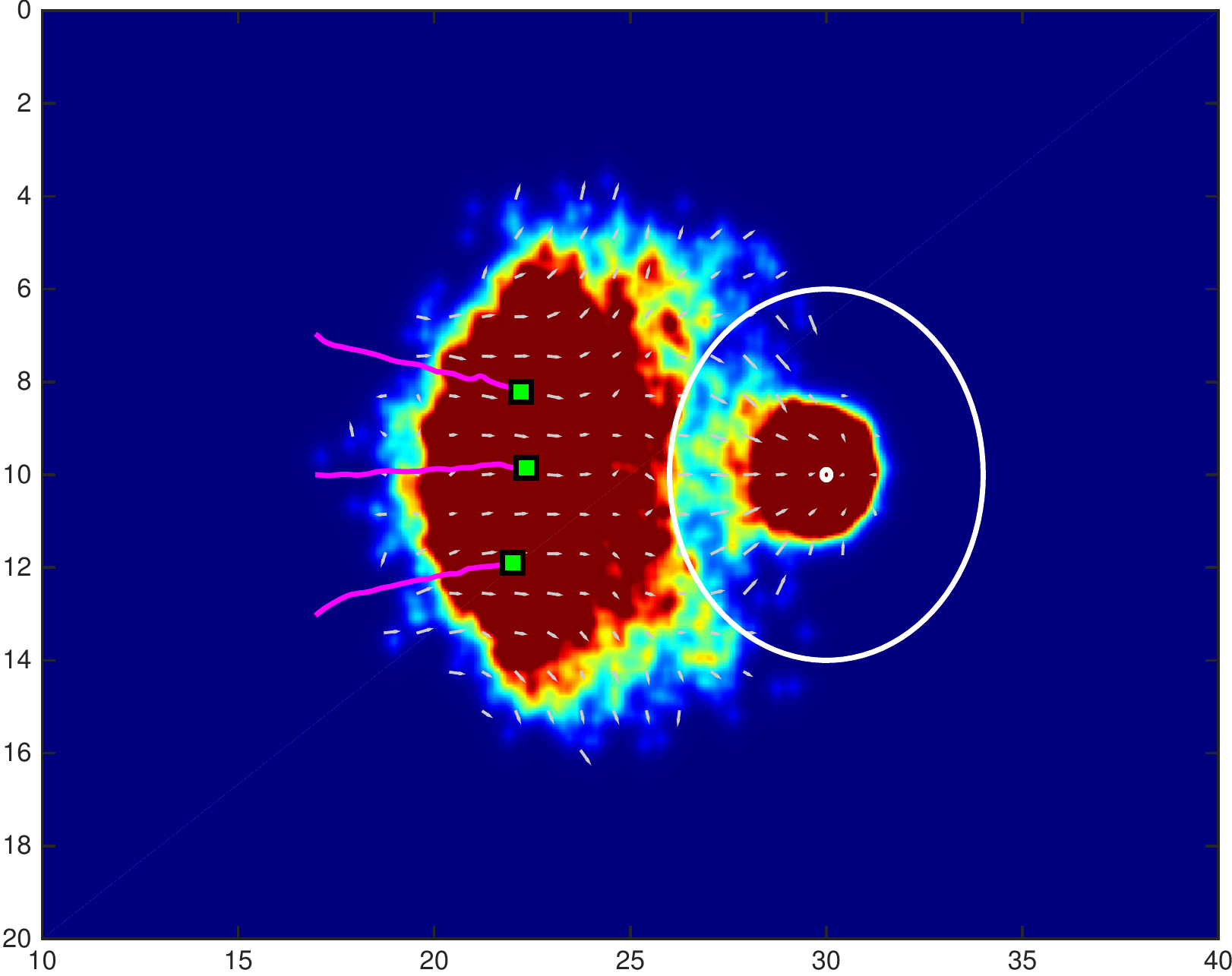}\quad
\includegraphics[width=0.3\textwidth]{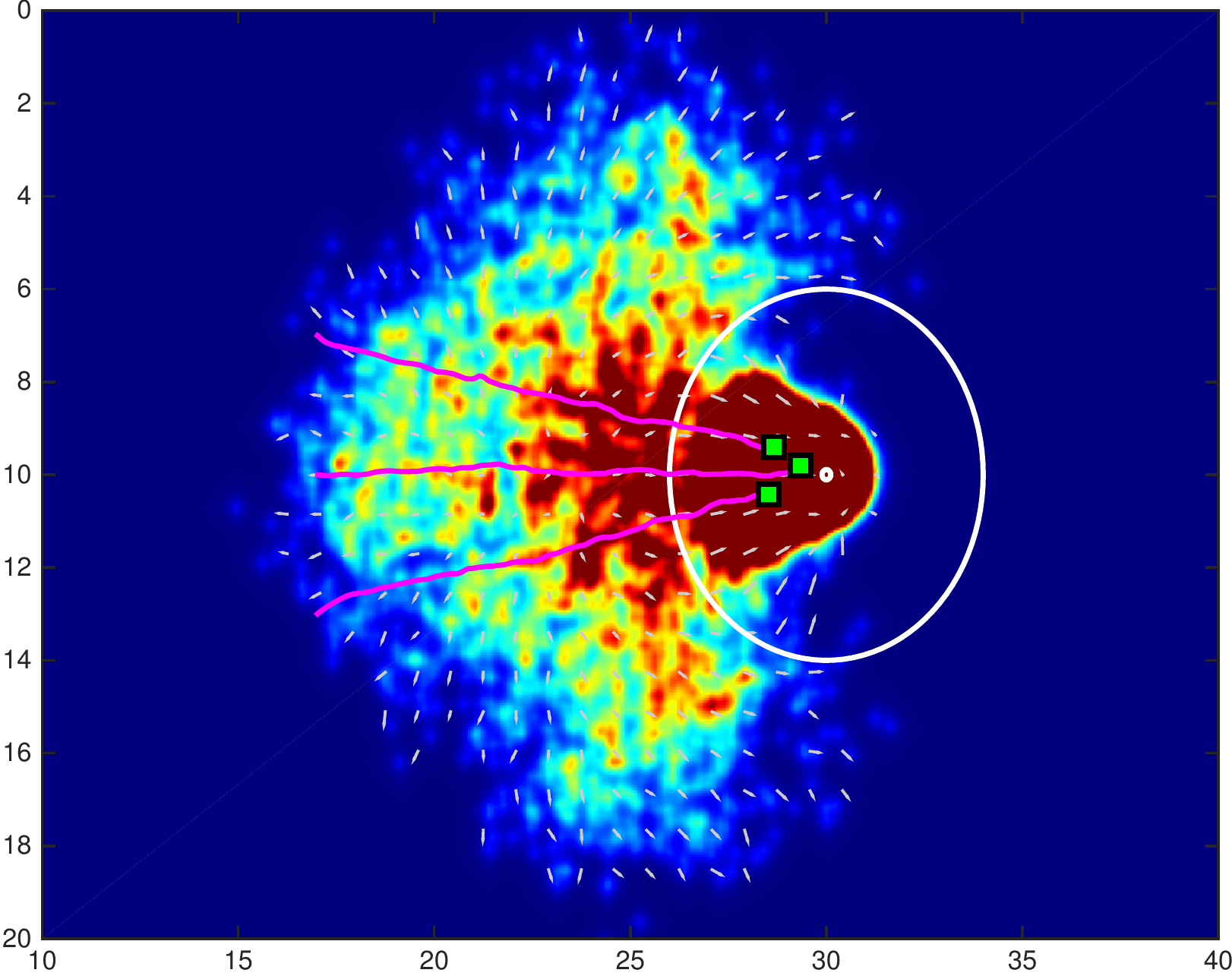}\quad
\includegraphics[width=0.3\textwidth]{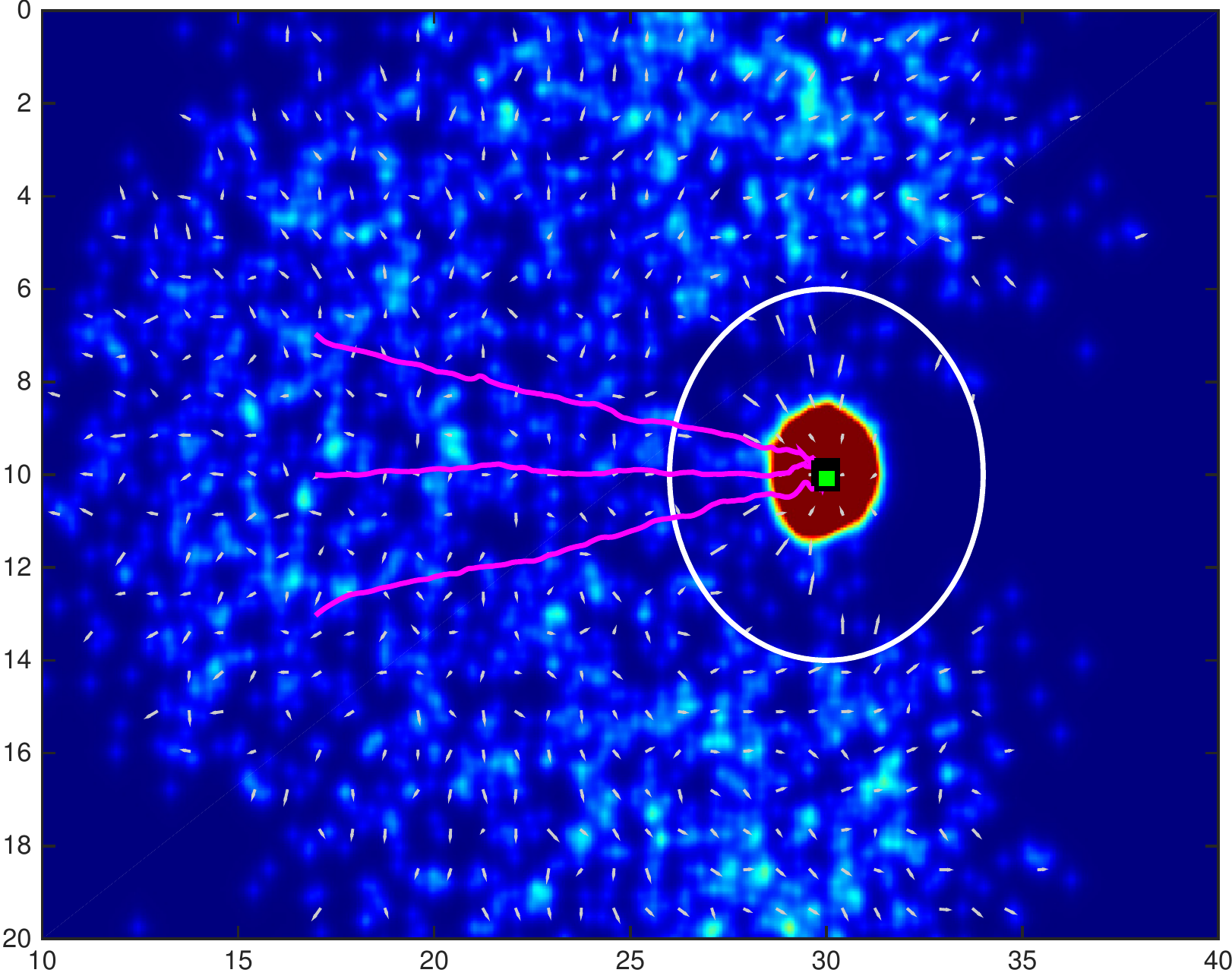}\\
\includegraphics[width=0.3\textwidth]{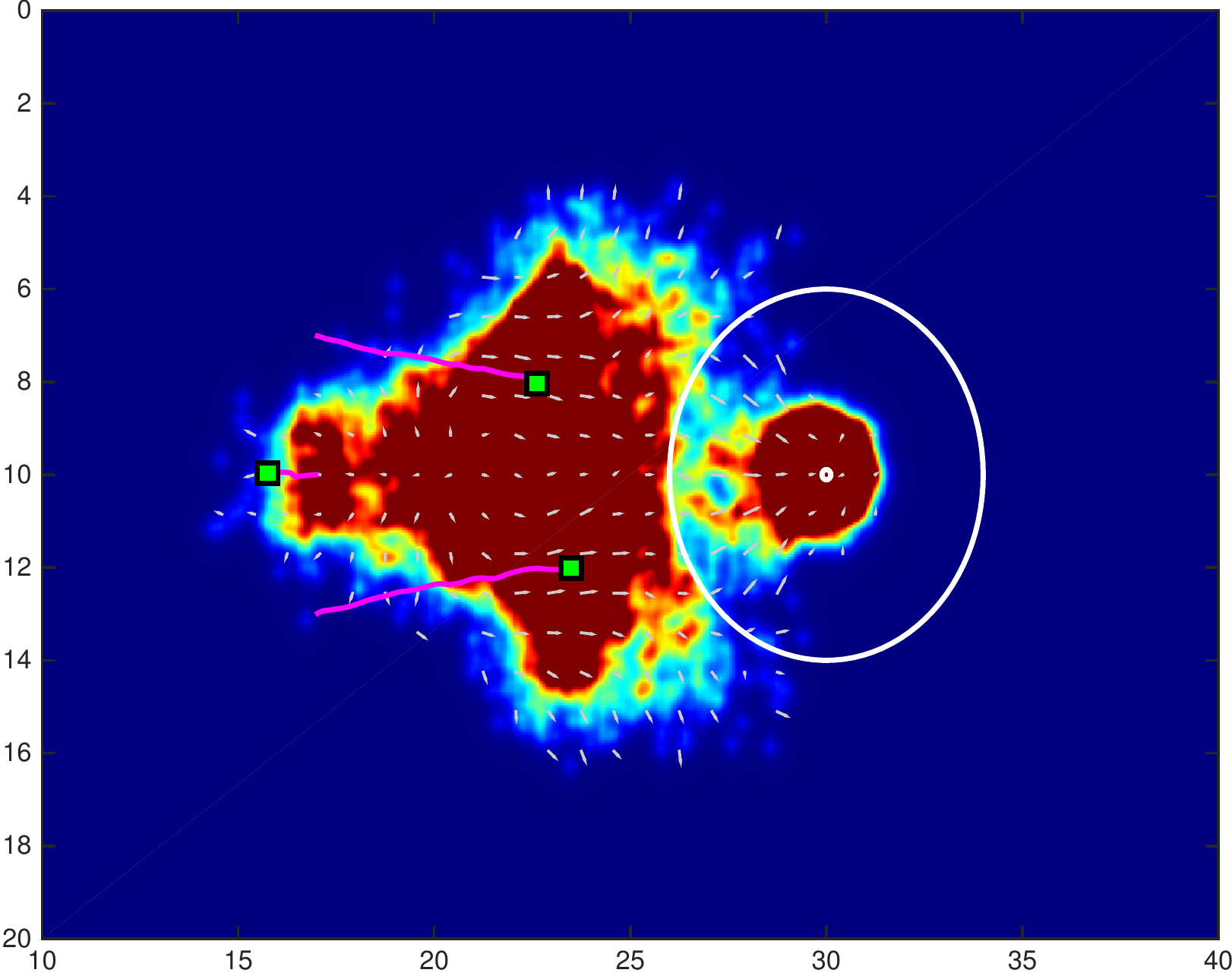}\quad
\includegraphics[width=0.3\textwidth]{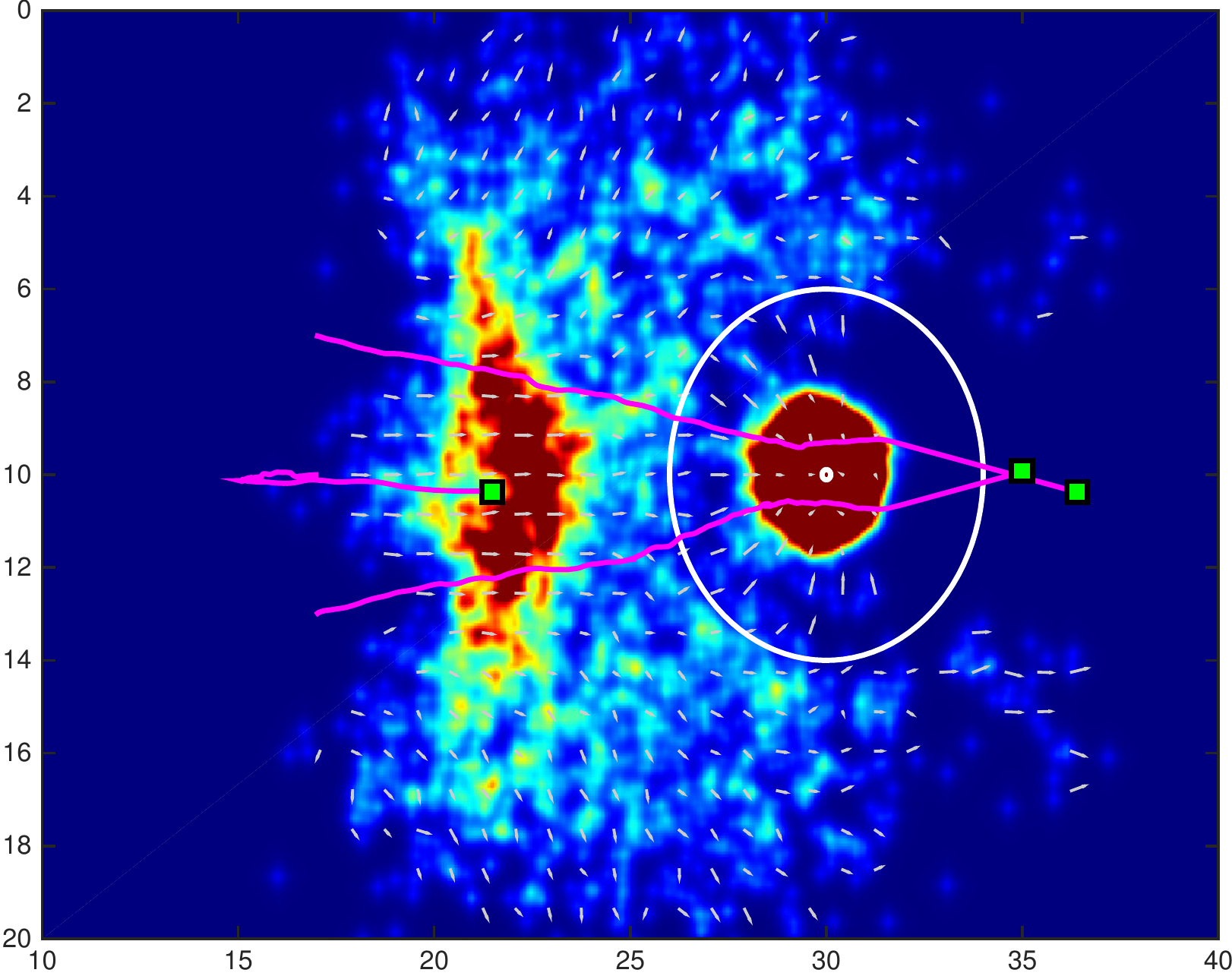}\quad
\includegraphics[width=0.3\textwidth]{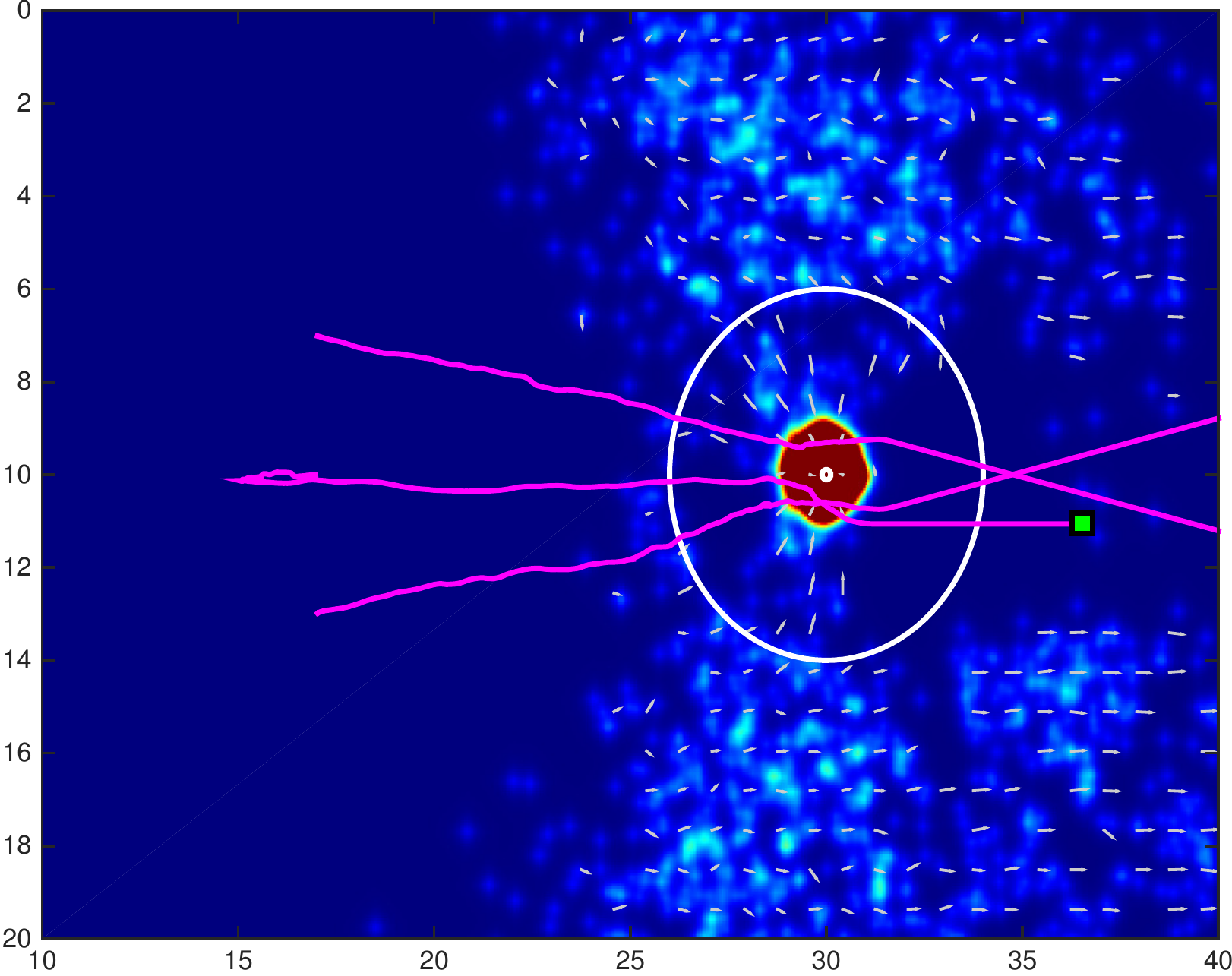}
\caption{{\em S\#2-Mesoscopic dynamics}. First row: no leaders. Second row: three leaders, go-to-target strategy. Third row: three leaders, optimal strategy (compass search).}
\label{fig:S1_meso}
\end{figure}

We consider here the case of a continuous density of followers. Figure \ref{fig:S1_meso}(first row) shows the evolution of the uncontrolled system of followers. Due to the diffusion term and the topological alignment, large part of the mass spreads around the domain and is not able to reach the target exit.

In Fig.\ \ref{fig:S1_meso}(second row) we account the action of three leaders, driven by a go-to-target strategy defined as in the microscopic case. It is clear that also in this case the action of leaders is able to influence the system and promote the evacuation, but the presence of the diffusive term causes the dispersion of part of the continuous density. The result is that part of the mass is not able to evacuate, unlike the microscopic case. 

In order to improve the go-to-target strategy we rely on the compass search, where, differently from the microscopic case, the optimization process accounts the objective functional \eqref{defJ_evacmass}, i.e., the total mass evacuated at final time.
Figure \ref{fig:S1_meso}(third row) sketches the optimal strategy found in this way: on one hand, the two external leaders go directly towards the exit, evacuating part of the density; on the other hand the central leader moves slowly backward, misleading part of the density and only later it moves forwards towards the exit. The efficiency of the leaders' strategy is due in particular by the latter movement of the last leader, which is able to gather the followers' density left behind by the others, and to reduce the occupancy of the exit's visibility area by delaying the arrival of part of the mass. 

\begin{figure}[h!]
\begin{center}
\includegraphics[width=4.85cm]{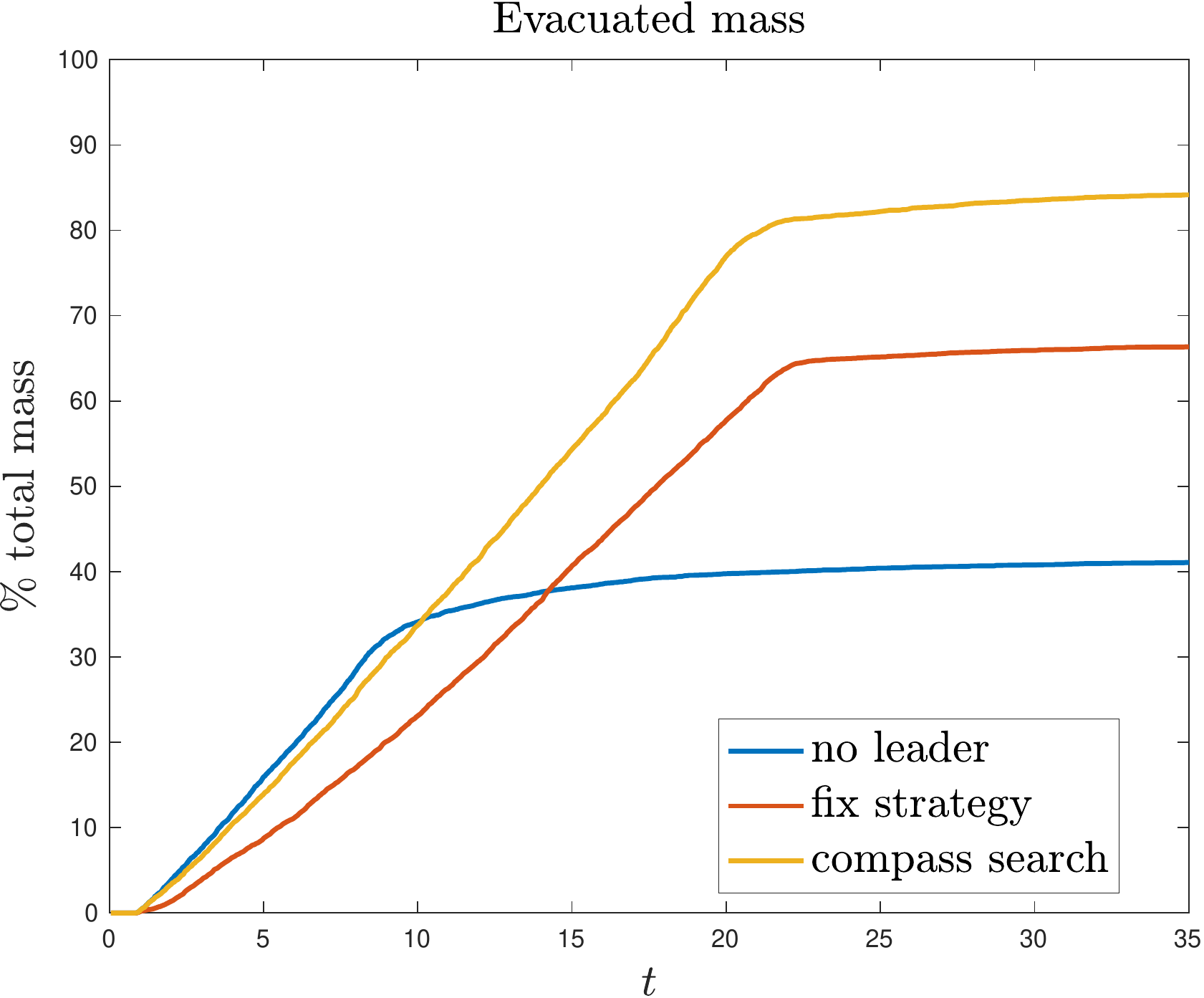}
\includegraphics[width=4.85cm]{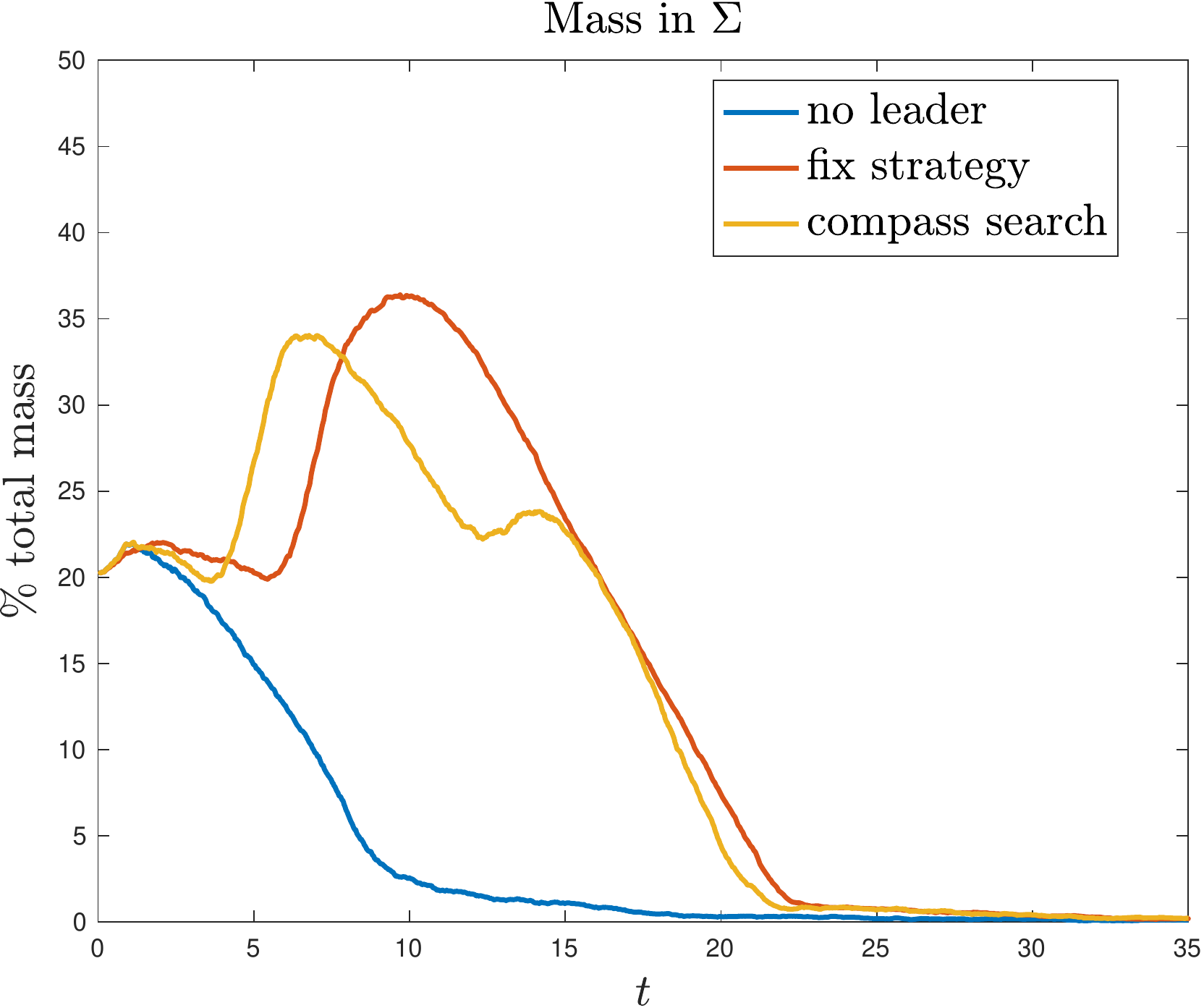}
\caption{{\em S\#2. Invisible leaders control}. On the left: Percentage of mass evacuated in time. On the right: Occupancy of the visibility area in terms of total mass percentage.}\label{occupancy2}
\end{center}
\end{figure}

In Figure \ref{occupancy2} we summarize, for the three numerical experiments, the evacuated mass and the occupancy of the exit's visibility area $\Sigma$ as functions of time.  In the right plot the occupancy of the exit's visibility area shows clearly the difference between the leaders' action: for the go-to-target strategy, the amount of mass occupying  $\Sigma$ concentrates and the evacuation is partially hindered by the clogging effect, as only 66.3\% of the total mass is evacuated. The optimal strategy (obtained after 30 iterations) is able to better distribute the mass arrival in $\Sigma$, and an higher efficiency is reached, evacuating up to 84.1\% of the total mass.

\subsubsection{S\#3: Invisible leaders in presence of obstacles}
Finally we test the microscopic and mesoscopic model in presence of obstacles. The crowd is initially confined in a rectangular room with three walls. In order to evacuate, people must first leave the room and then search for the exit point. We assume that walls are not visible, i.e., people can perceive them only by physical contact. This corresponds to an evacuation in case of null visibility (but for the exit point which is still visible from within $\Sigma$). Walls are handled as in \cite{cristiani2011MMS}.

\paragraph{\em Microscopic model}

\begin{figure}[h!]
\centering
\includegraphics[width=0.3\textwidth]{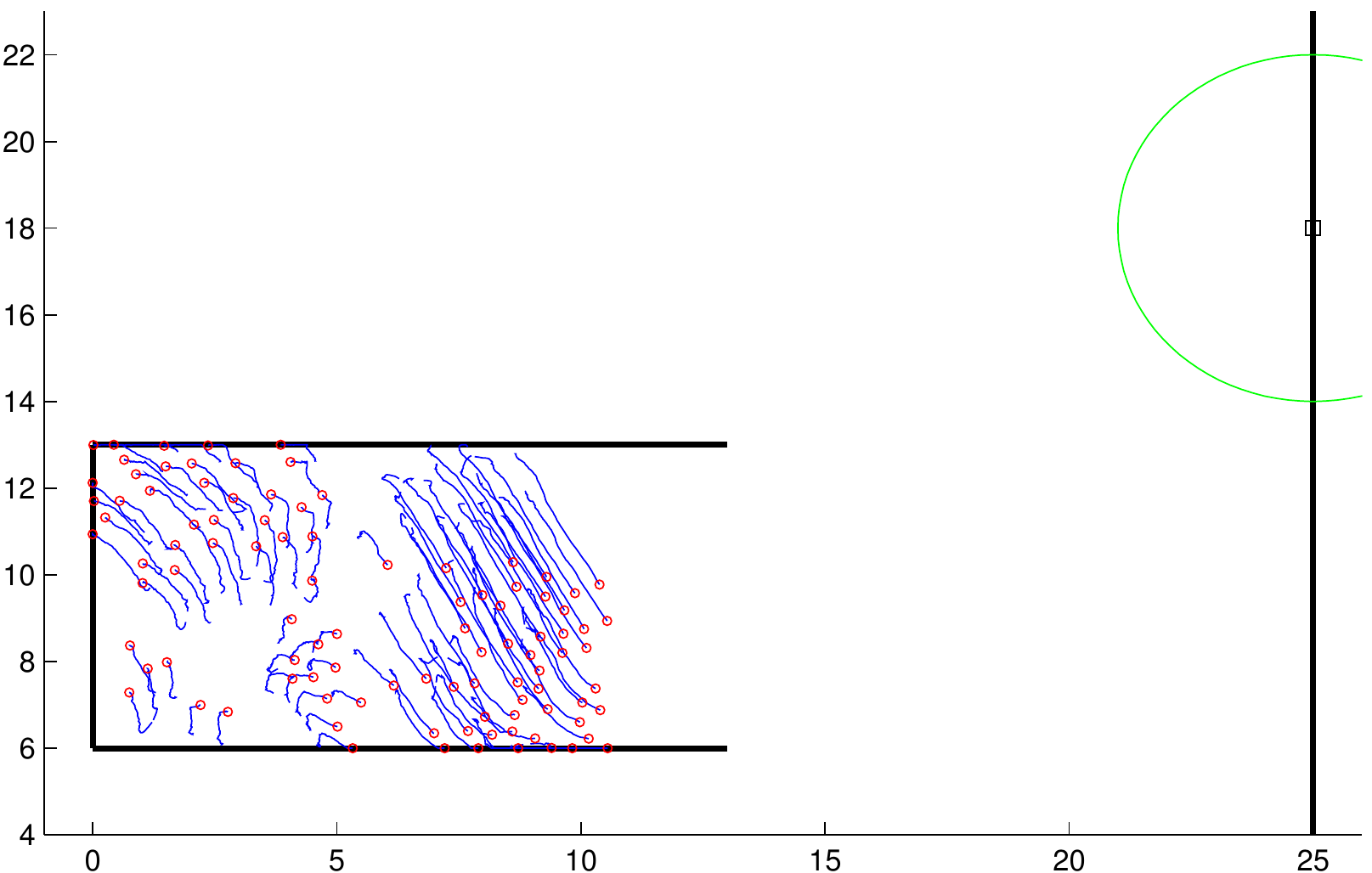}\quad
\includegraphics[width=0.3\textwidth]{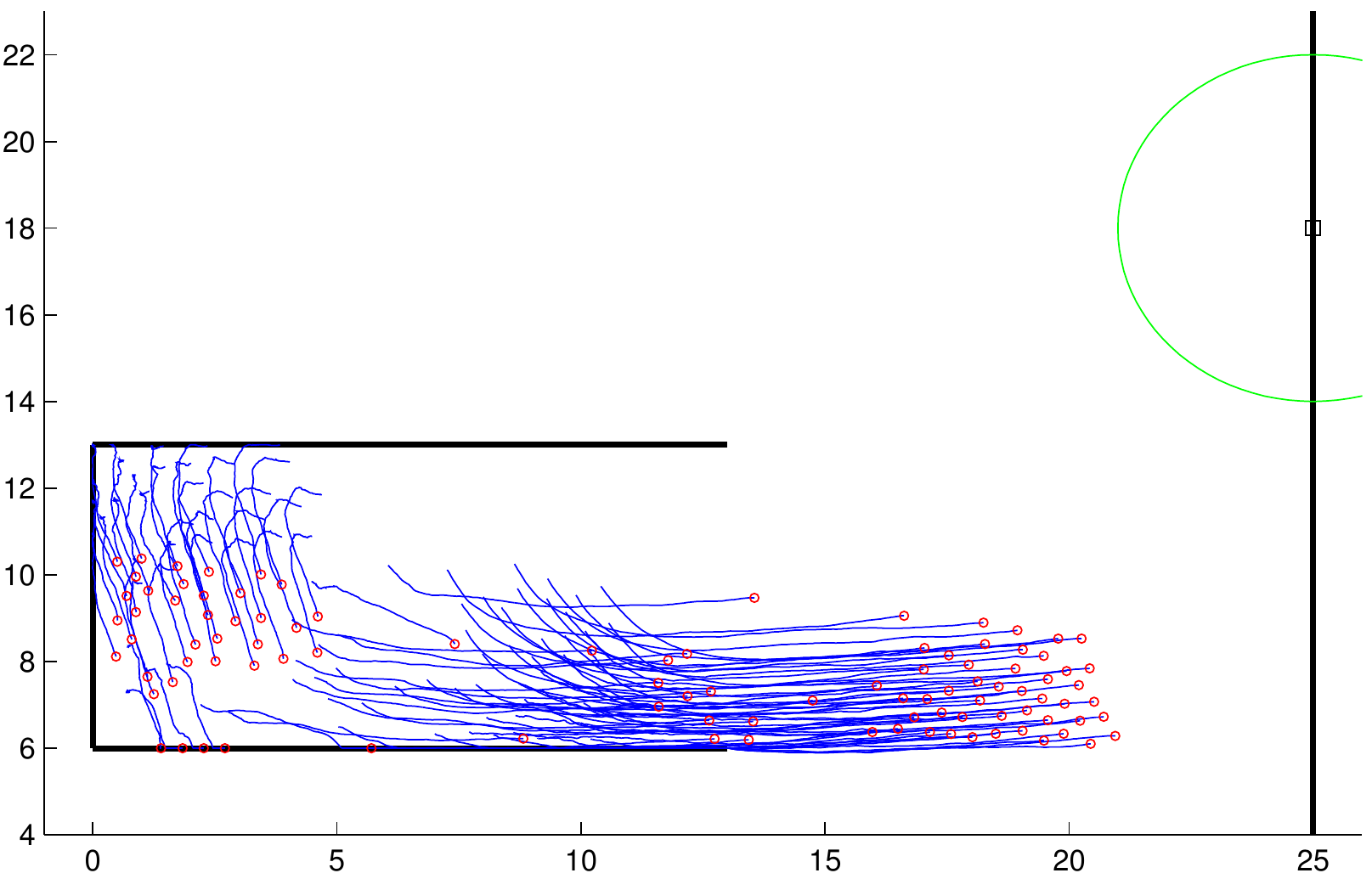}\quad
\includegraphics[width=0.3\textwidth]{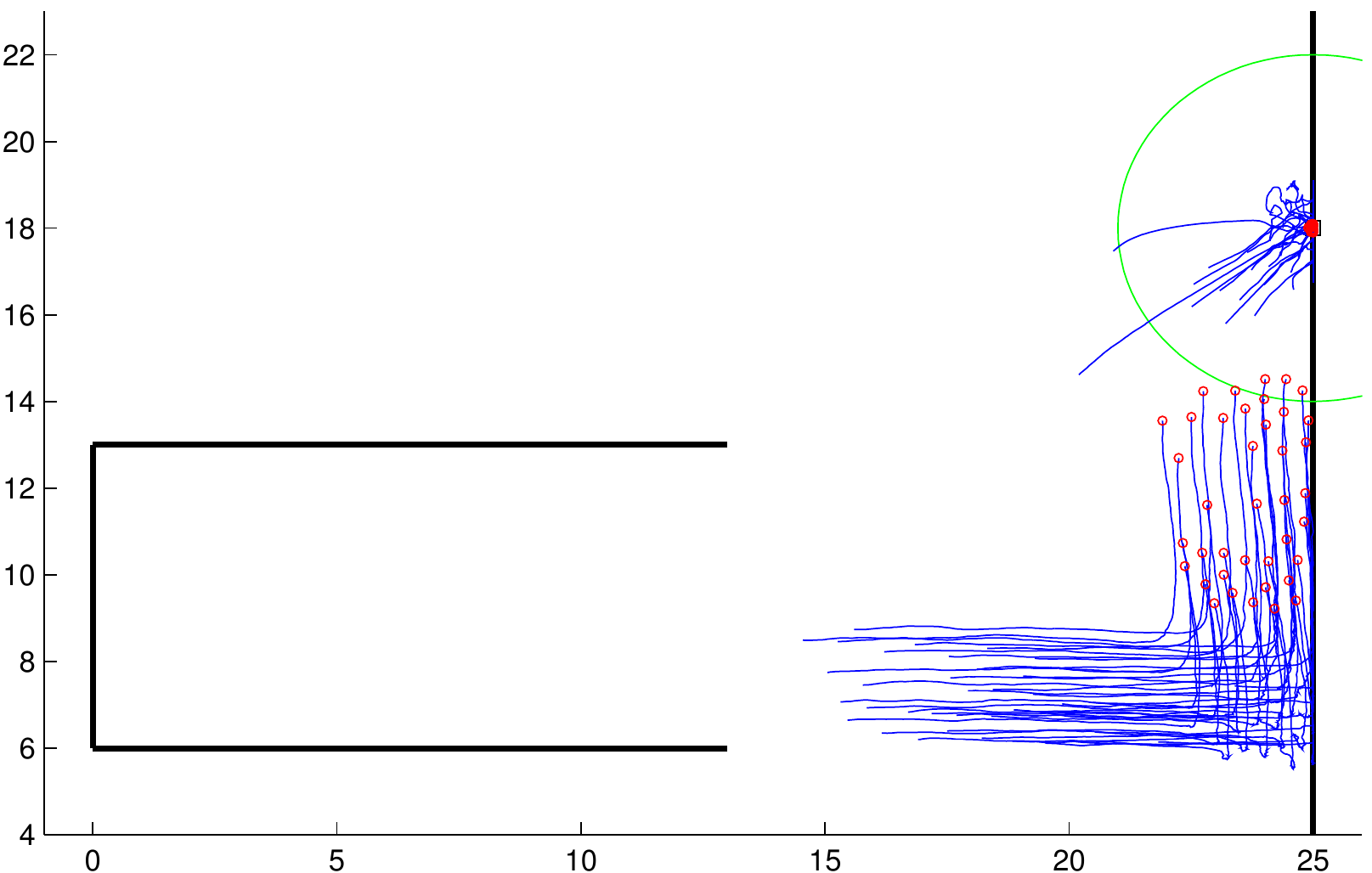}\\ 
\includegraphics[width=0.3\textwidth]{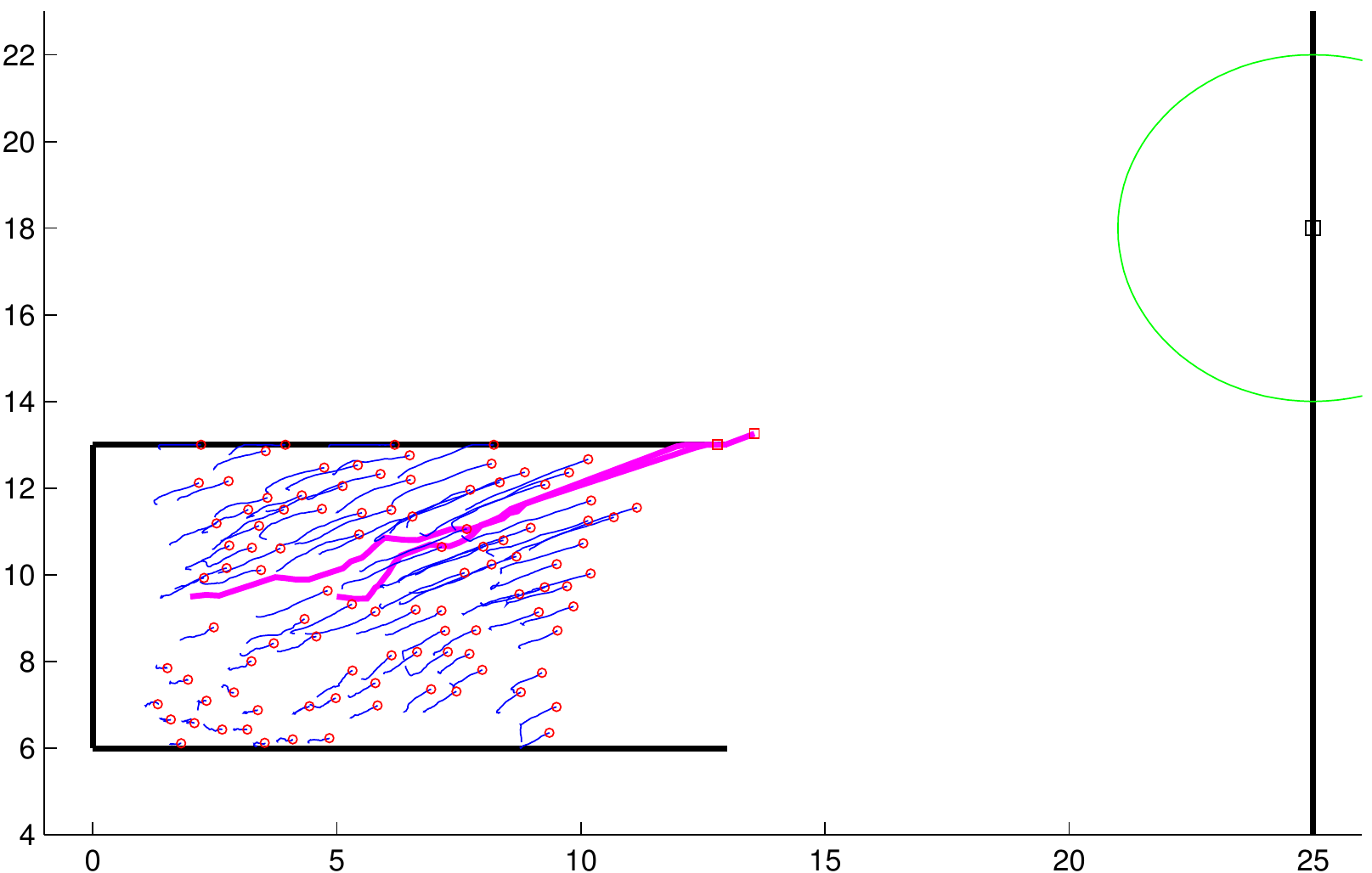}\quad
\includegraphics[width=0.3\textwidth]{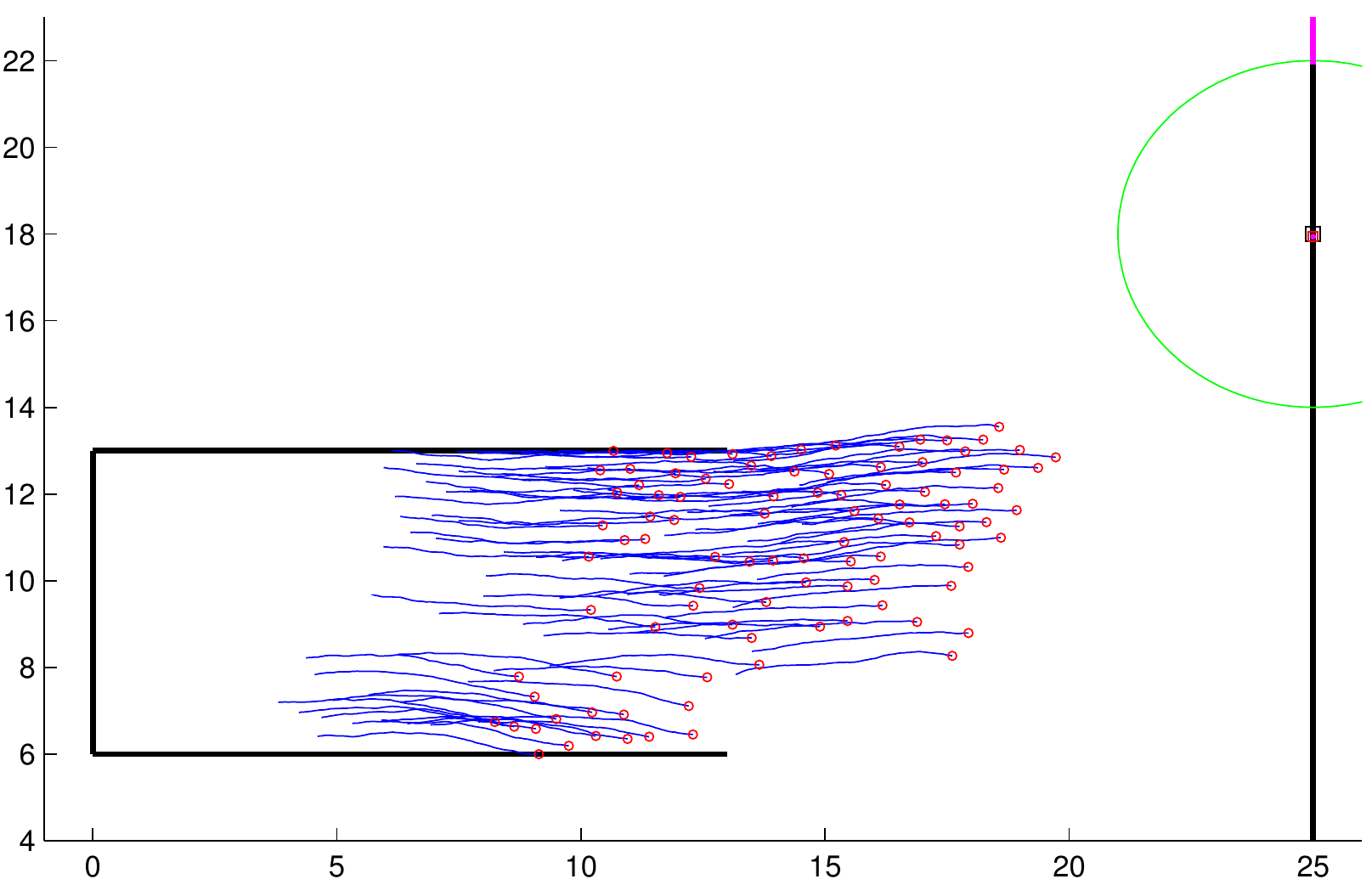}\quad
\includegraphics[width=0.3\textwidth]{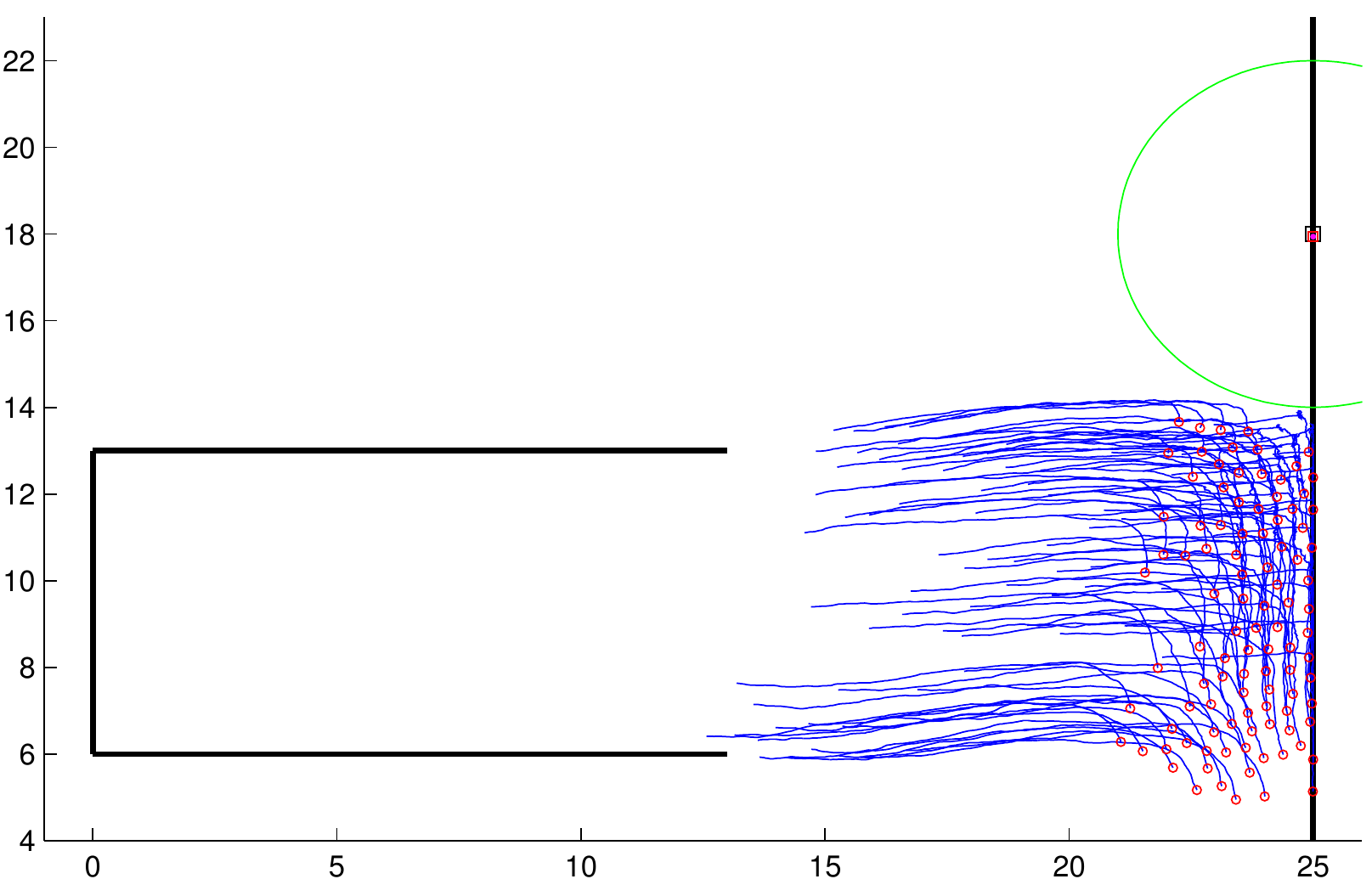}\\ 
\includegraphics[width=0.3\textwidth]{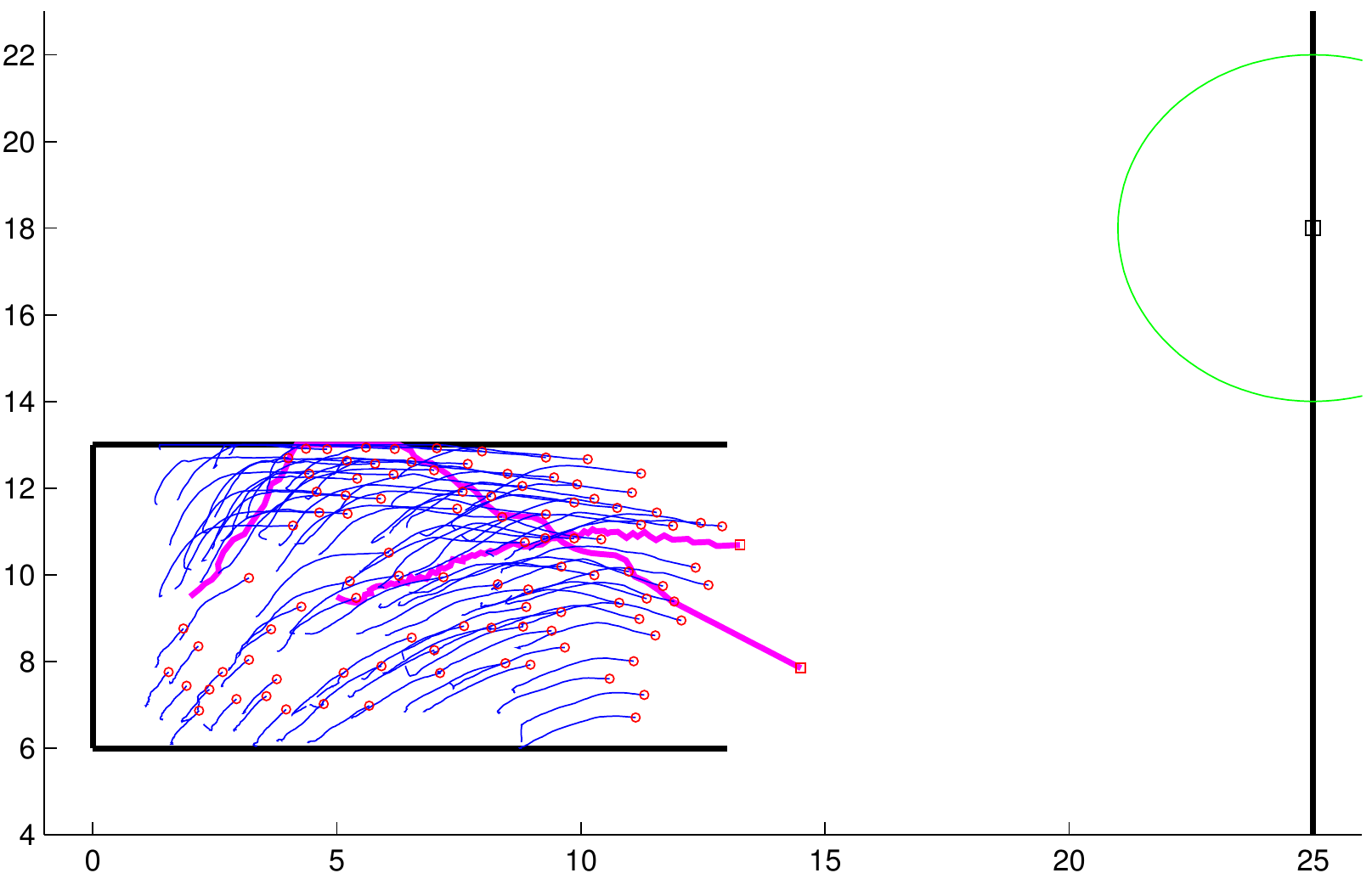}\quad
\includegraphics[width=0.3\textwidth]{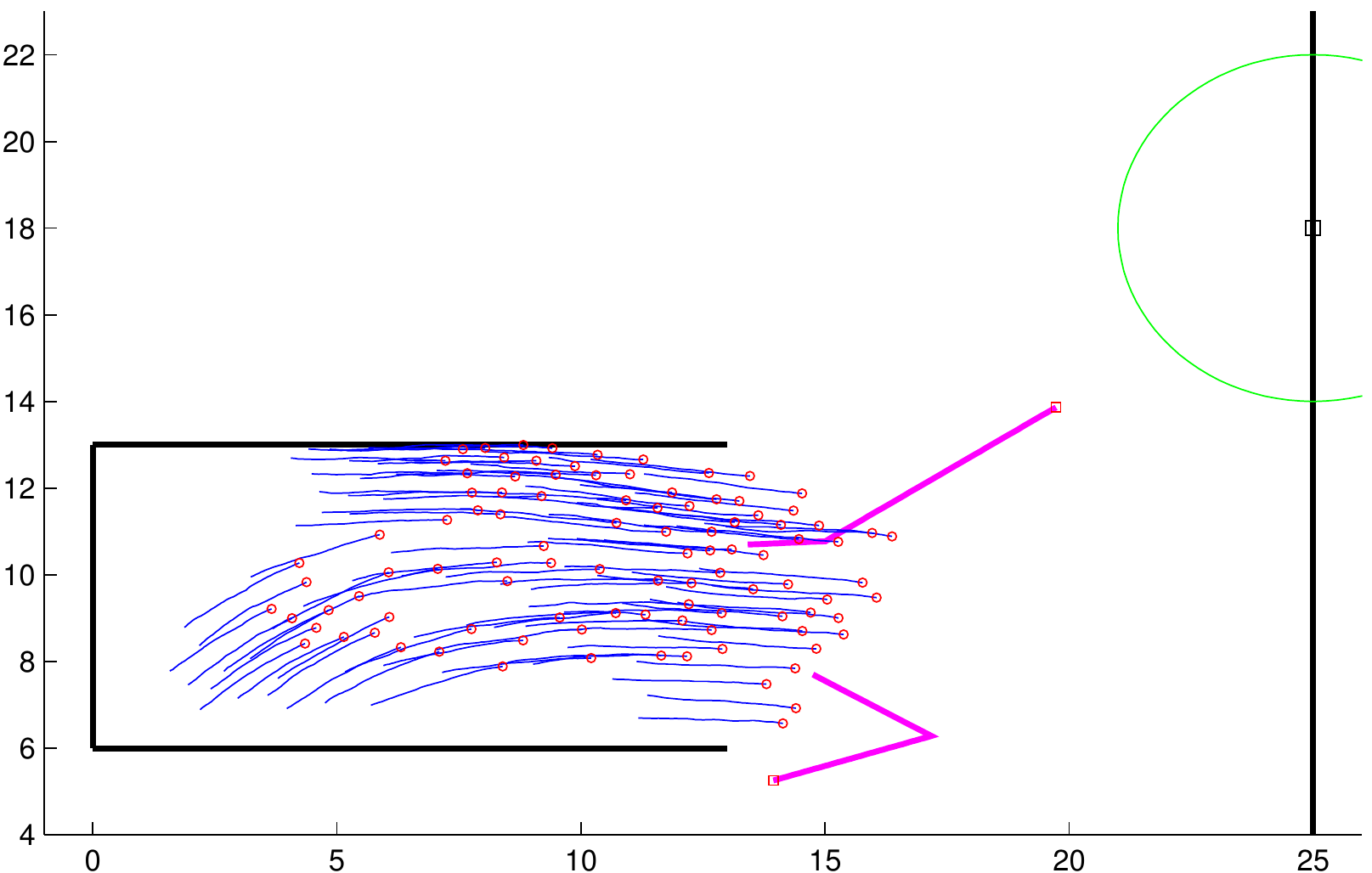}\quad
\includegraphics[width=0.3\textwidth]{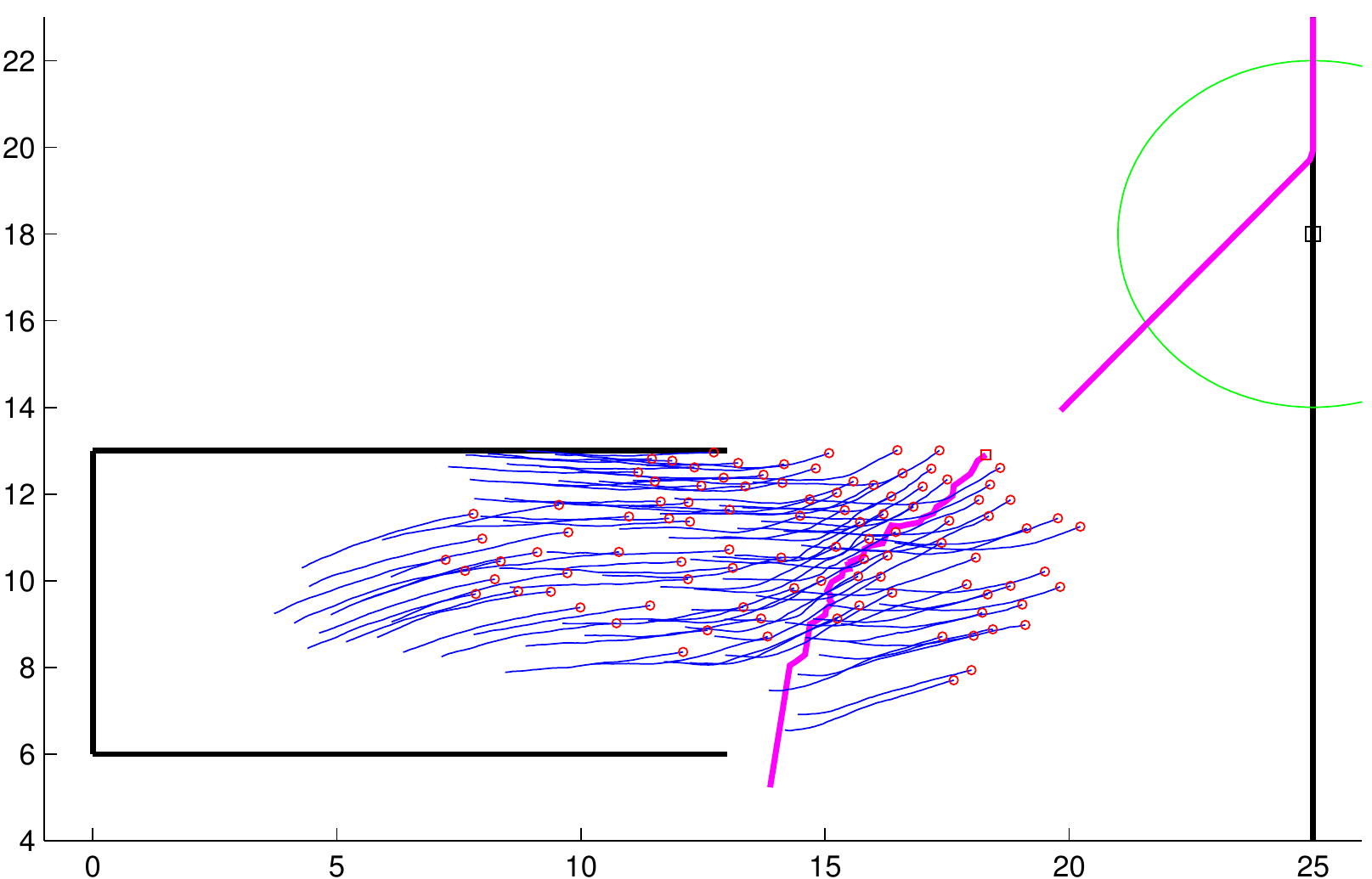}
\caption{{\em S\#3-Microscopic simulation}. 
First row: no leaders.
Second row: two leaders and go-to-target strategy.
Third row: two leaders and best strategy computed by the compass search.
}
\label{fig:S2}
\end{figure}

In Figure \ref{fig:S2}(first row) we observe the case where no leaders are present: the crowd splits in several groups and most of the people hit the wall. After some attempts the crowd finds the way out, and then it crashes into the right boundary of the domain. Finally, by chance people decide, \emph{en cascade}, to go upward. The crowd leaves the domain in 1162 time steps.

If instead we hide in the crowd two leaders who point fast towards the exit (Fig.\ \ref{fig:S2}(second row)), the evacuation from the room is completed in very short time, but after that, the influence of the leaders vanishes. Unfortunately, this time people decide to go downward after hitting the right boundary, and nobody leaves the domain.  Slowing down the two leaders helps keeping the leaders' influence for longer time, although it is quite difficult to find a good choice. 

Compass search optimization finds (after 30 iterations) a nice strategy for the two leaders which remarkably improves the evacuation time, see Fig.\ \ref{fig:S2}(third row). One leader behaves similarly to the previous case, while the other diverts the crowd pointing SE, then comes back to wait for the crowd, and finally points NE towards the exit. This strategy allows to bring everyone to the exit in 549 time steps, without bumping anyone against the boundary, and avoiding congestion near the exit. 

\paragraph{\em Mesoscopic model}

\begin{figure}[h!]
\begin{center}
\includegraphics[width=3.85cm]{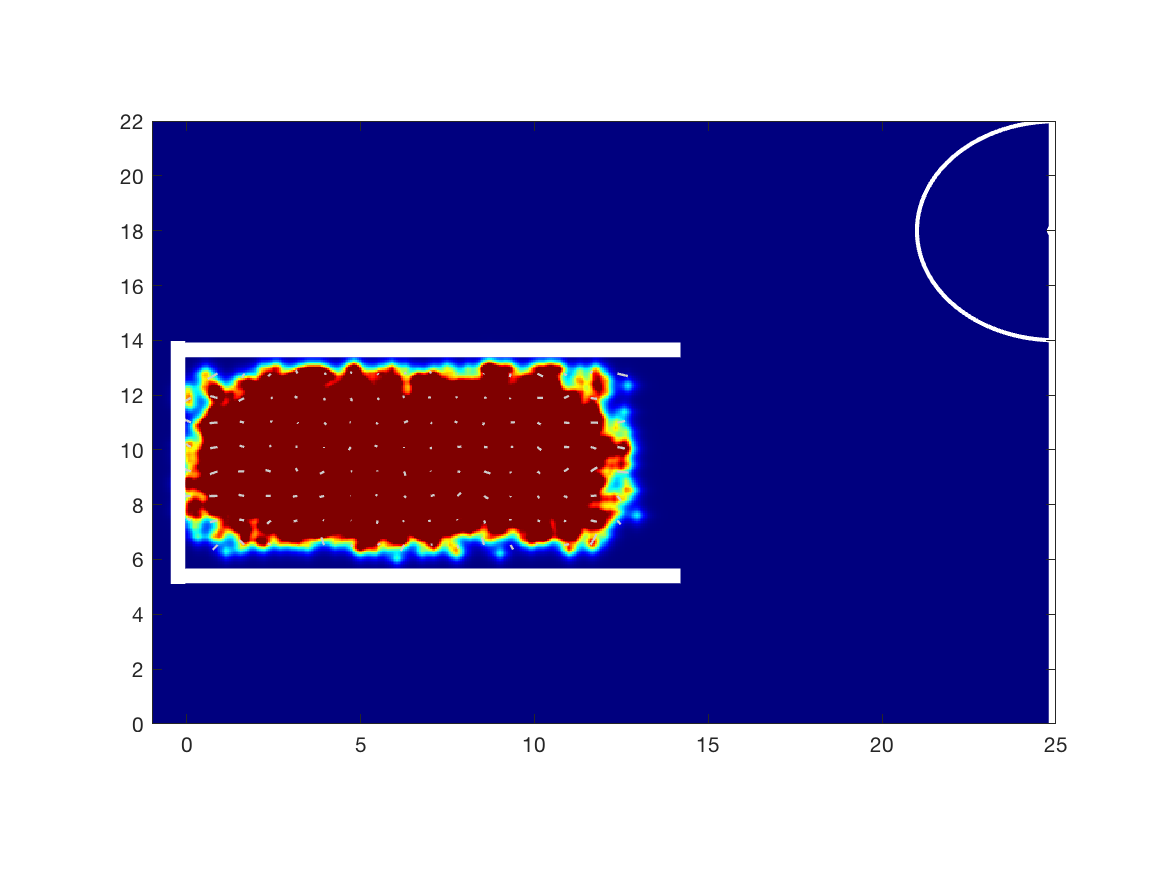}
\includegraphics[width=3.85cm]{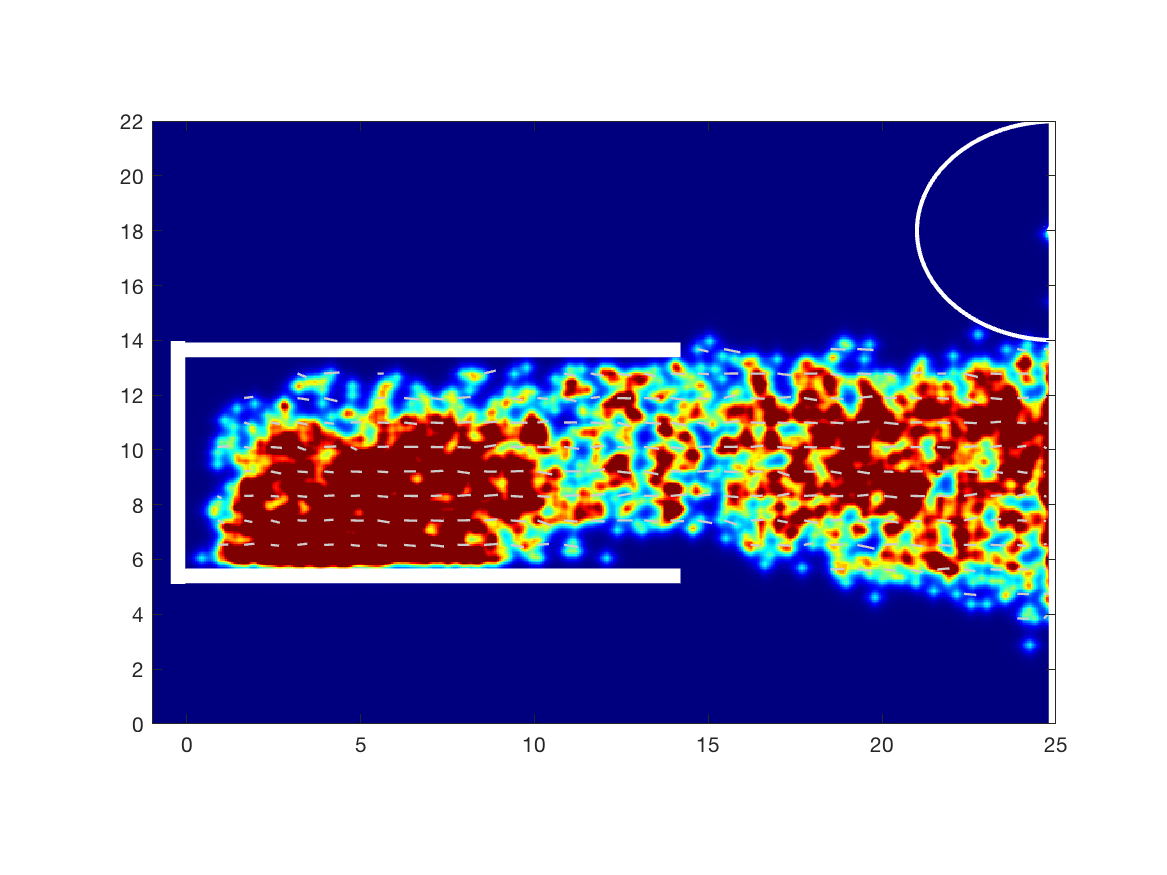}
\includegraphics[width=3.85cm]{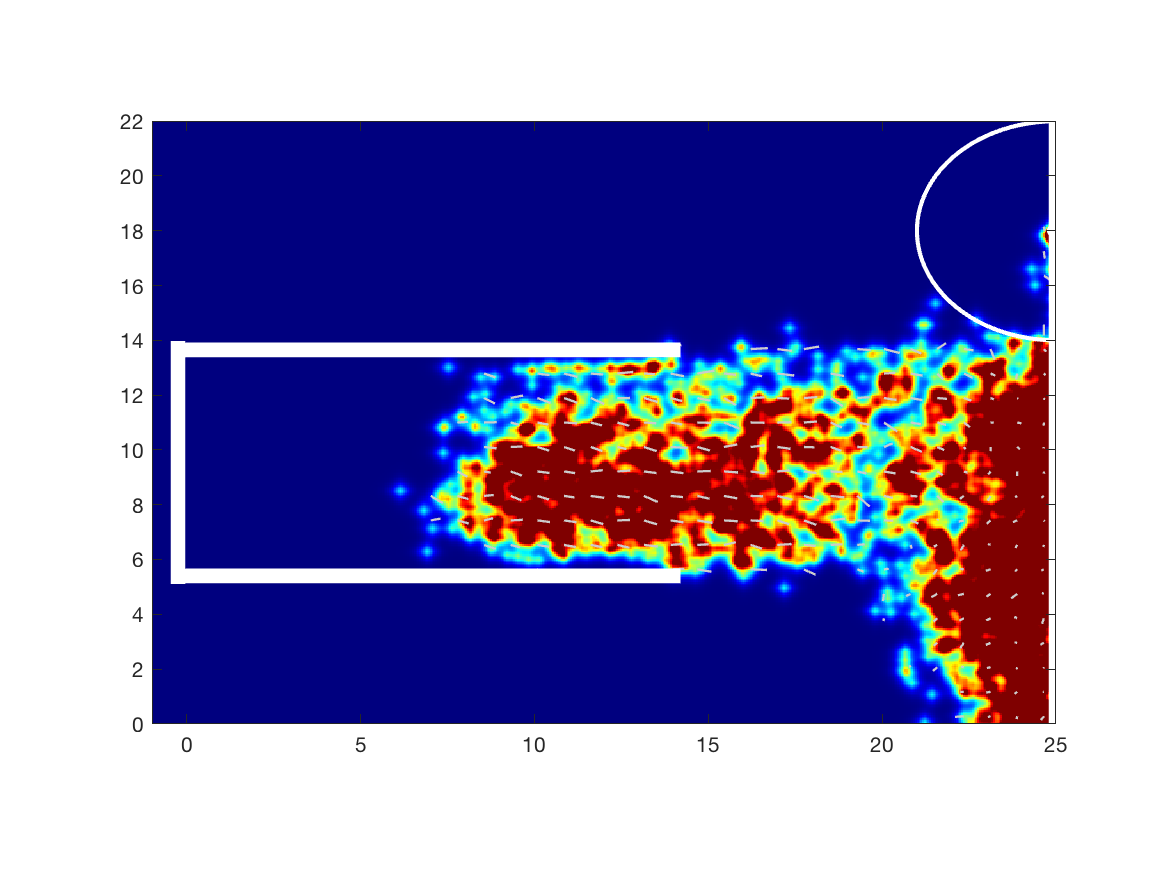}
\\
%
\includegraphics[width=3.85cm]{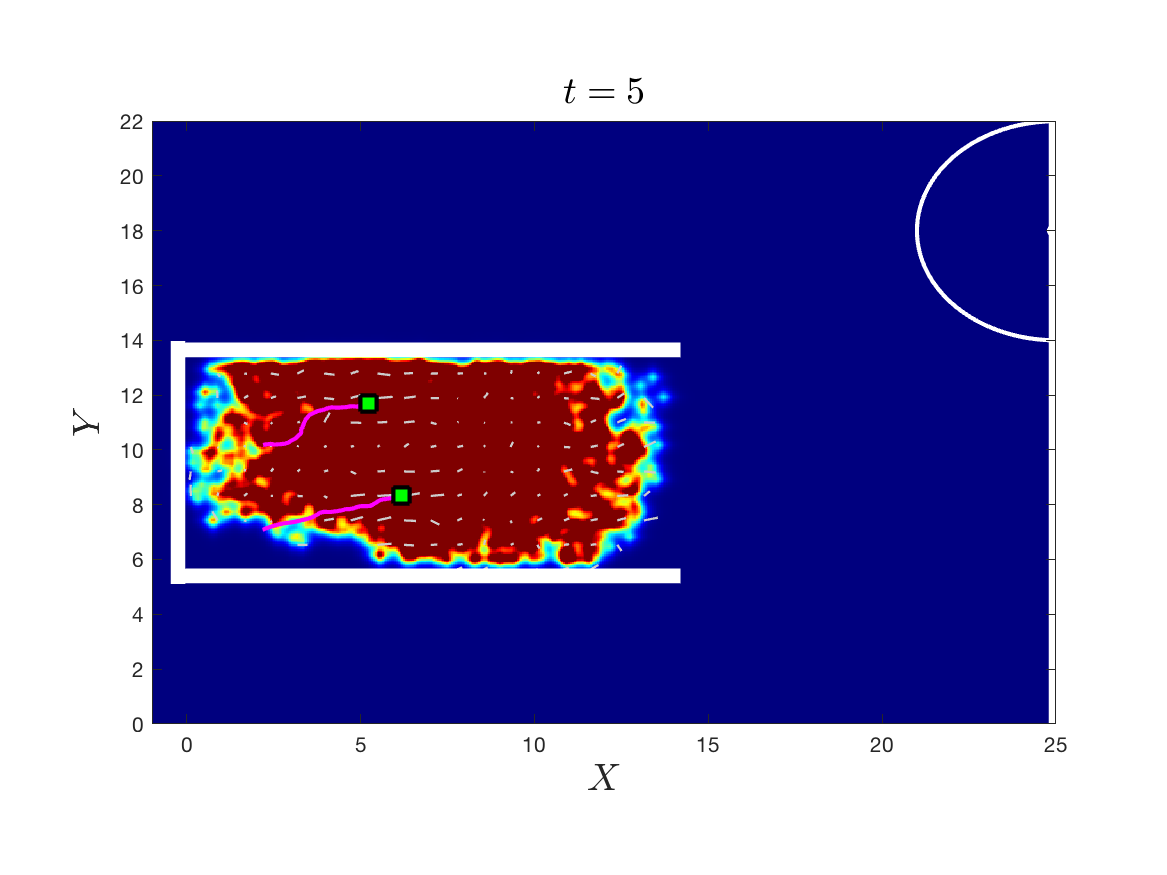}
\includegraphics[width=3.85cm]{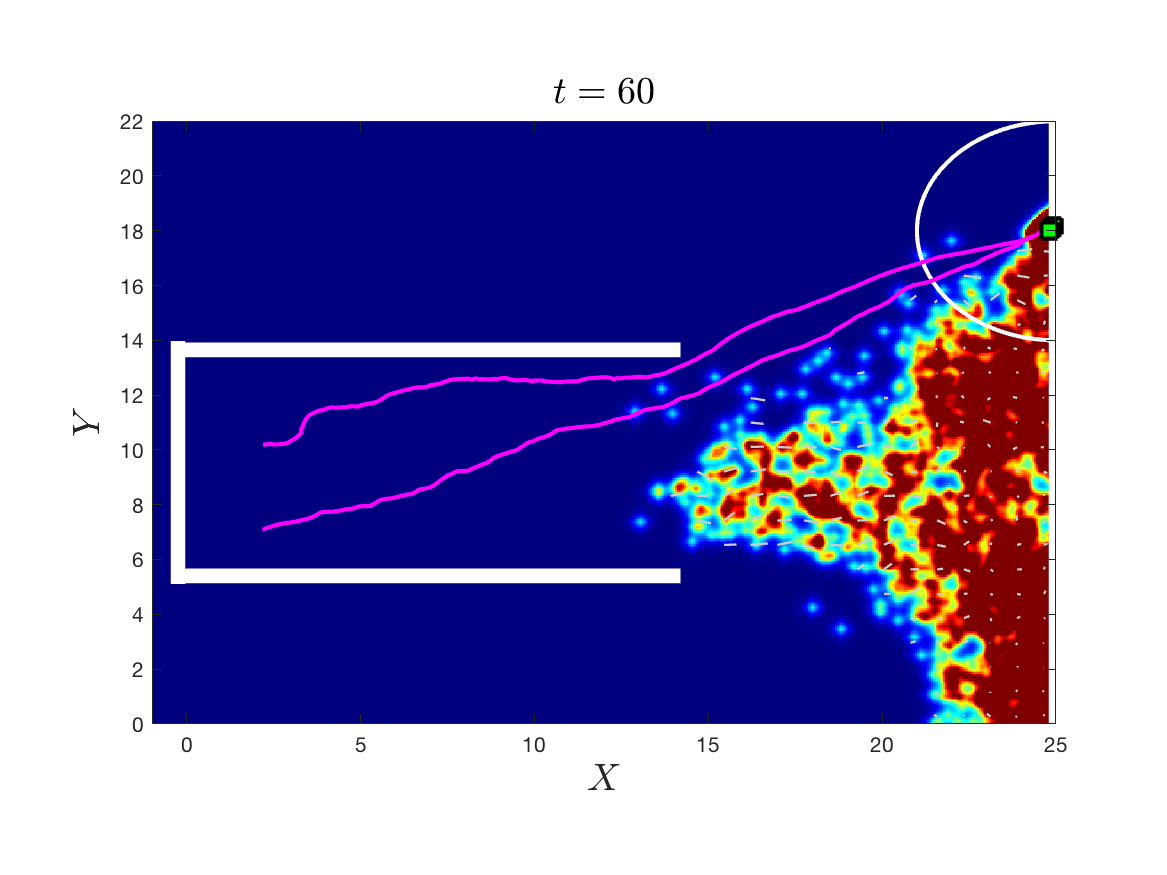}
\includegraphics[width=3.85cm]{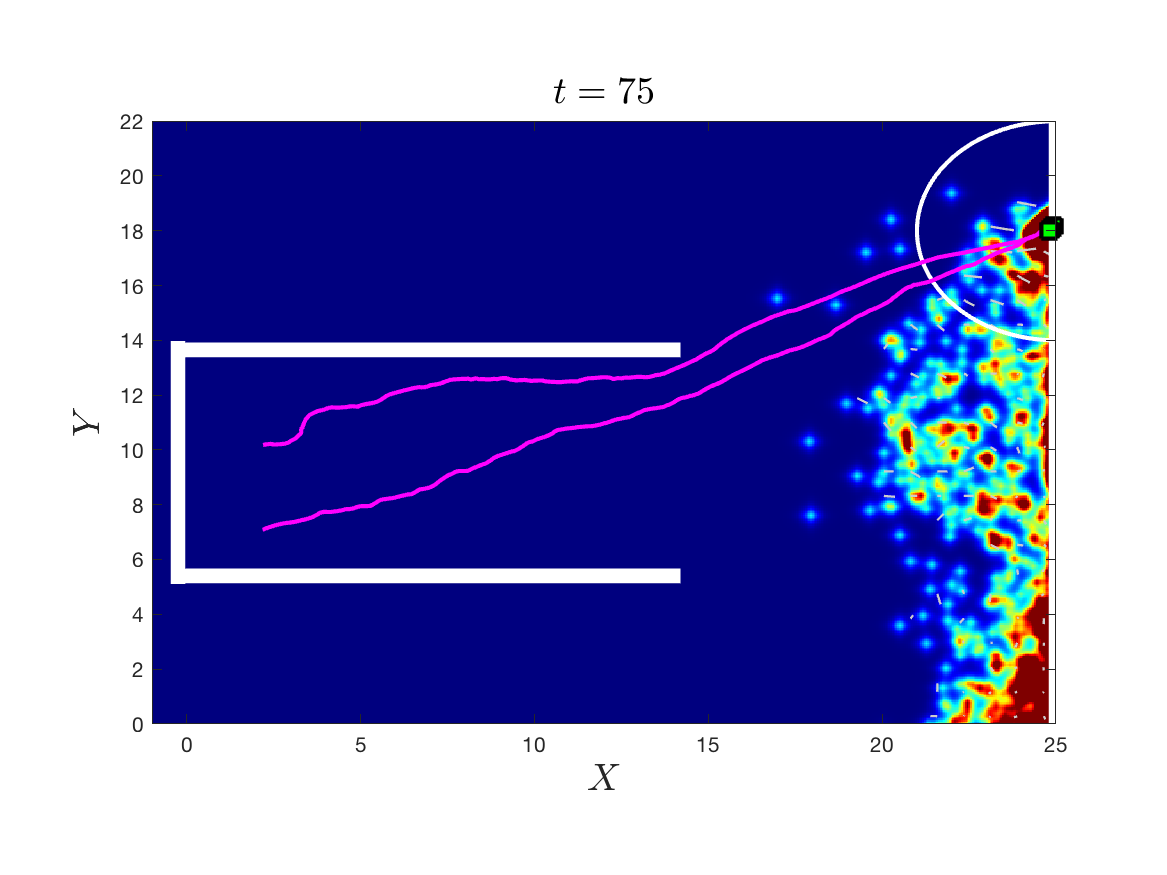}
\\
\includegraphics[width=3.85cm]{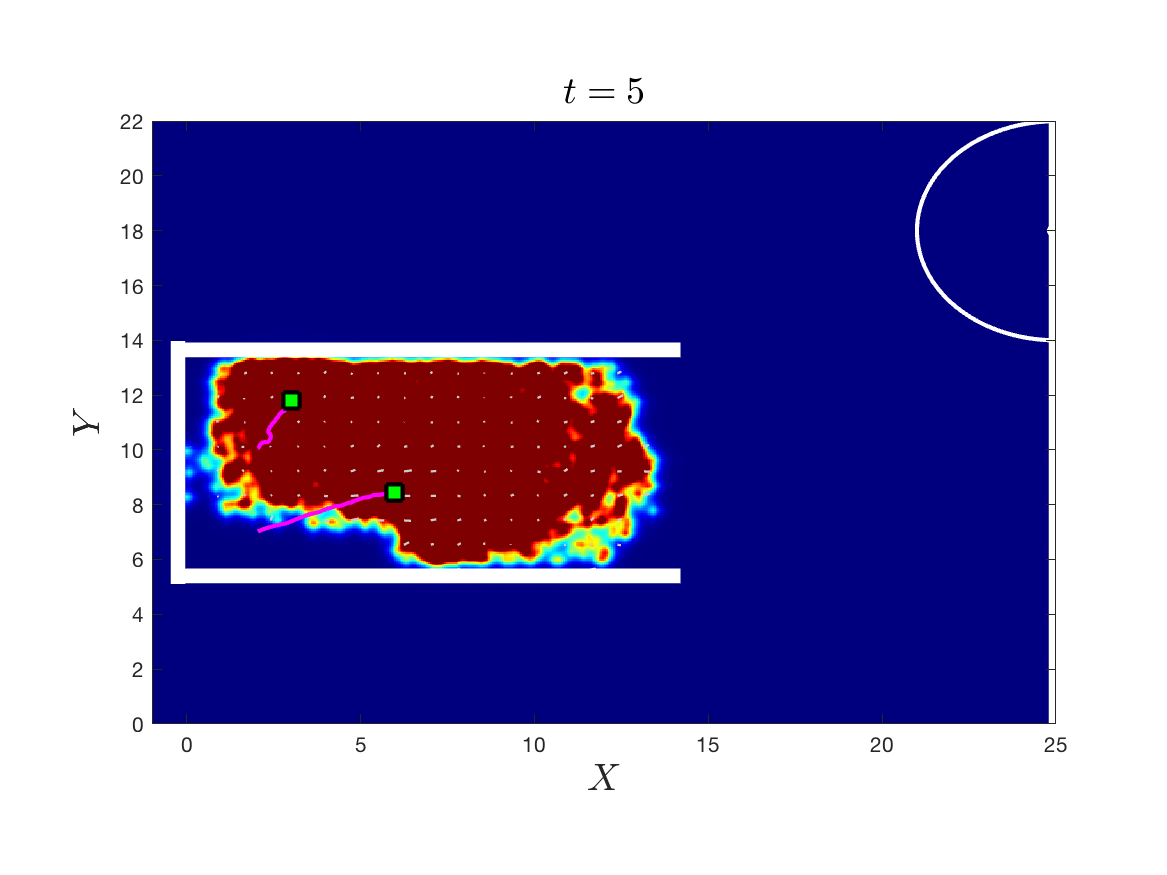}
\includegraphics[width=3.85cm]{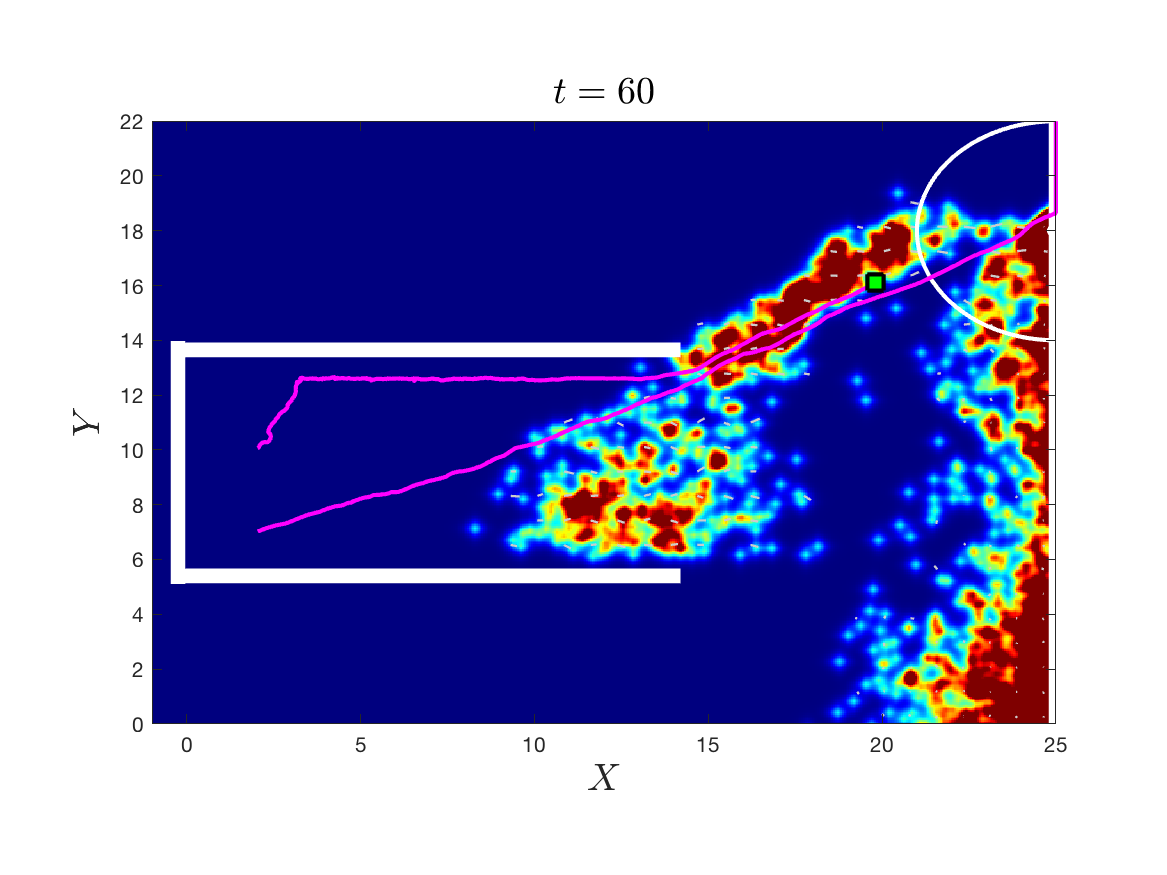}
\includegraphics[width=3.85cm]{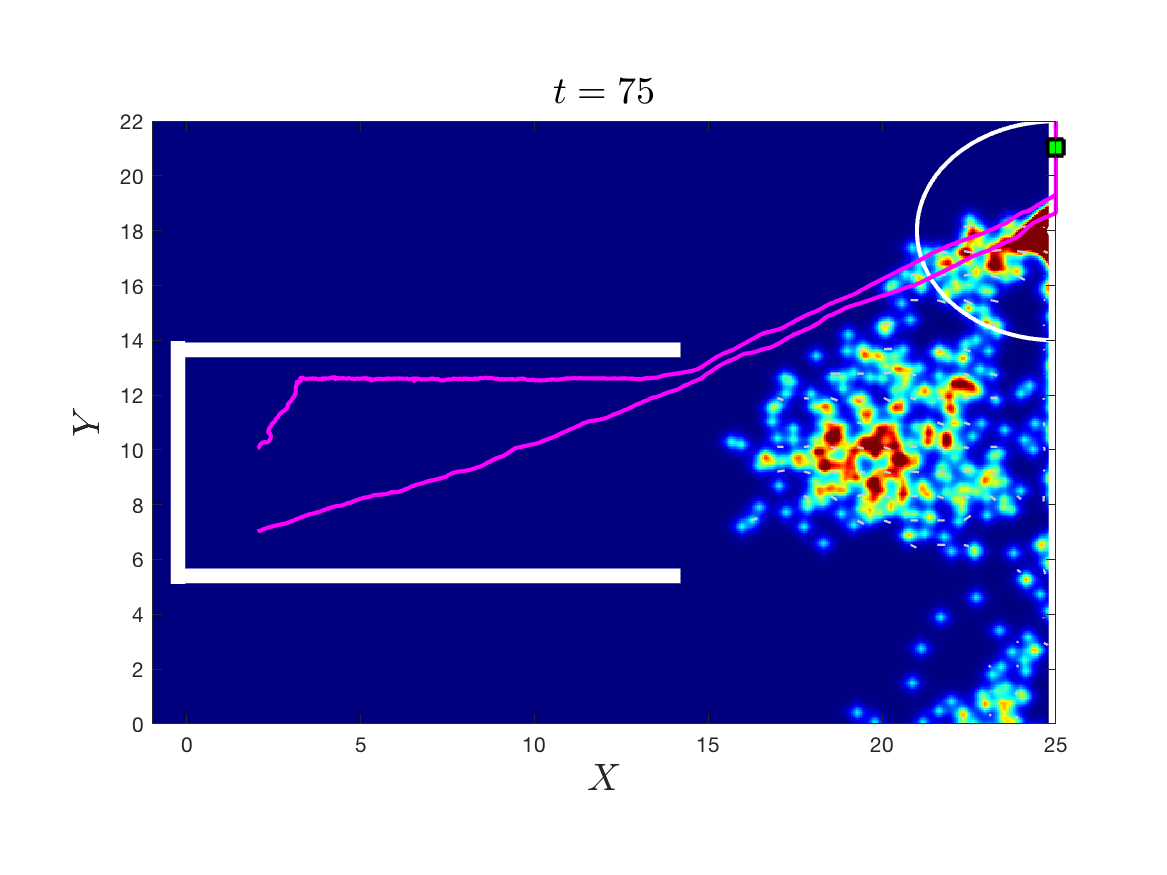}
\caption{{\em S\#3-Mesoscopic dynamics}.Top row: uncontrolled setting. Middle row:  two invisible leaders with go-to-target strategy. Bottom row: two invisible leaders with optimized strategy (compass search).}\label{setting3_leaders}
\end{center}
\end{figure}

In Figure \ref{setting3_leaders} we report the evolution of the mesoscopic density of followers. 
First row shows the evolution  of the uncontrolled case, contrary to the microscopic in this case evacuation is not reached: the mass slowly diffuse outside the corridor and move in the opposite direction with respect the target exit, only a small percentage of the mass is able to evacuate.

Second rows depicts the case with two leaders and a go-to-target strategy, positioned at the end of the corridor. Their movements are able to influence large part of the crowd, at final time 67.2\% of the mass is evacuated.
 
Employing the compass search method we show in the bottom row of Figure \ref{setting3_leaders} an improvement of the go-to-target strategy (after 9 iterations). In this case evacuation of the 72.4\% of the total mass is reached: one of the two leaders deviates from the original direction, slowing down part of the mass. Similarly to the optimal strategy retrieved in S\#2, the optimization suggests to avoid congestion around the exit.

\begin{figure}[h!]
\begin{center}
\includegraphics[width=4.85cm]{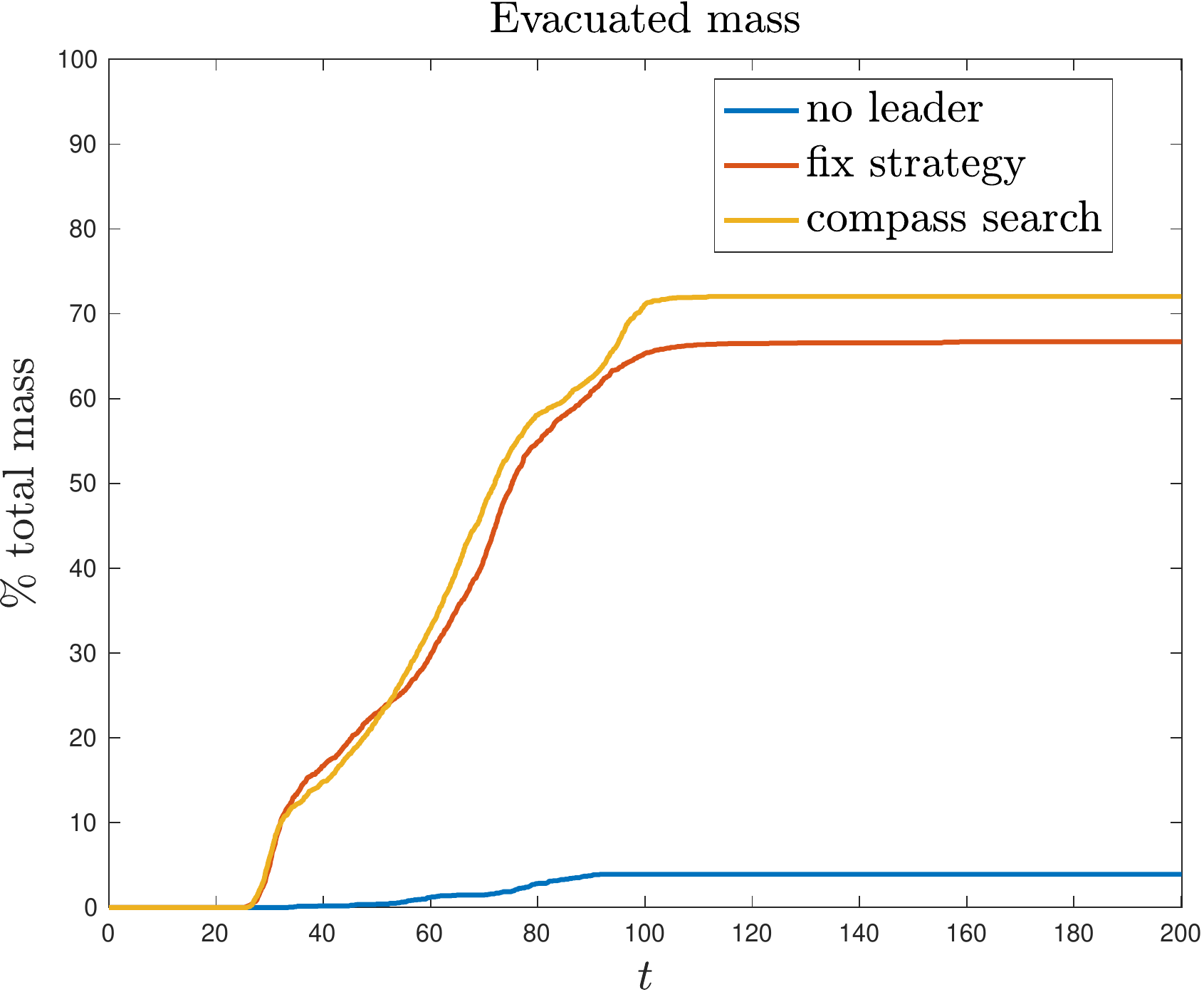}
\includegraphics[width=4.85cm]{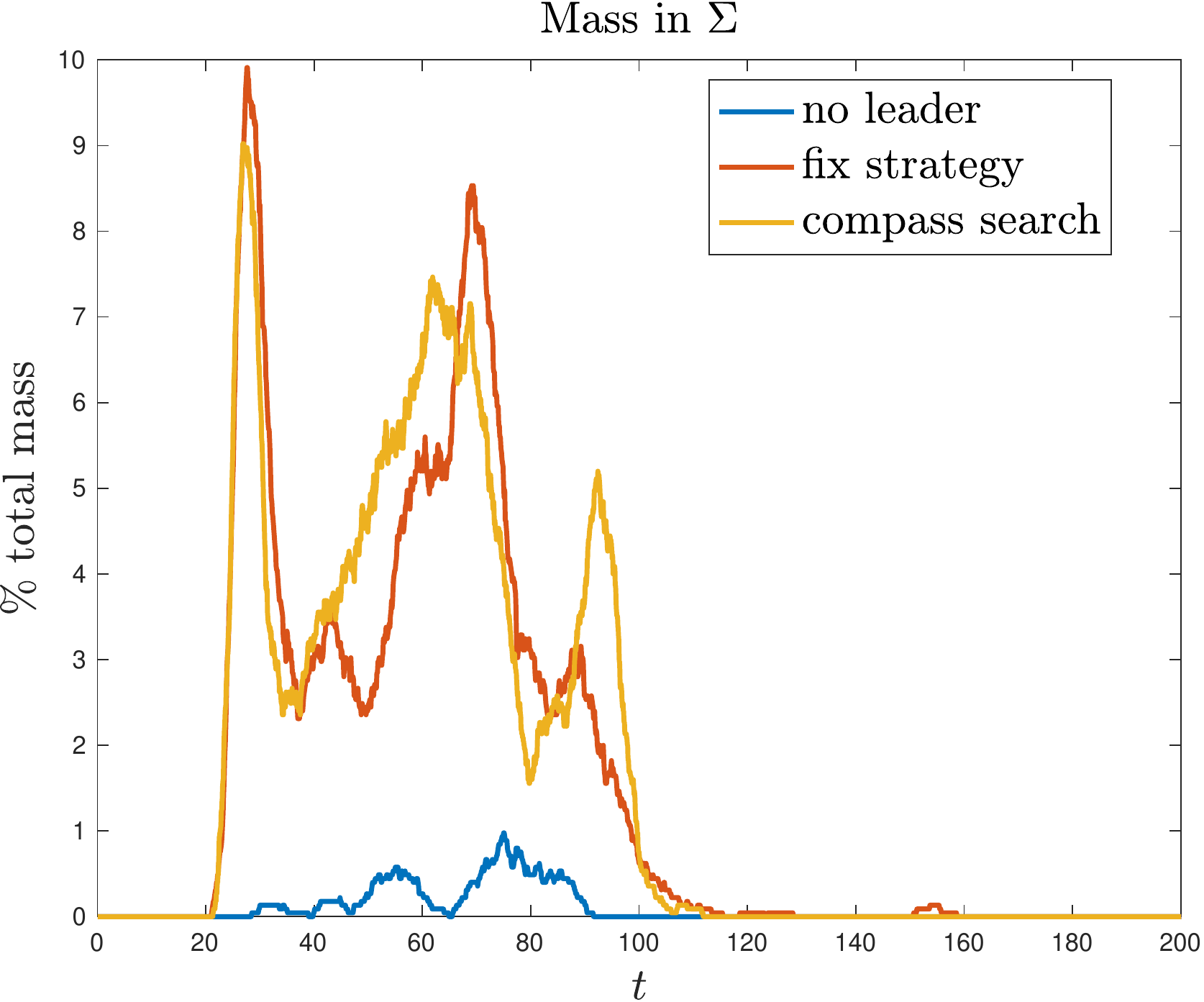}
\caption{{\em S\#3. Invisible leaders control}. On the left: Percentage of mass evacuated in time. On the right: Occupancy of the visibility area in terms of total mass percentage.}\label{occupancy}
\end{center}
\end{figure}

We compare in Figure \ref{occupancy} the outcomes of the three different situations. The left plot reports the percentage of evacuated mass as a function of time, On the right we depict the occupancy of the visibility area. 
We observe that also in this case the optimal strategy suggests to decrease the congestion around the exit in order to increase the total mass evacuated.

\section{Crowd controls through smart obstacles}\label{sec:ctrl_obstacles}

The exploration activity of an unknown environment by a group of pedestrians may become crucial if the time of egress represents a critical variable. This could not be only connected with a specific state of danger, like in case of fire or earthquake, because even staying too long into an environment can be undesirable. For this reason, several signals and other indications need to be accurately located in order to correctly address a crowd entering a room. Unfortunately, in case of low visibility, the classical signage cannot be perceived, and other devices can be adopted to this aim, like lighting and sound effects.

Also the shape of the room can be  designed to facilitate the egress. Walls or obstacles can be shaped in order to operate a guidance of the crowd, and several studies gave evidence of the usefulness of this strategy \cite{cristiani2017AMM,cristiani2019AMM}. On the other hand, the number of obstacles cannot be too large, in order to preserve the original purpose of the room. As an extreme situation, if we lock every useless passage of a maze, we are minimizing the egress time, but we have no longer a maze.

In the following, we are illustrating the activity of optimization of the position of a number of fixed walls to minimize the egress time of a crowd from a simple square room with four entrance and four exit. The simulation of the crowd movement is performed by means of the {\em micro-scale} model previously described in Section \ref{sec:models}.

\subsection{Selection of the objective function}\label{sec:ctrl_obs_fo}

Due to the presence of a random component in the speed of the single pedestrian, the final egress time of the crowd estimated by the numerical model is a stochastic outcome. If we want to utilize this quantity as the objective function of an optimization problem, a statistical approach is essential. We can compute the expected value if we repeat the simulation a large number of times, but this data is, in our opinion, still not sufficiently representative. In fact, the variability of the egress time is also fundamental. For this reason, we are here considering as objective function the sum of expected value plus their variance,
\[
 {\cal F} = EV(x) + \sigma(x).
\]
With this definition, we can assure with a probability of 80.15\% that the egress time is lower than ${\cal F}$ (if the egress time follows a normal distribution).

\begin{figure}[h!]
	\begin{center}
		\includegraphics[width=0.95\textwidth]{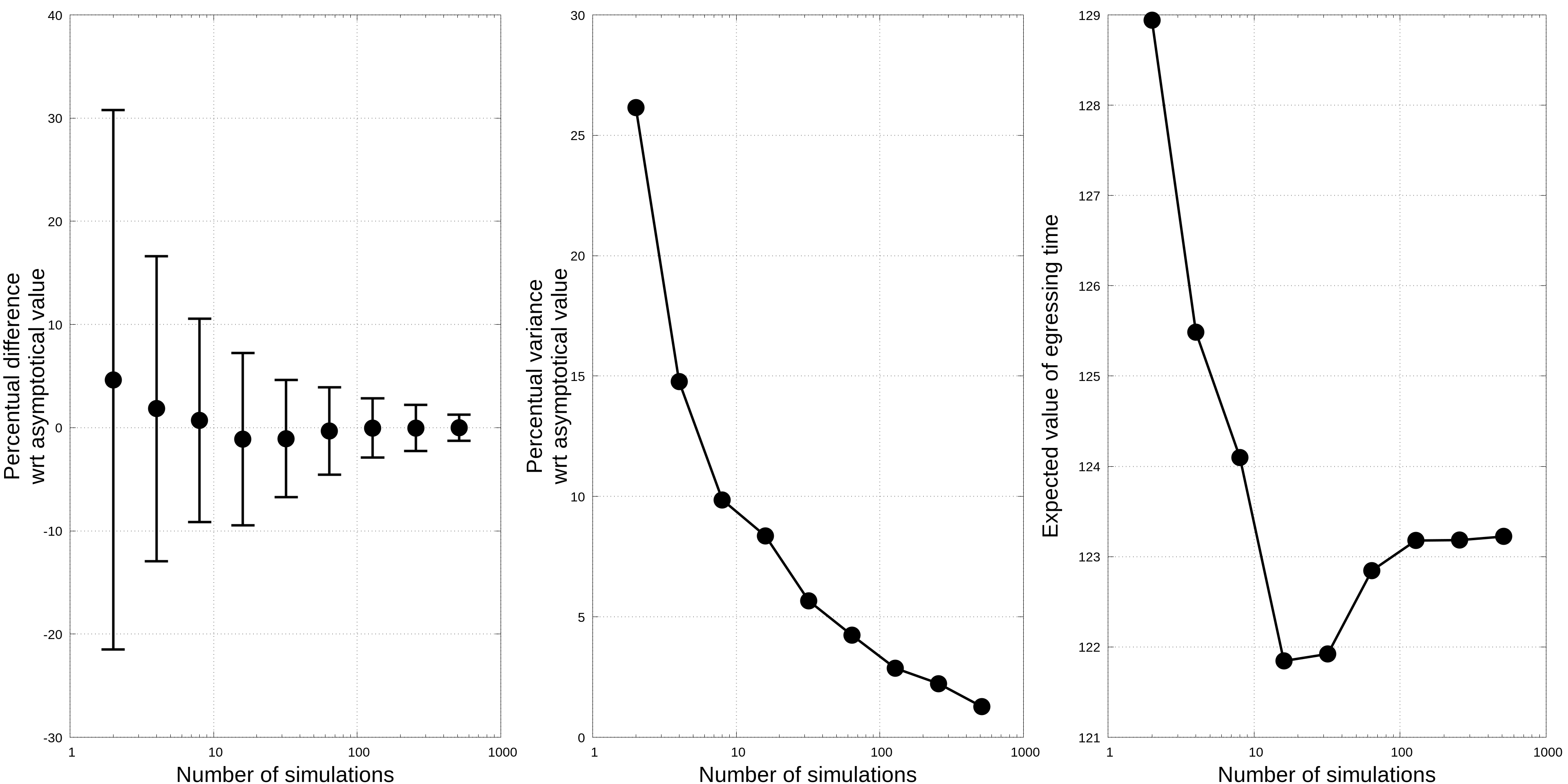}
		\caption{Expected value and variance as a function of the number of simulations.
		}\label{medie}
	\end{center}
\end{figure}

Now we need to have an estimate of the number of times we need to repeat the simulation in order to have a stable value of the statistical indicators. To do that, some numerical tests have been produced. The simulator has been run for a number $M$ of times, and this block of $M$ simulations have been repeated for 256 times. For each block, we can compute the expected value and the variance: after that, we can also analyze statistically the 256 blocks, computing the effective value and variance of the elementary expected value. Results are reported in Figure \ref{medie}.

\begin{figure}[h!]
	\begin{center}
		\includegraphics[width=0.95\textwidth]{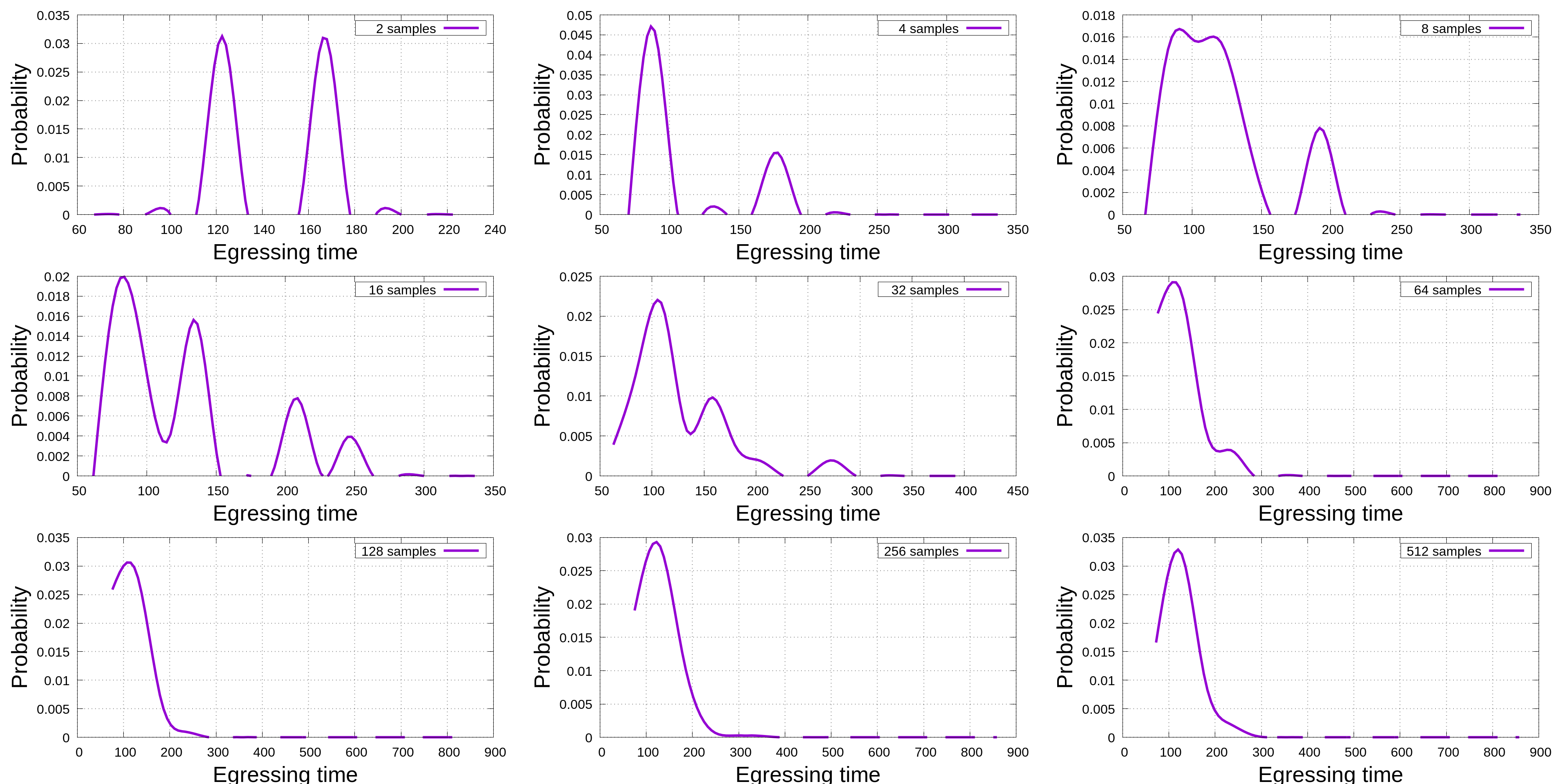}
		\caption{Experimental probability distribution of the expected value for the different blocks of simulations.
		}\label{prob}
	\end{center}
\end{figure}

The experimental probability distribution (EPD) of the expected value for the different blocks of simulations is reported in Figure \ref{prob}. From this picture, it is evident that a stable value of the expected value cannot be obtained if the number of simulations for each block is lower than 128. This information is also deducible observing the right sub-graph of Figure \ref{medie}: the expected value becomes stable for the indicated number of simulations. As a consequence, in the following 256 simulations will be applied in order to evaluate the qualities of a room configuration. 

\subsection{Optimization algorithm}\label{sec:ctrl_obs_GO}

In this study, an heuristic optimization algorithm is adopted, namely the Imperialist Competitive Algorithm ({\tt ICA}). In its original formulation \cite{Atashpaz2007}, the {\tt ICA} is described as an evolutionary algorithm. A number of trial vectors of parameters, each defining a different configuration of the system ({\em county}), are distributed onto the design space and assigned to different groups: each group is called {\em empire}. The {\em county} presenting the most convenient value of the objective function inside a {\em empire} is called {\em imperialist}, and each {\em county} is placed under the control of a single {\em imperialist}. Since here we are referring to a minimization problem, the {\em most convenient value} is represented by the lower value of the objective function. More details about the algorithm can be found in \cite{Atashpaz2007}. In the original formulation, the initialization of the {\em counties} is performed randomly. The {\em counties} are then assigned to an {\em imperialist} on the base of their relative power, so that, at the beginning, the most powerful {\em imperialist} have the control of a larger {\em empire}. At each iteration, three main actions are performed:

\begin{itemize}
	\item {\bf Shifting the {\em counties}:} each {\em county} is moved toward the {\em imperialist} according to the equation
	\begin{equation}\label{eq:basic}
	{\bf X}^k_{t+1} = {\bf X}^k_t + r \beta ({\bf X}^i-{\bf X}^k_t)
	\end{equation}
	where ${\bf X}$ is the generic vector of the coordinates of a point in the design variable space, ${\bf X}^i$ is the position of the (fixed) {\em imperialist} controlling the moving $k^{th}$ {\em county} ${\bf X}^k$, $t$ is the current iteration, $k$ is identifying the {\em county}, $\beta$ is the so-called {\em assimilation coefficient}, controlling the atractive action of the {\em imperialist} on the {\em county} and $r$ is a random number in between 0 and 1. If the product $r\beta$ is greater than the unit value, the {\em county} will overpass the {\em imperialist}, changing the side from which the {\em county} observes the {\em imperialist}. The displacement vector (${\bf X}^k_{t+1}-{\bf X}^k_t$) is further deviated from the indicated direction by a random angle in between -$\theta$ and +$\theta$, $\theta$ to be fixed.
	\item {\bf Change of the {\em imperialist}:} if a {\em county} finds a value of the objective function smaller than the value owned by the referenced {\em imperialist}, the positions of the {\em county} and the {\em imperialist} are swapped.
	\item {\bf Imperialistic competition:} the power of each {\em empire} is computed as the power of the {\em imperialist} plus a fraction $\xi$ of the sum of the powers of the single {\em counties} of the {\em empire}. The worst {\em county} of the worst {\em empire} is re-assigned to the best {\em empire}. In a minimization process, the average value of the power of the {\em counties} is summed up to the power of the {\em imperialist}: lower value means higher power.
	\item {\bf {\em Empire} elimination:} if, after the {\em Imperialistic Competition}, an {\em Imperialist} have no more {\em counties} under hits control, the {\em empire} is eliminated.
\end{itemize}

Looking at equation \eqref{eq:basic}, we can observe the full equivalence with the one dimensional, first-order, autonomous, linear differential equation that governs the evolution of a state variable
\begin{equation}\label{eq:linear}
y_{t+1} = a y_t + b.
\end{equation}
In fact, equation \eqref{eq:linear} is absolutely equivalent to equation \eqref{eq:basic} once we rewrite it in the reference frame of the {\em imperialist}. Since the value of the state variables are assigned at the beginning, that is, the relative position of the {\em county} with respect to the {\em imperialist}, at the first step we have
\begin{equation}
y_1 = a y_0 + b.
\end{equation}
Applying the equation \eqref{eq:linear}, we have that at step $t$
\begin{equation}
y_t = a^t y_0 + b \sum_{i=0}^{t-1} a^i. 
\end{equation}
If $b=0$, we can demonstrate that, if $0 < a < 1$, the series converges to the zero value (the origin of the reference frame) \cite{Galor:2007}. Since the local reference frame of an {\em Empire} coincides with an {\em imperialist}, each {\em county} converges toward the corresponding {\em imperialist}. To be more explicit, we can simplify the equation \eqref{eq:basic} by rewriting it in the reference frame of the {\em imperialist}: the term ${\bf X^i}$ disappears, and the equation now reads
\[
{\bf X}_{t+1} = {\bf X}_t - r \beta {\bf X}_t = (1 - r \beta) {\bf X}_t.
\]
We are clearly in the case of equation \eqref{eq:linear} where $b=0$ and $a=(1-r\beta)$. The motion is developing along the direction connecting the initial position of the {\em county} and the origin of the reference frame (that is, the {\em imperialist}). The {\em county} is converging on the corresponding {\em imperialist}: convergence is monotone or not depending on the value of $\beta$. Since the coefficient $a$ needs to be positive and smaller than the unit value in order to have convergence \cite{Galor:2007}, we have convergence if
\[
0 < 1 - r \beta < 1 \Rightarrow \beta < \frac{1}{r} ; \quad r \beta > 0.
\]
The different terminology adopted for the description of the algorithm is hiding a substantial similarity between {\tt ICA} and the multi-swarm Particle Swarm Optimization ({\tt PSO}) formulation \cite{Hendtlass}. With respect to the original formulation of {\tt PSO}, the {\tt ICA} has not a personal memory, so that the new position of a {\em county} is not influenced by the positions previously visited by itself, while {\tt PSO} is using this information. 
Conversely, both {\tt PSO} and {\tt ICA} show a limited interaction with the other elements of the {\em empire}: in fact, the {\em Imperialist} is the equivalent of the best element of {\tt PSO}, that is, the best visit of the whole swarm/{\em empire}. The great difference with {\tt PSO} is the aforementioned proof of convergence of {\tt ICA}, while for {\tt PSO} an incomplete proof of convergence can be obtained \cite{Clerc}. 

An improvement of the original {\tt ICA} is proposed in \cite{peri2019CAIE}, namely {\tt hICA}, and this version of the algorithm is here applied. The improvements obtained by {\tt ICA} can be addressed mainly to the following modifications:

\begin{enumerate}
	\item The initial distribution of the {\em counties} is not random, but it is produced using an Uniformly Distributed Sequence.
	\item The coefficients in equation \eqref{eq:basic} have been optimized.
	\item The {\em empires} are re-initialized if only a single {\em empire} is survived.
	\item A local search algorithm (Simplex method) is applied if we have no improvements of the current best solution after a certain number of iterations. This is surely one of the main improvements of the algorithm.
\end{enumerate}

\subsection{Test case}\label{sec:ctrl_obs_test}

In order to empathize the ability of a fixed obstacle to efficiently redirect the flow of a group of pedestrians, such that the evacuation time is minimized, a very simple test case has been designed. Four entrance and four exit are symmetrically placed in a square room. The entrances are at the corners, the exit at the center of each side of the room. The role of the obstacle(s), in this case, is to break the symmetry of the flow, avoiding indecision (and the subsequent dead time) when different sub-groups are colliding, exploiting also all the exits.

The full number of pedestrians has been fixed to 100, in order to have a good balance between interactions and computational time. Simulation has been repeated 256 times for each configuration of the room in order to derive statistical variables, as from the indications collected previously (Figure \ref{medie}).

The only constraint is related to the distance between the wall and the obstacle, in order to avoid blockage effects (and also the exclusion of an exit or an entrance).

Regarding the selection of the obstacle(s), two different cases have been considered: one or two linear walls. A single wall is defined using four variables: two for the barycentre of the wall, one for its full length and one for the orientation (in between 0 and 90 degrees). The width of the wall is fixed. As a consequence, we have 4 design variables for a single wall and 8 design variables for a couple of walls. The design variables are selected in order to reduce the possibility of violation of the constraint, so that the barycentre of the wall cannot stay on the border of the room. Minimum and maximum length of the walls are also fixed.

Stopping criterion for the optimization algorithm is represented only by the full number of evaluations of the objective function: in order to balance the opportunities of the two optimization problems, in accordance with \cite{peri2019CAIE}, the maximum number has been fixed at 1000$\times NDV$, where $NDV$ is the number of design variables. Consequently, the problem with more design variables takes longer to complete.

Due to the symmetry of entrance and exit, the solution of the problem is cyclical, since four configurations can be obtained by a rotation of 90 degrees around the center of the room.

\subsection{Numerical results}\label{sec:ctrl_obs_res}

\begin{figure}[h!]
	\begin{center}
		\includegraphics[width=0.95\textwidth]{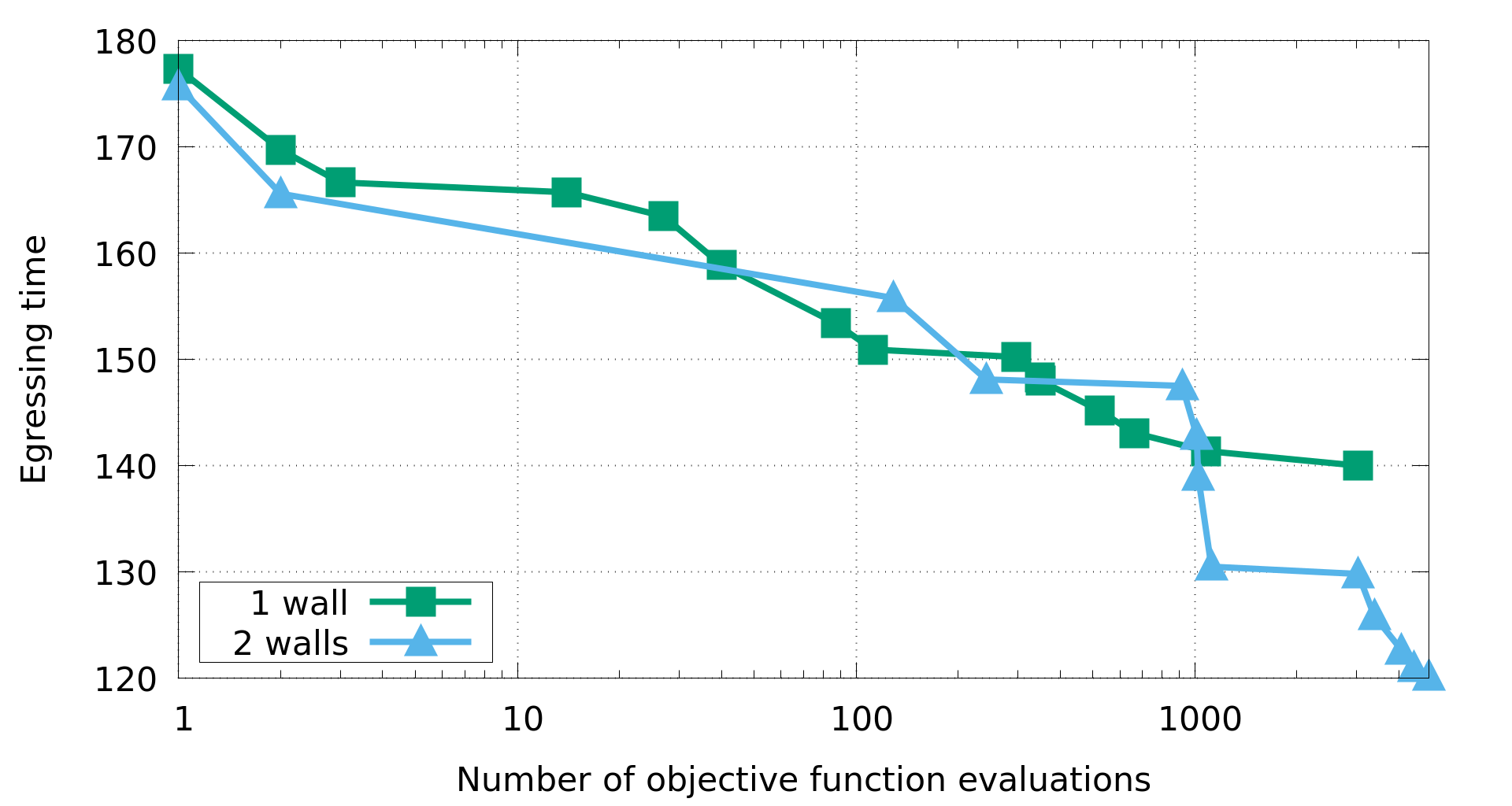}
		\caption{Convergence history of the optimization problems: case of the room with one obstacle and room with two obstacles.
		}\label{ConvOpt}
	\end{center}
\end{figure}

In Figure \ref{ConvOpt}, the convergence history of the two optimization problems is reported. The rate of convergence of the two problems is very similar as soon as, in the case of two walls, the optimizer is able to identify a new solution improving largely the egress time, further refined in the last part of the optimization problem solution. On the contrary, the identification of the optimal solution for the problem with a single wall appears to be pretty fast, and only marginal improvements are obtained after a couple of iterations: this is probably connected with the simplicity of the shape of the obstacle, unable to create a great variety of convenient situations.
%
\begin{table}\caption{Expected value, variance and objective function for the different optimal configuration plus the case of empty room. Statistics are obtained performing 100,000 simulations.
	}\label{TabOpt}
	\begin{center}
		\begin{tabular}{|l|g|g||l|l||g|g||l|l|}  \hline
			& $EV$   & $\Delta$\% & Most probable  & $\Delta$\% & $\sigma$ & $\Delta$\% & $EV + \sigma$ & $\Delta$\% \\ \hline\hline
			Empty  & 133.22 &            & 114.43         &            & 39.99    &            & 173.21        &            \\ \hline
			1 Obs. & 114.66 & -14.10\%   & 100.73         & -12.28\%   & 36.94    & -7.63\%    & 151.60        & -12.48\%   \\ \hline
			2 Obs. & 93.50  & -29.82\%   & 59.71          & -48.00 \%  & 37.22    & -6.93\%    & 130.72        & -24.53\%   \\ \hline
		\end{tabular}
	\end{center}
\end{table}

The objective function value, the most probable egress time, the expected value and the variance of the optimal configurations are reported in Table \ref{TabOpt}. The variance is substantially unchanged at the end of the optimization process: this is a partly unexpected result. One might therefore imagine that the regularization of the pedestrian flow, obtained through the obstacles, would also reduce the variability of the dwell time into the room. The reason of the small reduction of the variance can be linked with the constraint on the distance between the wall and the sides of the room: there is still a quite large gap between the obstacles and the borders. This gap has been introduced considering the fact the main function the room is designed for must be preserved after the insertion of the walls, so that their impact on the environment should be limited. As a consequence, the pedestrian, although driven toward the exit, have still a quite large space to explore, and the random component of the individual speed plays a not negligible role. As a consequence, the variance of the egress time is substantially not changing.

It is really interesting to compare the EPD of the egress times for the case of empty room with the ones of the optimized solutions. For these three configurations, 100,000 simulations have been produced in order to increase the stability and credibility of the statistic indicators. The most probable value of the egress time is shifted to the lower values passing from empty room to one obstacle to two obstacles, as it is also evident from Table \ref{TabOpt}. In this last case, the higher probability is very close to the minimum egress time, representing a very good feature of the optimal configuration.

\begin{figure}[h!]
	\begin{center}
		\includegraphics[width=0.95\textwidth]{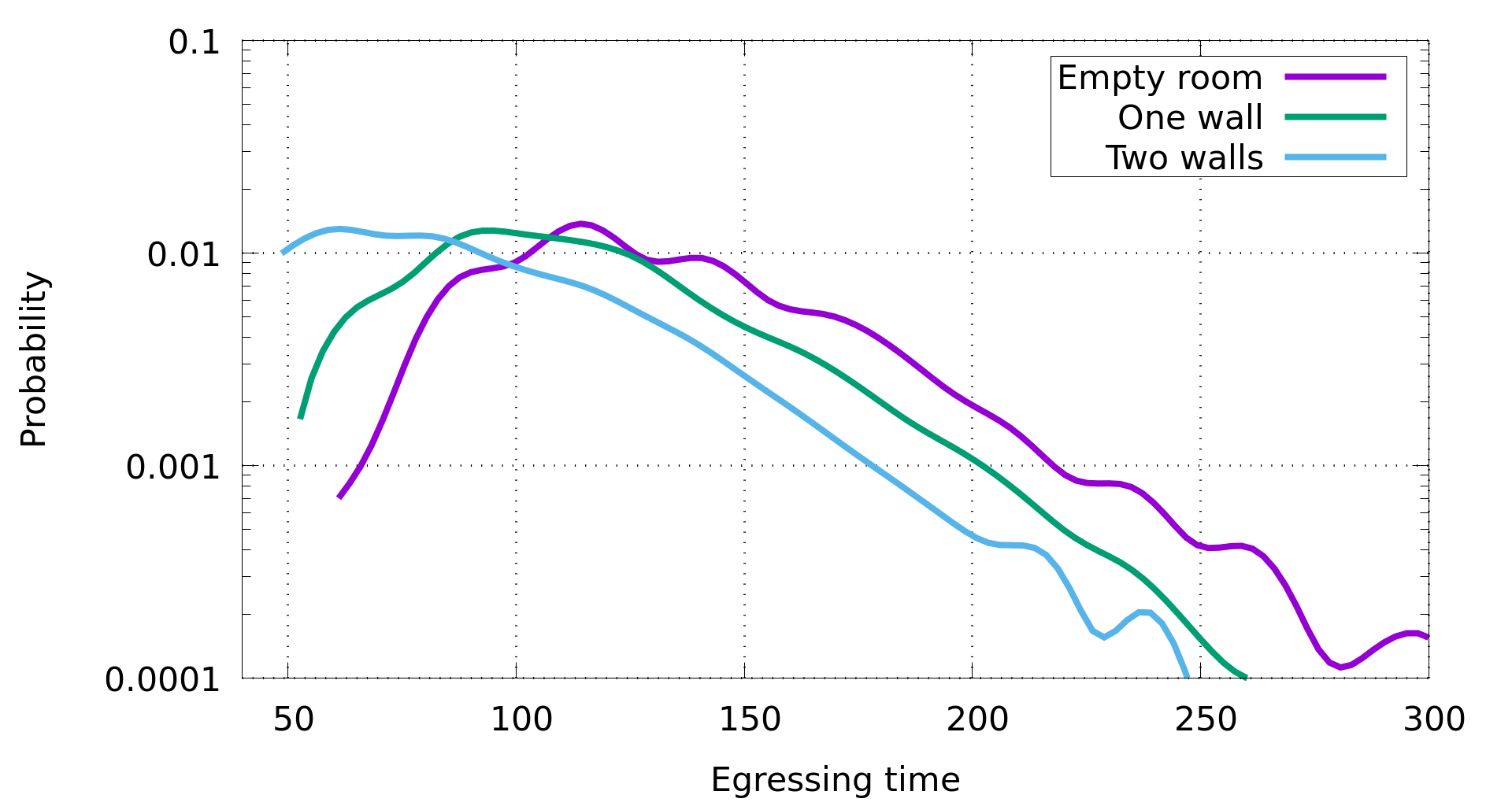}
		\caption{Experimental probability distribution of the egress time in the case of empty room, room with one obstacle and room with two obstacles.
		}\label{ProbOpt}
	\end{center}
\end{figure}

\begin{figure}[h!]	
	\begin{center}
		\includegraphics[width=0.24\textwidth]{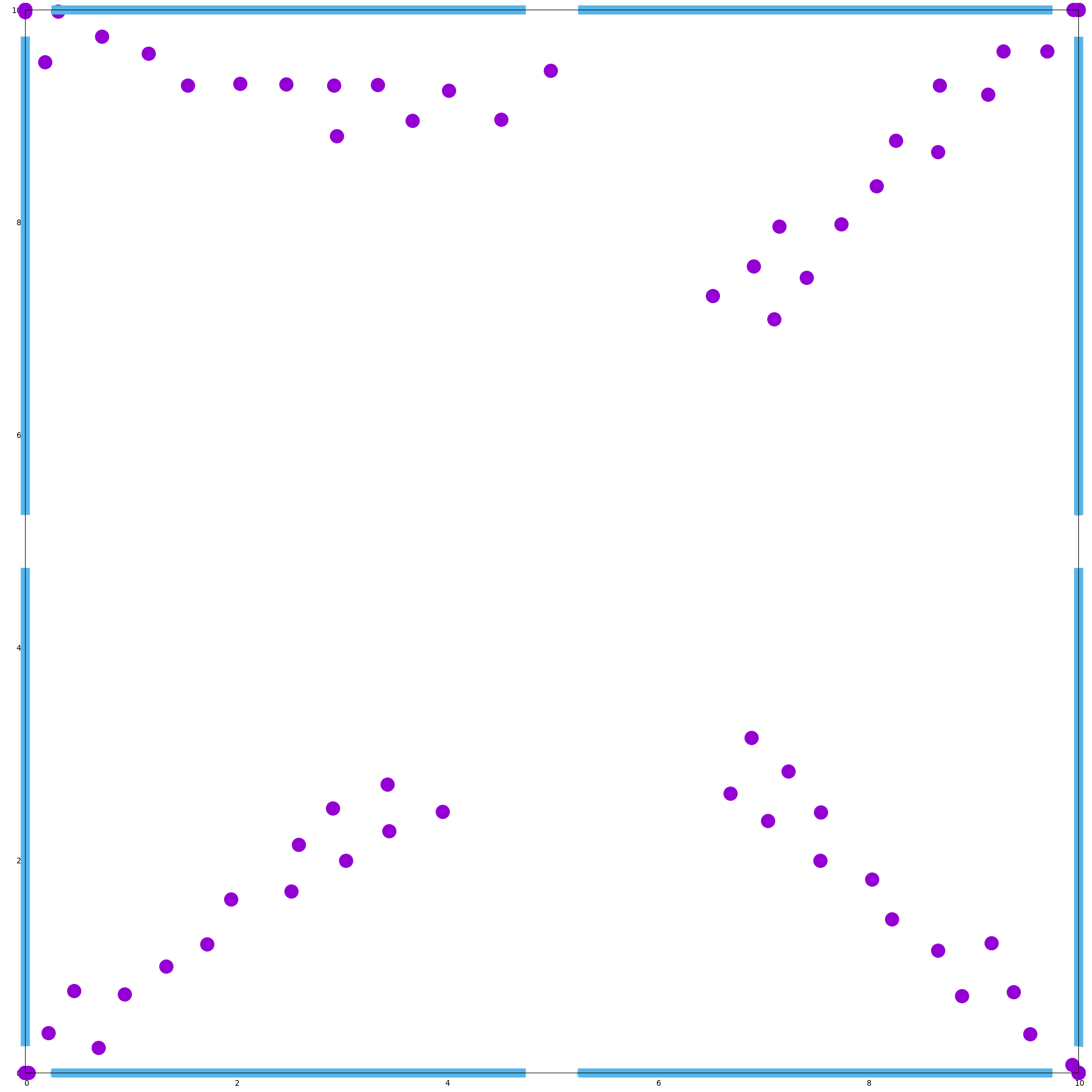}
       \includegraphics[width=0.24\textwidth]{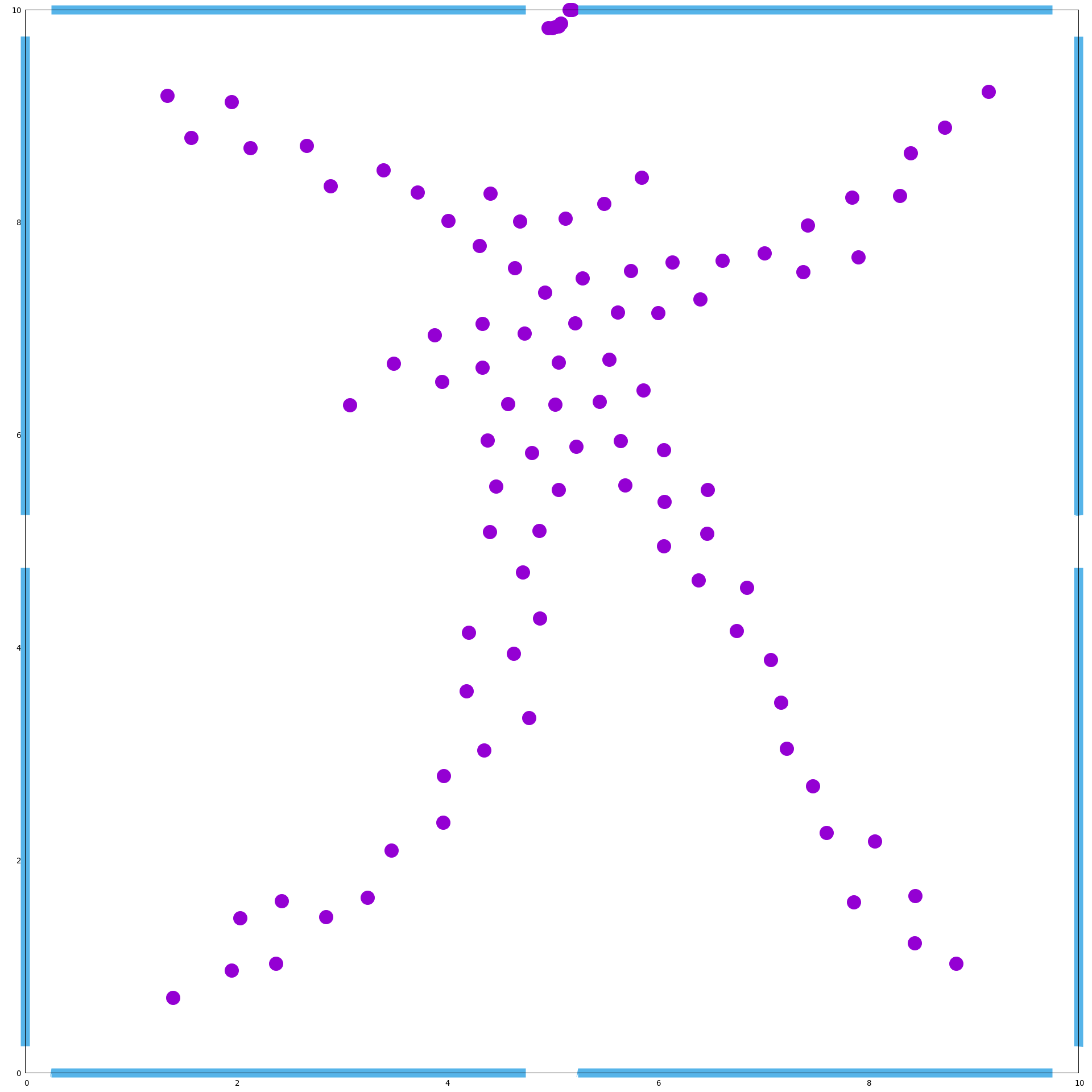}
       \includegraphics[width=0.24\textwidth]{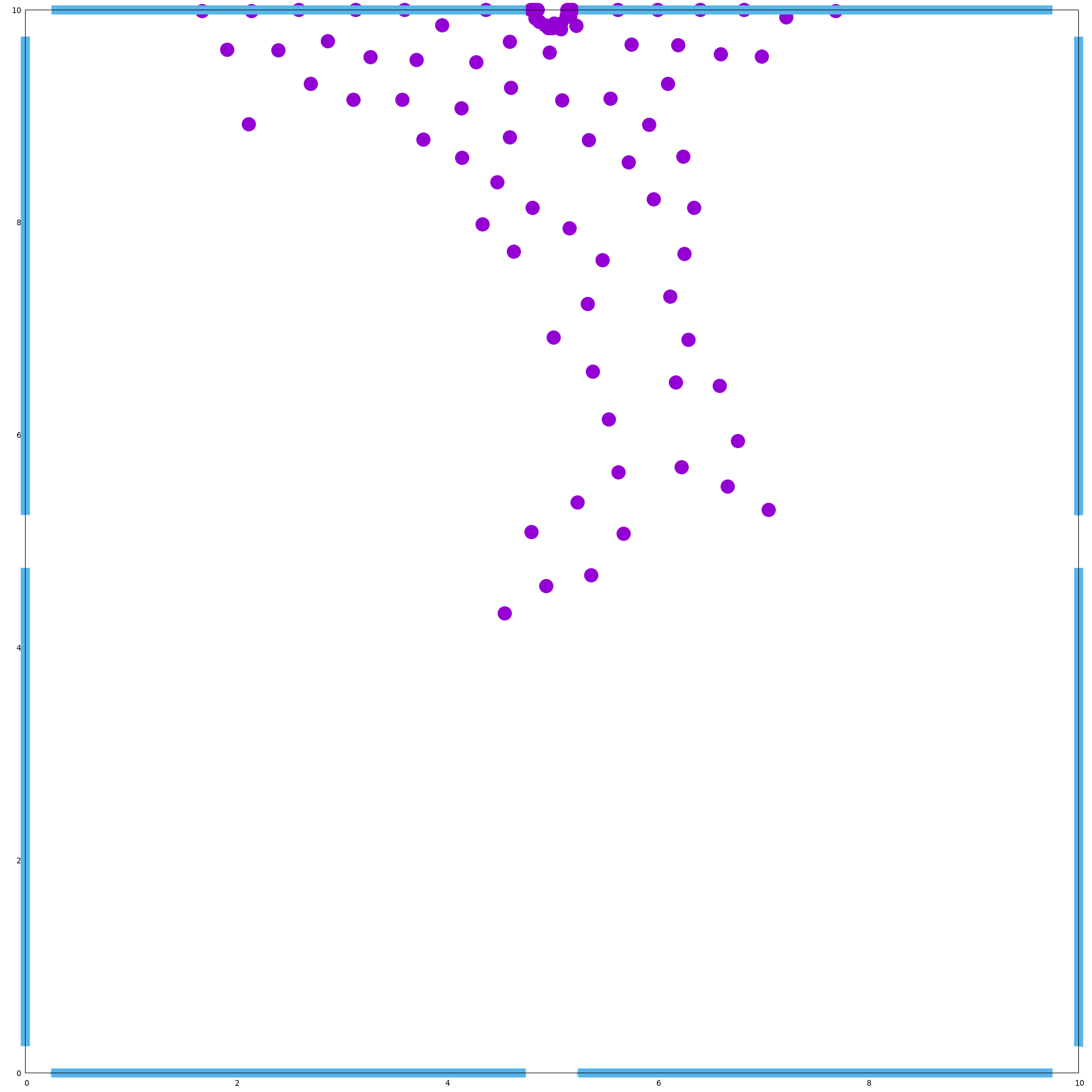}
       \includegraphics[width=0.24\textwidth]{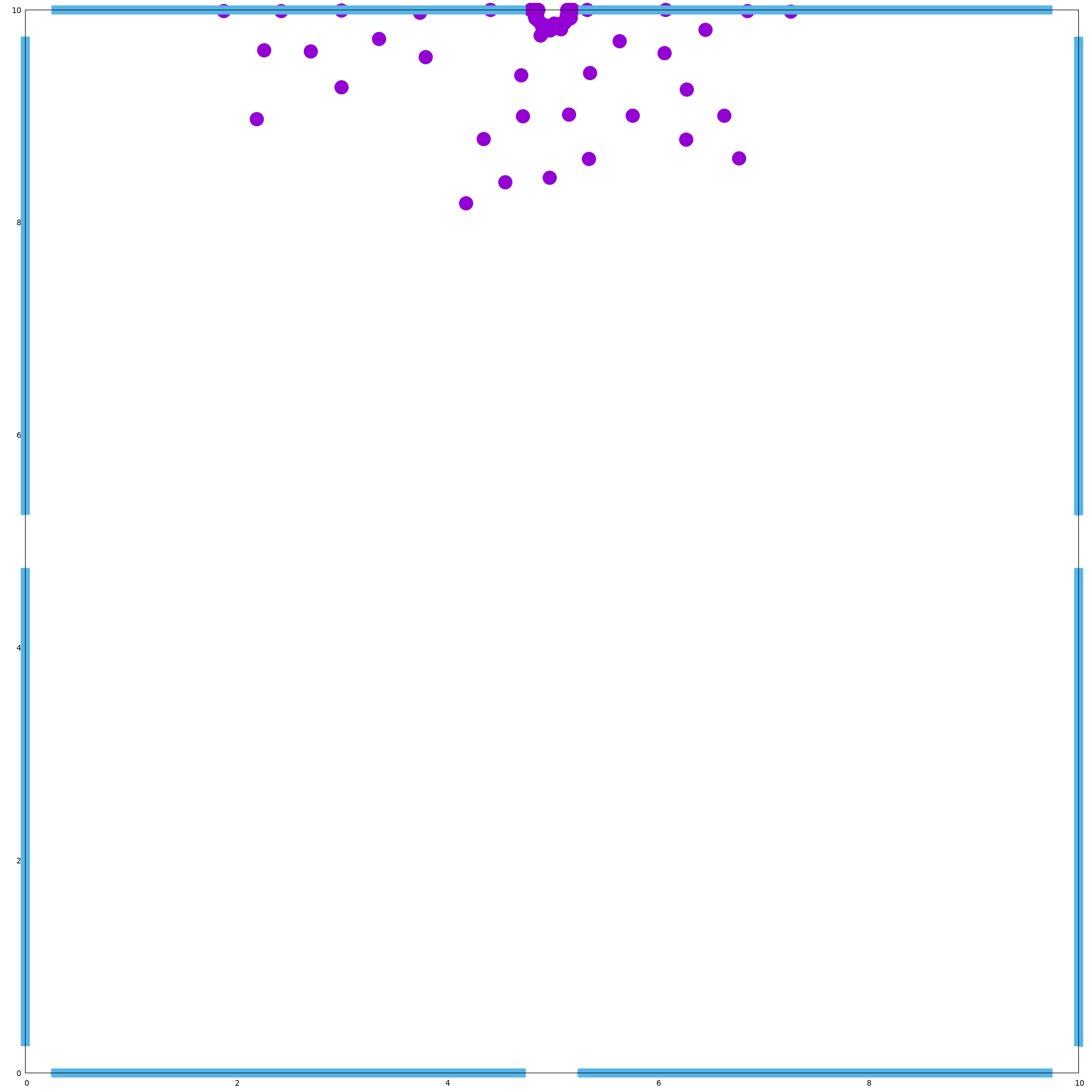} \\
		\includegraphics[width=0.24\textwidth]{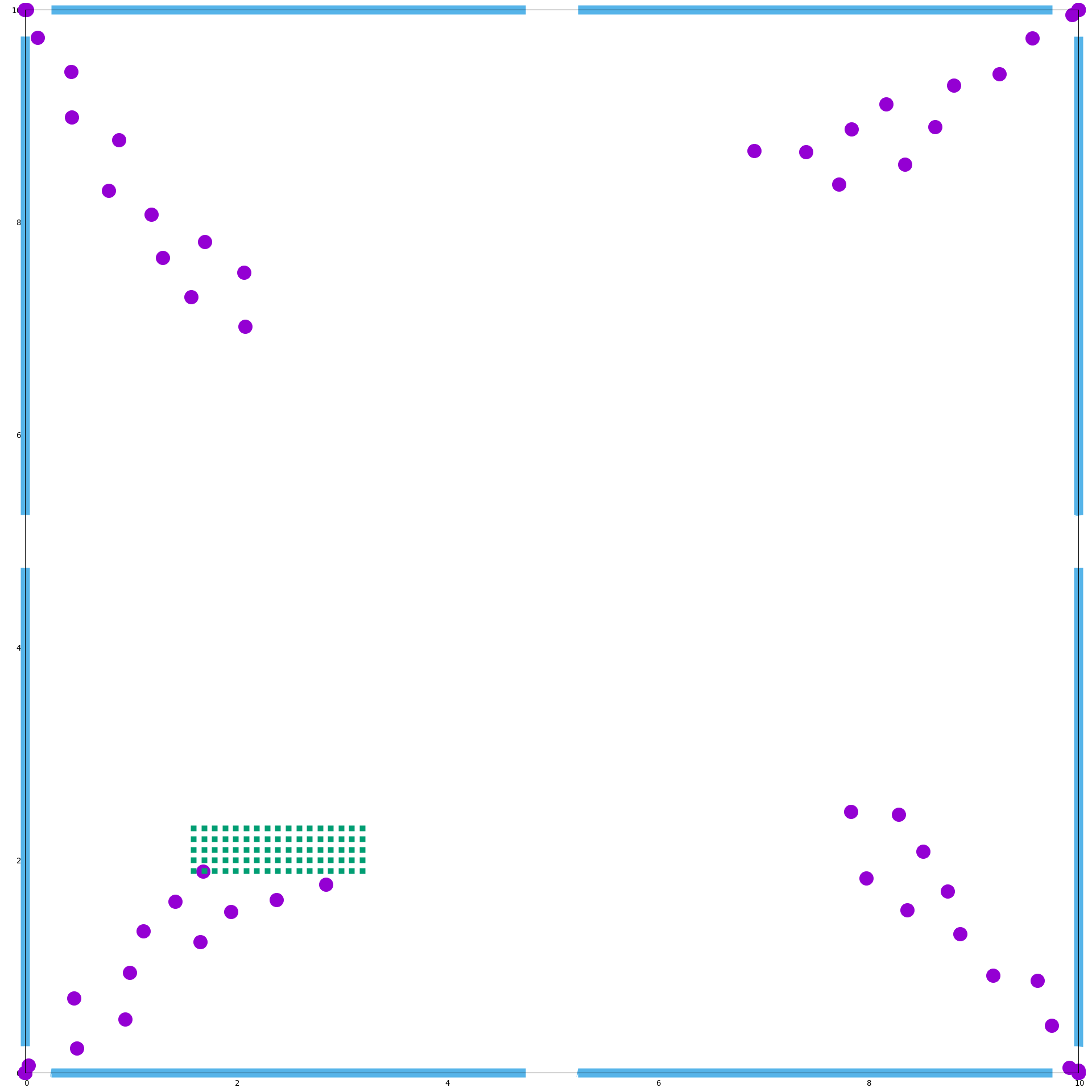}
       \includegraphics[width=0.24\textwidth]{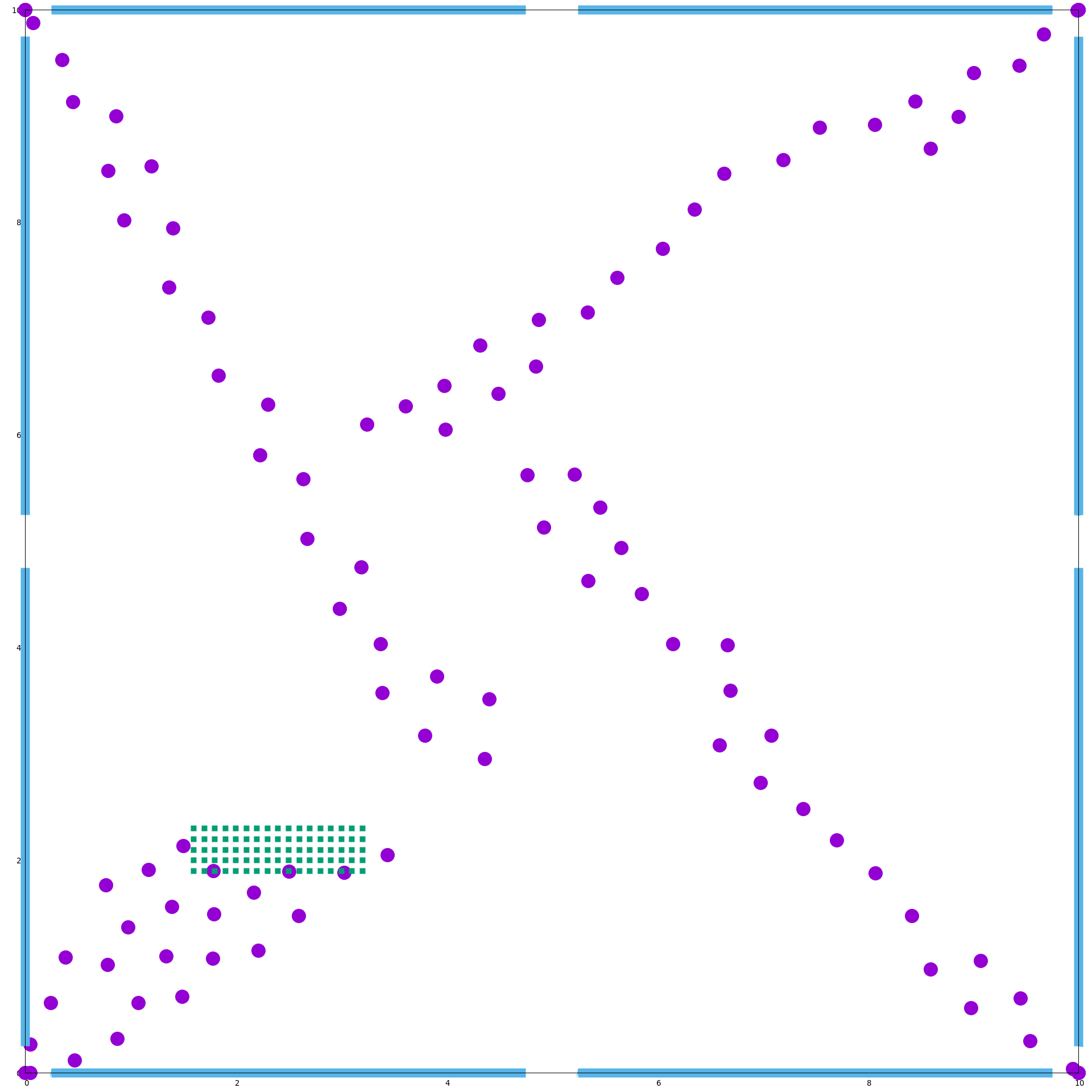}
       \includegraphics[width=0.24\textwidth]{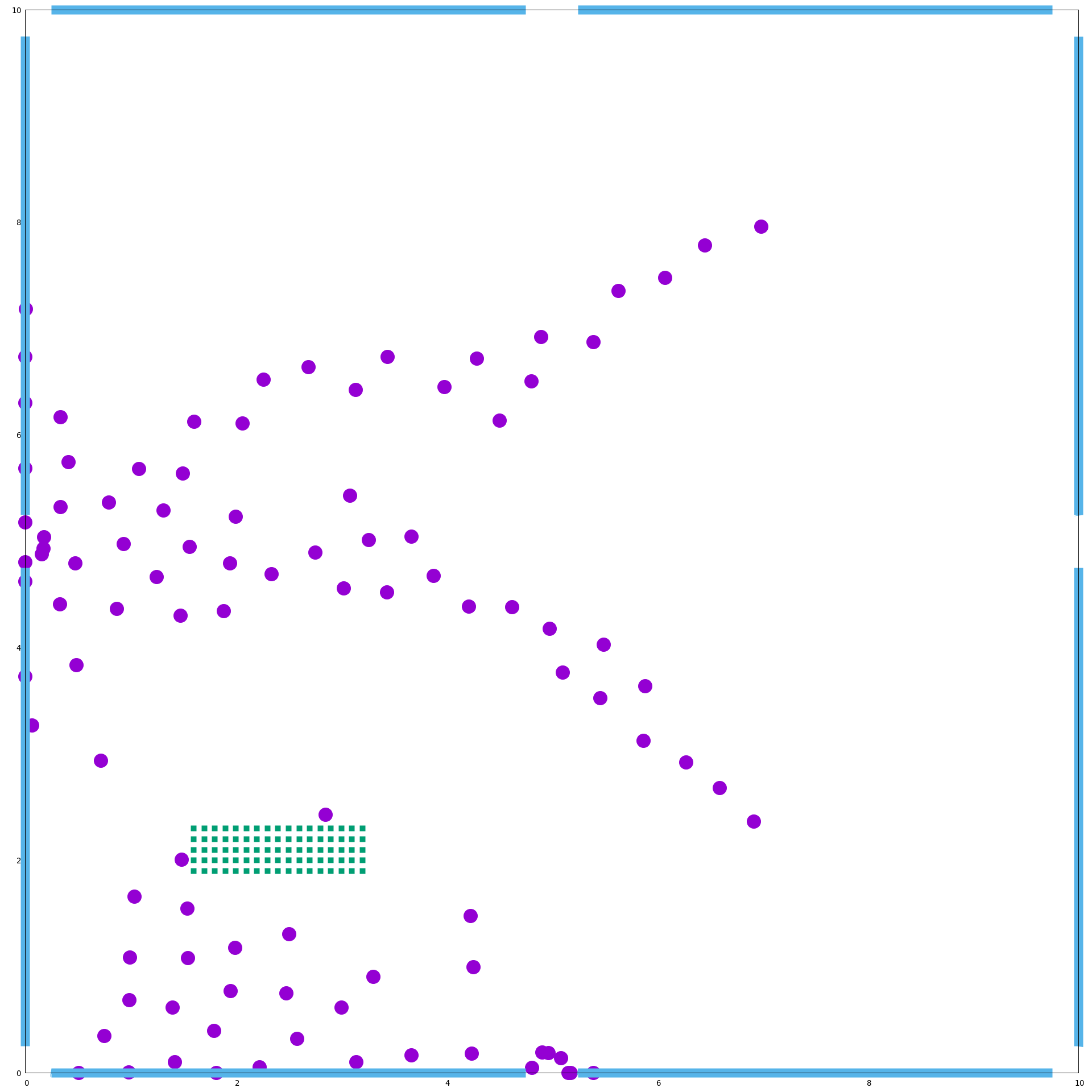}
       \includegraphics[width=0.24\textwidth]{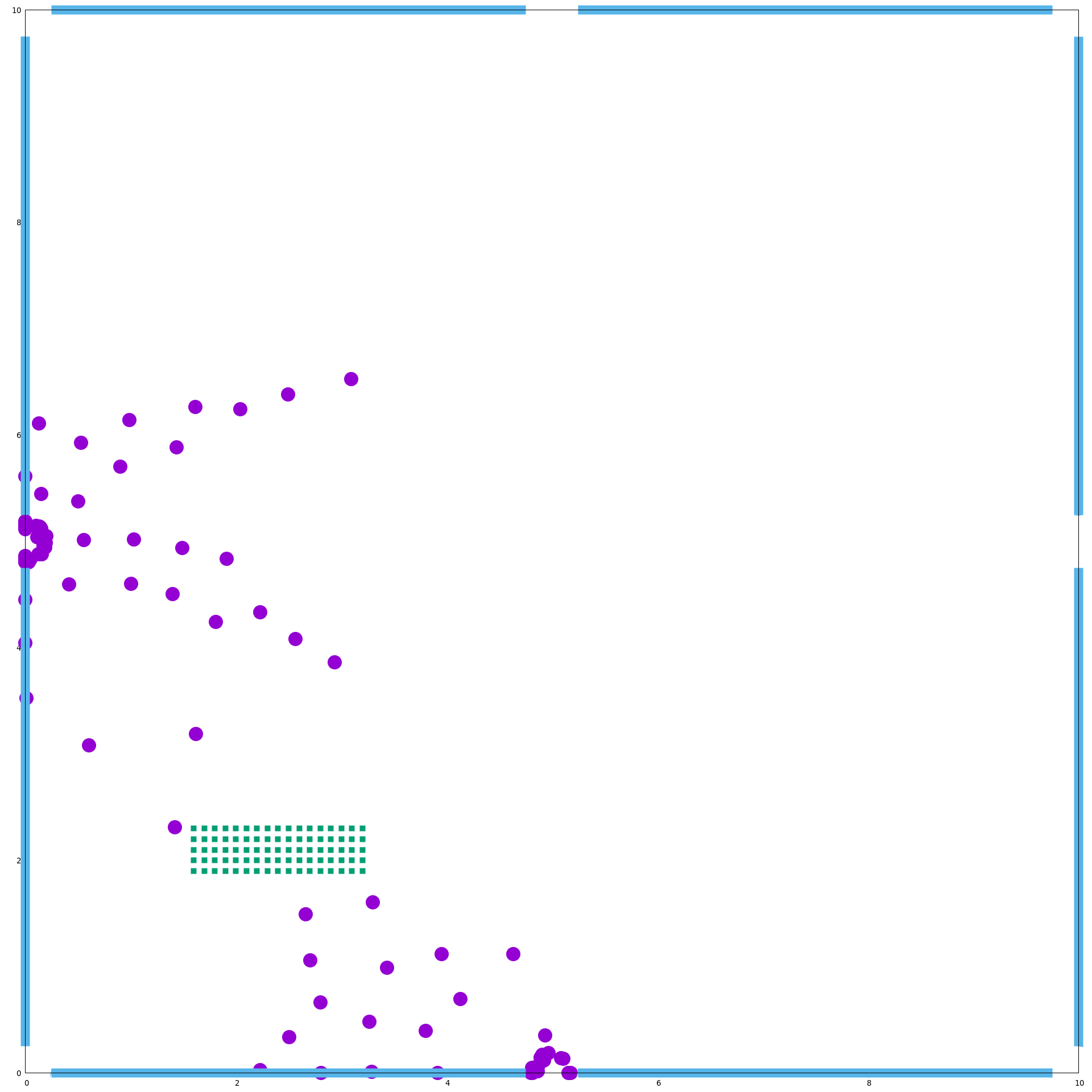} \\
		\includegraphics[width=0.24\textwidth]{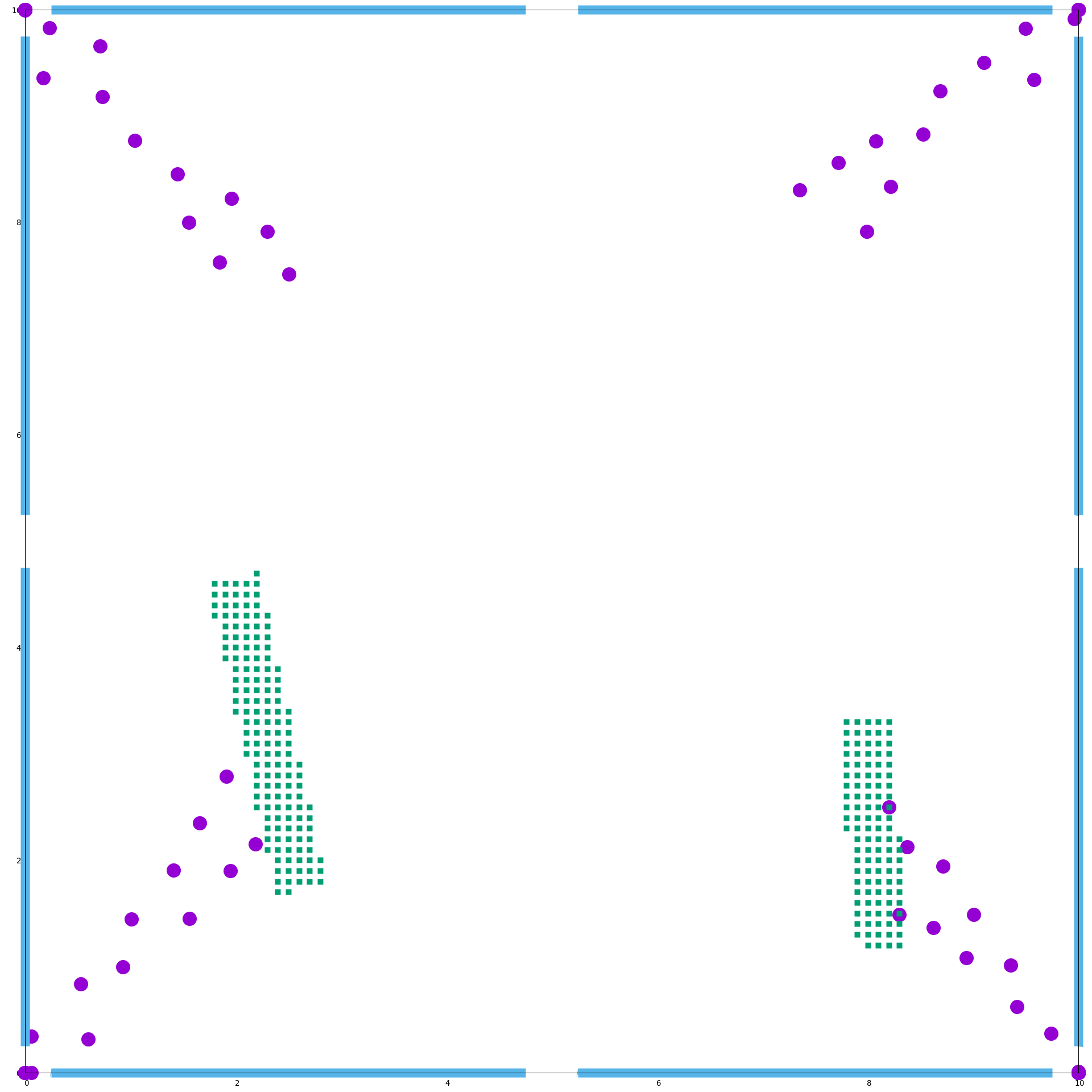}
		\includegraphics[width=0.24\textwidth]{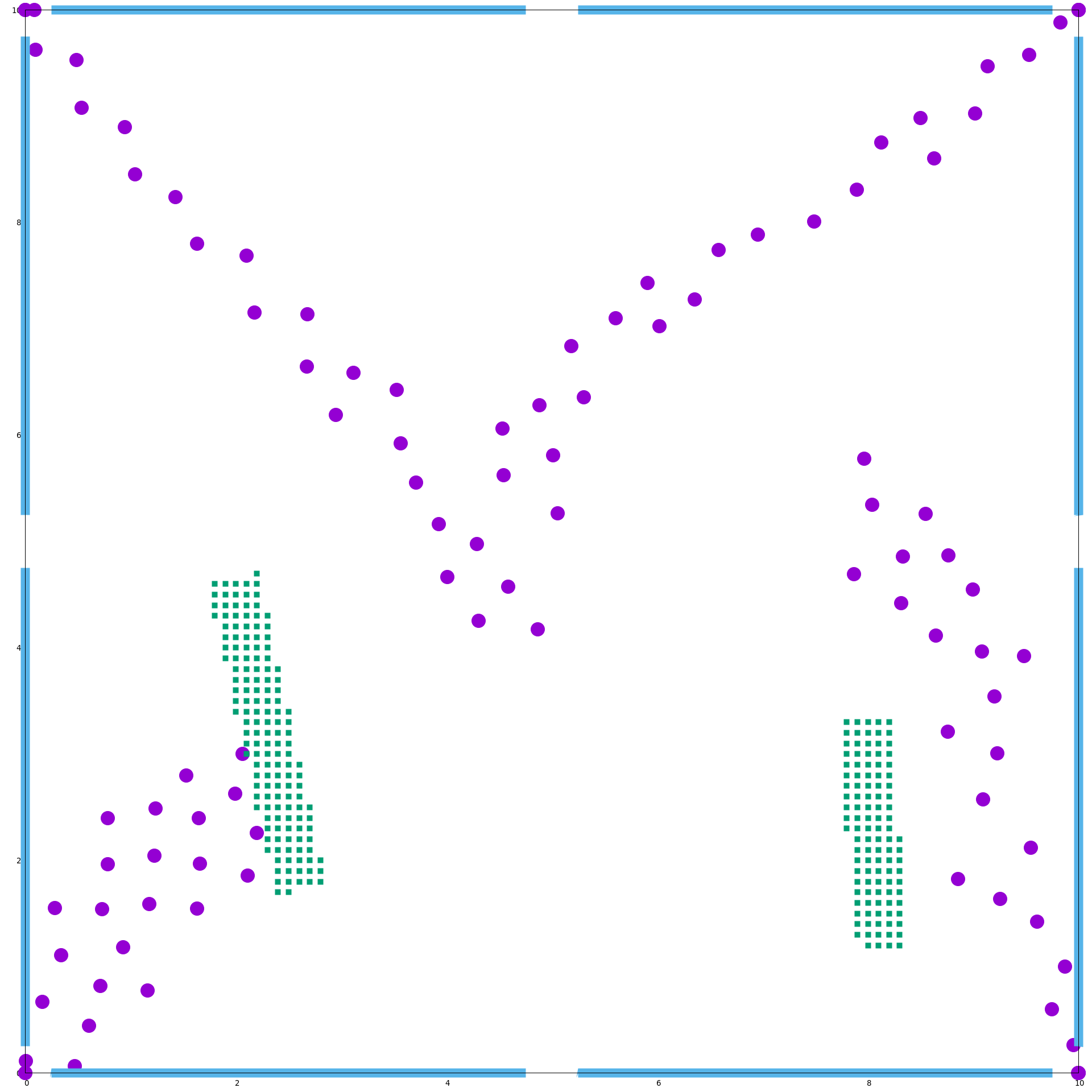}
		\includegraphics[width=0.24\textwidth]{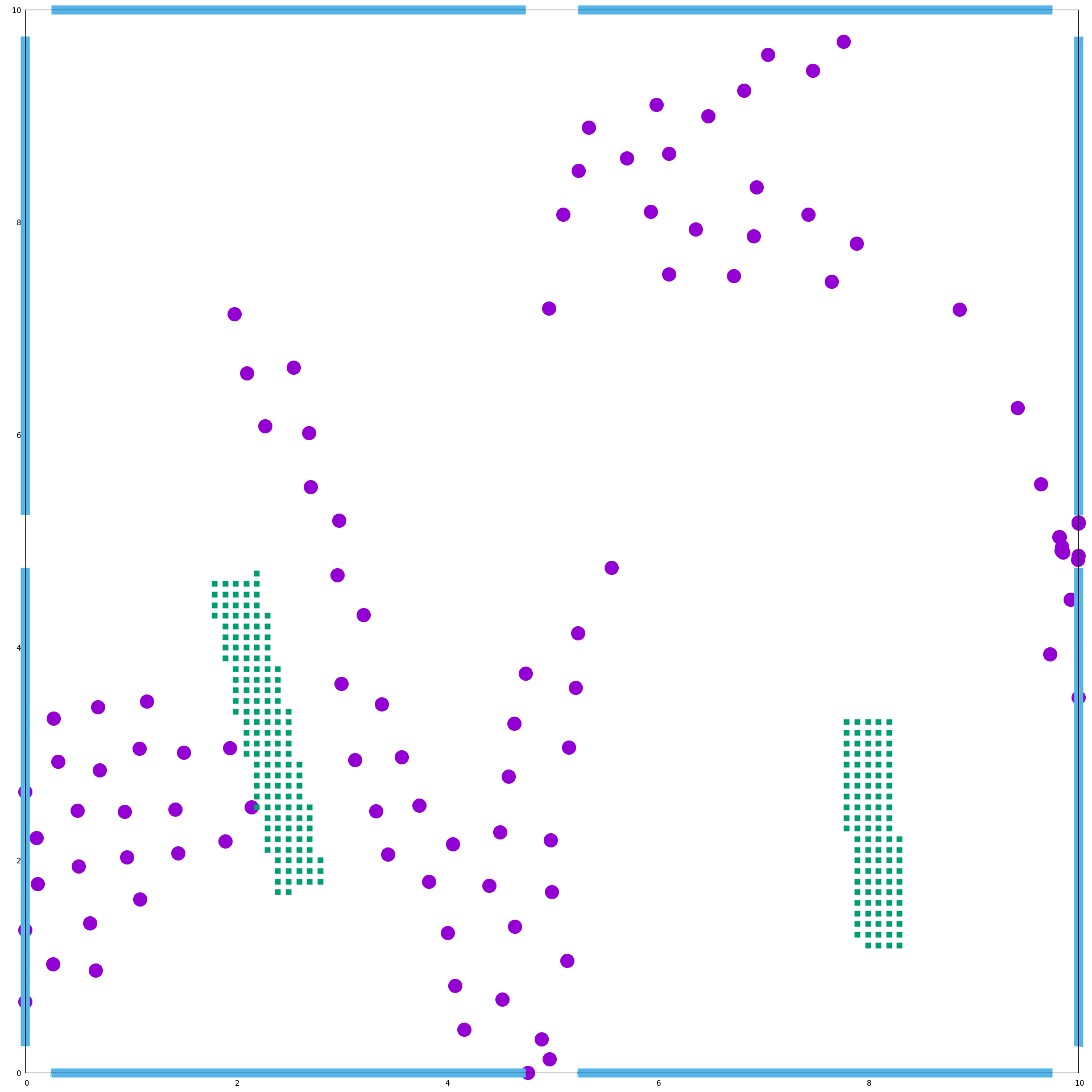}
		\includegraphics[width=0.24\textwidth]{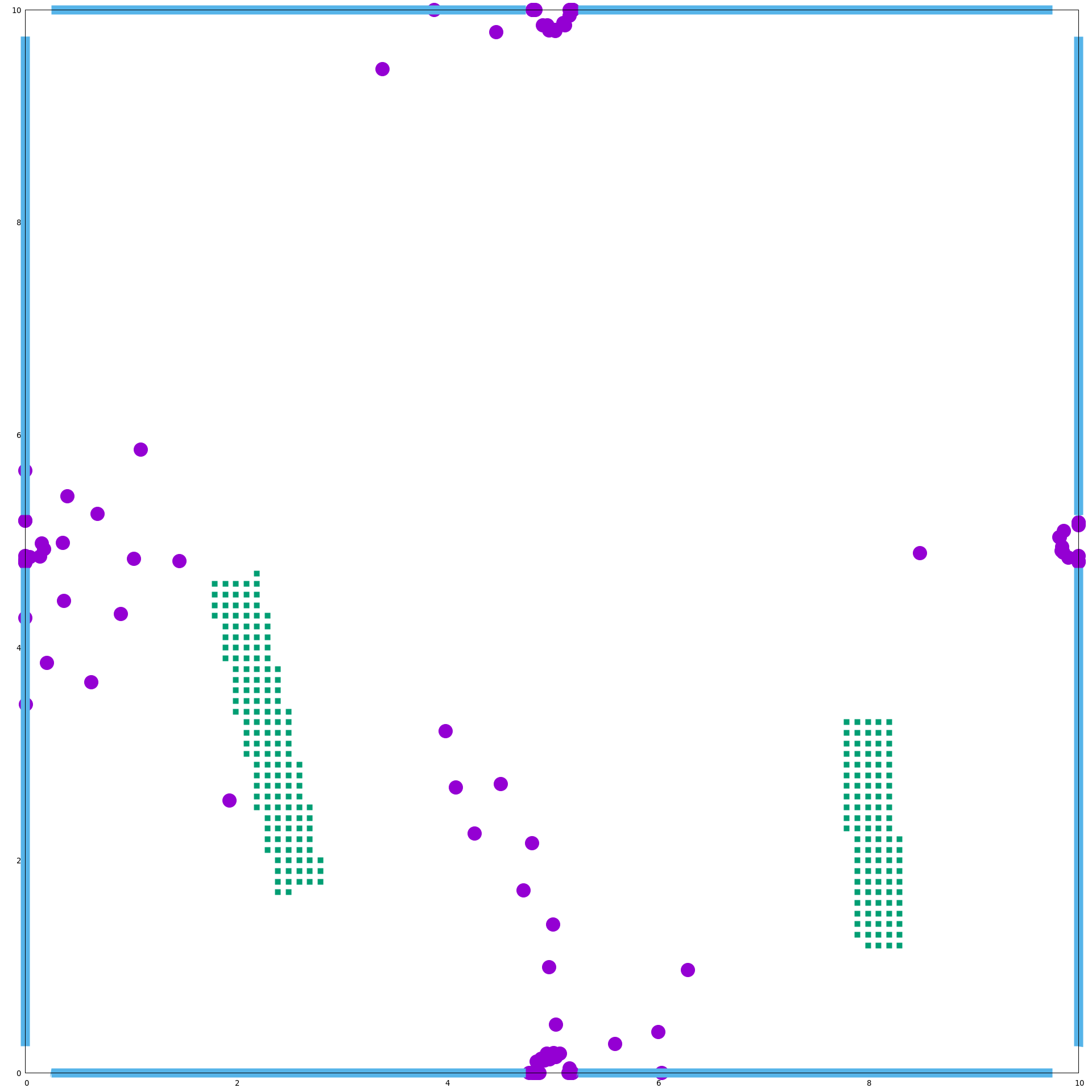} \\
		\caption{Flow of the pedestrian in the case of empty room, room with a single wall and room with two walls.
		}\label{flussi}
	\end{center}
\end{figure}

In Figure \ref{flussi}, the trajectories of the pedestrians are reported: from top to bottom the number of walls is increasing, while the time of the simulation is running from left to right. Best solution among 100,000 simulations is reported, and this is particularly advantageous for the case of the empty room, since the probability of the reported configuration is relatively small. The final outcome from this study is that a single wall is redirecting a single group of pedestrians. Naturally, if no wall is adopted, all the pedestrians are converging at the center of the room. This is connected with this specific configuration: in fact, when a pedestrian is entering the room, in order not to hit the boundaries s/he move toward the center of the room following the bisector of the corner. 
The following pedestrians have a further attraction, that is the trajectory of the preceding pedestrian(s), so that typically all the groups are moving (on average) along the bisector of the angle between the walls. 
The elimination of one or more groups from this path is facilitating the deviation of the converging part of the group to one of the exit. When a single obstacle is used, one group is segregated, and it moves toward the closest exit. 
If two obstacles are utilized, two groups are eliminated from the central area and the remaining two groups are moving together toward the opposite exit. In the particular case reported in Figure \ref{flussi}, the tail of the right upper group is shifted to the top by a subset of the group entering from the lower right corner: this way, all the exits are exploited, and the congestion at the exits is reduced, also lowering the overall egress time.
%

We can then conclude that the use of an optimizer for the determination of the best configuration is essential for at least two distinct reasons. Firstly, although the final configuration appears to be logical, it is not easy to be identified without an aid (the paradox of the egg of Columbus). Secondly, the fine tuning of the general layout of the walls is providing an advantage impossible to obtain by a simple manual positioning. Finally, these two aspects have been pointed out in the case of a relatively simple room geometry: a much more complex geometry would include more and more difficulties in the determination of the optimal configuration, and the use of an optimization algorithm becomes vital.
\section{Conclusions and research directions}\label{sec:conclusions}
This survey has been devoted to present some recent results in the mathematical modeling and control of crowd dynamics. We discussed the various level of modeling, from the microscopic scale of agent based systems to the macroscopic scale of the crowd density and bulk velocity, through the mesoscopic scale based on a statistical description of the system. Several corresponding control problems, aimed at minimizing the escape time of the crowd from a given environment, have been illustrated and solved by numerical methods.

These results allow us to draw some conclusions. First of all, we can say that while the \emph{modeling} (i.e.\ the mathematical description) of pedestrian flows has now reached a stage of maturity, the same cannot be said for the \emph{optimization} of pedestrian flows. In this field there is still room for many experiments, both virtual and real. 

In the case of control through leaders, their impact on the crowd is not yet completely understood. In particular, in crowd management it is of paramount importance to be able to secure crowd evacuation through minimal intervention in order to avoid adversarial behaviors against authorities, and de-escalate tensions.  Indeed, we have shown that few agents may change completely the behavior of the whole system, breaking initial uncertainties. A further research direction concerns the optimal positioning and amount of leaders within the crowd at the time of the first movement. 

In the case of optimization through obstacles, basically no experiment was conducted on real people (we do not consider here the experiments investigating the effect of small obstacles in front of exit doors). 
Although the simulations suggest the existence of multiple optimal configurations of the obstacles, and it is therefore not easy to choose which to put into practice, virtual experiments all lead in the same, clear direction: \textit{breaking of symmetry} is beneficial to pedestrian flows. This means, e.g., that clogging can be avoided by redirecting people through asymmetric paths, which lead people at exits at different times. Moreover, the perception of the walking area can be completely upset by using smart obstacles, in such a way that naturally chosen exit paths are rebuilt for a more efficient exit usage.

\bibliographystyle{spmpsci} 
\bibliography{biblio_crowd}

\begin{thebibliography}{10}
\providecommand{\url}[1]{{#1}}
\providecommand{\urlprefix}{URL }
\expandafter\ifx\csname urlstyle\endcsname\relax
  \providecommand{\doi}[1]{DOI~\discretionary{}{}{}#1}\else
  \providecommand{\doi}{DOI~\discretionary{}{}{}\begingroup
  \urlstyle{rm}\Url}\fi

\bibitem{abdelghany2014EJOR}
Abdelghany, A., Abdelghany, K., Mahmassani, H., Alhalabi, W.: Modeling
  framework for optimal evacuation of large-scale crowded pedestrian
  facilities.
\newblock European J. Oper. Res. \textbf{237}(3), 1105--1118 (2014).
\newblock \doi{10.1016/j.ejor.2014.02.054}

\bibitem{agnelli2015M3AS}
Agnelli, J.P., Colasuonno, F., Knopoff, D.: A kinetic theory approach to the
  dynamic of crowd evacuation from bounded domains.
\newblock Math. Models Methods Appl. Sci. \textbf{25}(1), 109--129 (2015).
\newblock \doi{10.1142/S0218202515500049}

\bibitem{ABFHKPPS}
Albi, G., Bellomo, N., Fermo, L., Ha, S.Y., Kim, J., Pareschi, L., Poyato, D.,
  Soler, J.: Vehicular traffic, crowds, and swarms: from kinetic theory and
  multiscale methods to applications and research perspectives.
\newblock Math. Models Methods Appl. Sci. \textbf{29}(10), 1901--2005 (2019).
\newblock \doi{10.1142/S0218202519500374}

\bibitem{albi2016SIAP}
Albi, G., Bongini, M., Cristiani, E., Kalise, D.: Invisible control of
  self-organizing agents leaving unknown environments.
\newblock SIAM J. Appl. Math. \textbf{76}(4), 1683--1710 (2016).
\newblock \doi{10.1137/15M1017016}

\bibitem{AlHePa15}
Albi, G., Herty, M., Pareschi, L.: Kinetic description of optimal control
  problems and applications to opinion consensus.
\newblock Commun. Math. Sci. \textbf{13}(6), 1407--1429 (2015).
\newblock \doi{10.4310/CMS.2015.v13.n6.a3}

\bibitem{albi2013MMS}
Albi, G., Pareschi, L.: Binary interaction algorithms for the simulation of
  flocking and swarming dynamics.
\newblock Multiscale Model. Simul. \textbf{11}(1), 1--29 (2013).
\newblock \doi{10.1137/120868748}

\bibitem{albi2013AML}
Albi, G., Pareschi, L.: Modeling of self-organized systems interacting with a
  few individuals: from microscopic to macroscopic dynamics.
\newblock Appl. Math. Lett. \textbf{26}, 397--401 (2013)

\bibitem{AP18a}
Albi, G., Pareschi, L.: Selective model-predictive control for flocking
  systems.
\newblock Commun. Appl. Ind. Math. \textbf{9}(2), 4--21 (2018).
\newblock \doi{10.2478/caim-2018-0009}

\bibitem{albi2014Phil}
Albi, G., Pareschi, L., Zanella, M.: Boltzmann-type control of opinion
  consensus through leaders.
\newblock Phil. Trans. R. Soc. A \textbf{372}, 20140138/1--18 (2014).
\newblock \doi{10.1098/rsta.2014.0138}

\bibitem{AlPaZa19}
Albi, G., Pareschi, L., Zanella, M.: Boltzmann games in heterogeneous consensus
  dynamics.
\newblock J. Stat. Phys. \textbf{175}(1), 97--125 (2019).
\newblock \doi{10.1007/s10955-019-02246-y}

\bibitem{andersenhard}
Andersen, H.C., Weeks, J.D., Chandler, D.: Relationship between the hard-sphere
  fluid and fluids with realistic repulsion force.
\newblock Phys. Rev. A \textbf{4}(4), 1597--1607 (1971)

\bibitem{arechavaleta2008IEEETR}
Arechavaleta, G., Laumond, J.P., Hicheur, H., Berthoz, A.: An optimality
  principle governing human walking.
\newblock IEEE Trans. Robot. \textbf{24}(1), 5--14 (2008)

\bibitem{Atashpaz2007}
Atashpaz-Gargari, E., Lucas, C.: Imperialist competitive algorithm: {A}n
  algorithm for optimization inspired by imperialistic competition.
\newblock In: IEEE Congress on Evolutionary Computation, pp. 4661--4667. IEEE
  (2007).
\newblock \doi{10.1109/CEC.2007.4425083}

\bibitem{audet2014MPC}
Audet, C., Dang, K.C., Orban, D.: Optimization of algorithms with {OPAL}.
\newblock Math. Prog. Comp. \textbf{6}(3), 233--254 (2014).
\newblock \doi{10.1007/s12532-014-0067-x}

\bibitem{BBCK18}
Bailo, R., Bongini, M., Carrillo, J.A., Kalise, D.: Optimal consensus control
  of the cucker-smale model.
\newblock IFAC-PapersOnLine \textbf{51}(13), 1--6 (2018)

\bibitem{bellomo2011SR}
Bellomo, N., Dogb\'e, C.: On the modeling of traffic and crowds: {A} survey of
  models, speculations, and perspectives.
\newblock SIAM Rev. \textbf{53}, 409--463 (2011)

\bibitem{BFK15}
Bongini, M., Fornasier, M., Kalise, D.: ({U}n)conditional consensus emergence
  under perturbed and decentralized feedback controls.
\newblock Discrete Contin. Dyn. Syst. \textbf{35}(9), 4071--4094 (2015).
\newblock \doi{10.3934/dcds.2015.35.4071}

\bibitem{borzi2015M3ASa}
Borz\`i, A., Wongkaew, S.: Modeling and control through leadership of a refined
  flocking system.
\newblock Math. Models Methods Appl. Sci. \textbf{25}(2), 255--282 (2015).
\newblock \doi{10.1142/S0218202515500098}

\bibitem{braess2005TS}
Braess, D., Nagurney, A., Wakolbinger, T.: On a paradox of traffic planning.
\newblock Transport. Sci. \textbf{39}(4), 446--450 (2005)

\bibitem{CFPT13}
Caponigro, M., Fornasier, M., Piccoli, B., Tr\'{e}lat, E.: Sparse stabilization
  and optimal control of the {C}ucker-{S}male model.
\newblock Math. Control Relat. Fields \textbf{3}(4), 447--466 (2013).
\newblock \doi{10.3934/mcrf.2013.3.447}

\bibitem{carrillo2009double}
Carrillo, J.A., D'Orsogna, M.R., Panferov, V.: Double milling in self-propelled
  swarms from kinetic theory.
\newblock Kinet. Relat. Models \textbf{2}(2), 363--378 (2009)

\bibitem{carrillo2010particle}
Carrillo, J.A., Fornasier, M., Toscani, G., Vecil, F.: Particle, kinetic, and
  hydrodynamic models of swarming.
\newblock In: Mathematical modeling of collective behavior in socio-economic
  and life sciences, pp. 297--336. Springer (2010)

\bibitem{carrillo2016M3AS}
Carrillo, J.A., Martin, S., Wolfram, M.T.: An improved version of the {H}ughes
  model for pedestrian flow.
\newblock Math. Models Methods Appl. Sci. \textbf{26}(4), 671--697 (2016).
\newblock \doi{10.1142/S0218202516500147}

\bibitem{cercignani1994}
Cercignani, C., Illner, R., Pulvirenti, M.: The mathematical theory of dilute
  gases.
\newblock Springer (1994)

\bibitem{chitour2012SICON}
Chitour, Y., Jean, F., Mason, P.: Optimal control models of goal-oriented human
  locomotion.
\newblock SIAM J. Control Optim. \textbf{50}(1), 147--170 (2012)

\bibitem{cirillo2013PhysA}
Cirillo, E.N.M., Muntean, A.: Dynamics of pedestrians in regions with no
  visibility - {A} lattice model without exclusion.
\newblock Physica A \textbf{392}, 3578--3588 (2013)

\bibitem{Clerc}
Clerc, M., Kennedy, J.: The particle swarm - explosion, stability, and
  convergence in a multidimensional complex space.
\newblock IEEE Transactions on Evolutionary Computation \textbf{6}(1), 58--73
  (2002).
\newblock \doi{10.1109/4235.985692}

\bibitem{colombi2016CAM}
Colombi, A., Scianna, M., Alaia, A.: A discrete mathematical model for the
  dynamics of a crowd of gazing pedestrians with and without an evolving
  environmental awareness.
\newblock Comput. Appl. Math. pp. 1--29 (2016).
\newblock \doi{10.1007/s40314-016-0316-x}

\bibitem{colombo2012M3AS}
Colombo, R.M., Garavello, M., Lecureux-Mercier, M.: A class of nonlocal models
  of pedestrian traffic.
\newblock Math. Models Methods Appl. Sci. \textbf{22}, 1150023/1--34 (2012)

\bibitem{coscia2008M3AS}
Coscia, V., Canavesio, C.: First-order macroscopic modelling of human crowd
  dynamics.
\newblock Math. Models Methods Appl. Sci. \textbf{18}(suppl01), 1217--1247
  (2008)

\bibitem{couzin2005N}
Couzin, I.D., Krause, J., Franks, N.R., Levin, S.A.: Effective leadership and
  decision-making in animal groups on the move.
\newblock Nature \textbf{433}, 513--516 (2005)

\bibitem{cristiani2017AMM}
Cristiani, E., Peri, D.: Handling obstacles in pedestrian simulations: {M}odels
  and optimization.
\newblock Appl. Math. Model. \textbf{45}, 285--302 (2017)

\bibitem{cristiani2019AMM}
Cristiani, E., Peri, D.: Robust design optimization for egressing pedestrians
  in unkwnown environments.
\newblock Appl. Math. Model. \textbf{72}, 553--568 (2019)

\bibitem{cristiani2011MMS}
Cristiani, E., Piccoli, B., Tosin, A.: Multiscale modeling of granular flows
  with application to crowd dynamics.
\newblock Multiscale Model. Simul. \textbf{9}, 155--182 (2011)

\bibitem{cristiani2014book}
Cristiani, E., Piccoli, B., Tosin, A.: Multiscale Modeling of Pedestrian
  Dynamics.
\newblock {Modeling, Simulation \& Applications}. Springer (2014)

\bibitem{cristiani2015NHM}
Cristiani, E., Priuli, F.S.: A destination-preserving model for simulating
  {W}ardrop equilibria in traffic flow on networks.
\newblock Netw. Heterog. Media \textbf{10}, 857--876 (2015)

\bibitem{cristiani2015SIAP}
Cristiani, E., Priuli, F.S., Tosin, A.: Modeling rationality to control
  self-organization of crowds: {A}n environmental approach.
\newblock SIAM J. Appl. Math. \textbf{75}(2), 605--629 (2015).
\newblock \doi{10.1137/140962413}

\bibitem{cucker2007IEEE}
Cucker, F., Smale, S.: Emergent behavior in flocks.
\newblock IEEE Trans. Autom. Contr. \textbf{52}(5), 852--862 (2007)

\bibitem{duan2014SR}
Duan, H., Sun, C.: Swarm intelligence inspired shills and the evolution of
  cooperation.
\newblock Sci. Rep. \textbf{4}, 5210 (2014)

\bibitem{during2009PRSA}
D\"uring, B., Markowich, P., Pietschmann, J.F., Wolfram, M.T.: {B}oltzmann and
  {F}okker-{P}lanck equations modelling opinion formation in the presence of
  strong leaders.
\newblock Proc. R. Soc. A \textbf{465}, 3687--3708 (2009)

\bibitem{escobar2003LNCS}
Escobar, R., De~La~Rosa, A.: Architectural design for the survival optimization
  of panicking fleeing victims.
\newblock In: W.~Banzhaf, T.~Christaller, P.~Dittrich, J.T. Kim, J.~Ziegler
  (eds.) ECAL 2003, LNAI 2801, pp. 97--106. Springer-Verlag Berlin Heidelberg
  (2003)

\bibitem{etikyala2014M3AS}
Etikyala, R., G\"ottlich, S., Klar, A., Tiwari, S.: Particle methods for
  pedestrian flow models: from microscopic to nonlocal continuum models.
\newblock Math. Models Methods Appl. Sci. \textbf{24}, 2503--2523 (2014).
\newblock \doi{10.1142/S0218202514500274}

\bibitem{FTW18}
Festa, A., Tosin, A., Wolfram, M.T.: Kinetic description of collision avoidance
  in pedestrian crowds by sidestepping.
\newblock Kinet. Relat. Models \textbf{11}(3), 491--520 (2018).
\newblock \doi{10.3934/krm.2018022}

\bibitem{FS14}
Fornasier, M., Solombrino, F.: Mean-field optimal control.
\newblock ESAIM Control Optim. Calc. Var. \textbf{20}(4), 1123--1152 (2014).
\newblock \doi{10.1051/cocv/2014009}

\bibitem{frank2011PA}
Frank, G.A., Dorso, C.O.: Room evacuation in the presence of an obstacle.
\newblock Physica A \textbf{390}, 2135--2145 (2011).
\newblock \doi{10.1016/j.physa.2011.01.015}

\bibitem{Galor:2007}
Galor, O.: Discrete Dynamical Systems.
\newblock Springer Verlag (2007)

\bibitem{guo2012TRB}
Guo, R.Y., Huang, H.J., Wong, S.C.: Route choice in pedestrian evacuation under
  conditions of good and zero visibility: experimental and simulation results.
\newblock Transportation Res. B \textbf{46}(6), 669--686 (2012).
\newblock \doi{10.1016/j.trb.2012.01.002}

\bibitem{haghani2019JAT}
Haghani, M., Cristiani, E., Bode, N.W.F., Boltes, M., Corbetta, A.: Panic,
  irrationality, and herding: {T}hree ambiguous terms in crowd dynamics
  research.
\newblock J. Adv. Transport. \textbf{2019}, Article ID 9267643 (2019).
\newblock \doi{10.1155/2019/9267643}

\bibitem{han2006JSSC}
Han, J., Li, M., Guo, L.: Soft control on collective behavior of a group of
  autonomous agents by a shill agent.
\newblock Jrl. Syst. Sci. \& Complexity \textbf{19}(1), 54--62 (2006).
\newblock \doi{10.1007/s11424-006-0054-z}

\bibitem{han2013PLOSONE}
Han, J., Wang, L.: Nondestructive intervention to multi-agent systems through
  an intelligent agent.
\newblock PLoS ONE \textbf{8}(5), e61542 (2013).
\newblock \doi{10.1371/journal.pone.0061542}

\bibitem{helbing2001RMP}
Helbing, D.: Traffic and related self-driven many-particle systems.
\newblock Rev. Mod. Phys. \textbf{73}, 1067--1141 (2001)

\bibitem{helbing2005TS}
Helbing, D., Buzna, L., Johansson, A., Werner, T.: Self-organized pedestrian
  crowd dynamics: {E}xperiments, simulations, and design solutions.
\newblock Transport. Sci. \textbf{39}(1), 1--24 (2005)

\bibitem{helbing2000N}
Helbing, D., Farkas, I., Vicsek, T.: Simulating dynamical features of escape
  panic.
\newblock Nature \textbf{407}, 487--490 (2000)

\bibitem{helbing1995PRE}
Helbing, D., Moln\'{a}r, P.: Social force model for pedestrian dynamics.
\newblock Phys. Rev. E \textbf{51}, 4282--4286 (1995)

\bibitem{Hendtlass}
Hendtlass, T.: Wosp: A multi-optima particle swarm algorithm.
\newblock In: IEEE (ed.) Proceedings of the 2005 IEEE Congress on Evolutionary
  Computation, pp. 727--734. IEEE (2005)

\bibitem{HePaSt15}
Herty, M., Pareschi, L., Steffensen, S.: Mean-field control and {R}iccati
  equations.
\newblock Netw. Heterog. Media \textbf{10}(3), 699--715 (2015).
\newblock \doi{10.3934/nhm.2015.10.699}

\bibitem{hughes2002TRB}
Hughes, R.L.: A continuum theory for the flow of pedestrians.
\newblock Transportation Res. Part B \textbf{36}, 507--535 (2002)

\bibitem{hughes2003ARFM}
Hughes, R.L.: The flow of human crowds.
\newblock Annu. Rev. Fluid Mech. \textbf{35}, 169--182 (2003)

\bibitem{jiang2014PLOS}
Jiang, L., Li, J., Shen, C., Yang, S., Han, Z.: Obstacle optimization for panic
  flow - {R}educing the tangential momentum increases the escape speed.
\newblock PLoS ONE \textbf{9}(12), e115463 (2014).
\newblock \doi{10.1371/journal.pone.0115463}

\bibitem{johansson2007}
Johansson, A., Helbing, D.: Pedestrian flow optimization with a genetic
  algorithm based on boolean grids.
\newblock In: N.~Waldau, P.~Gattermann, H.~Knoflacher, M.~Schreckenberg (eds.)
  Pedestrian and Evacuation Dynamics 2005, pp. 267--272. Springer-Verlag Berlin
  Heidelberg (2007)

\bibitem{kachroo2008book}
Kachroo, P., Al-nasur, S.J., Wadoo, S.A., Shende, A.: Pedestrian dynamics.
  {F}eedback control of crowd evacuation.
\newblock Understanding Complex Systems. Springer-Verlag, Berlin Heidelberg
  (2008)

\bibitem{karper2015hydrodynamic}
Karper, T.K., Mellet, A., Trivisa, K.: Hydrodynamic limit of the kinetic
  {C}ucker--{S}male flocking model.
\newblock Math. Models Methods Appl. Sci. \textbf{25}(01), 131--163 (2015).
\newblock \doi{10.1142/S0218202515500050}

\bibitem{loehner2010AMM}
L\"ohner, R.: On the modeling of pedestrian motion.
\newblock Appl. Math. Model. \textbf{34}, 366--382 (2010)

\bibitem{matsuoka2015}
Matsuoka, T., Tomoeda, A., Iwamoto, M., Suzuno, K., Ueyama, D.: Effects of an
  obstacle position for pedestrian evacuation: {SF} model approach.
\newblock In: M.~Chraibi, M.~Boltes, A.~Schadschneider, A.~Seyfried (eds.)
  Traffic and Granular Flow '13, pp. 163--170. Springer International
  Publishing (2015)

\bibitem{mayne2000A}
Mayne, D.Q., Rawlings, J.B., Rao, C.V., Scokaert, P.O.M.: Constrained model
  predictive control: {S}tability and optimality.
\newblock Automatica \textbf{36}(6), 789 -- 814 (2000).
\newblock \doi{10.1016/S0005-1098(99)00214-9}

\bibitem{motsch2011JSP}
Motsch, S., Tadmor, E.: A new model for self-organized dynamics and its
  flocking behavior.
\newblock J. Stat. Phys. \textbf{144}(5), 923--947 (2011).
\newblock \doi{10.1007/s10955-011-0285-9}

\bibitem{NaPaTo10}
Naldi, G., Pareschi, L., Toscani, G. (eds.): Mathematical modeling of
  collective behavior in socio-economic and life sciences.
\newblock Modeling and Simulation in Science, Engineering and Technology.
  Birkh\"{a}user Boston, Inc., Boston, MA (2010).
\newblock \doi{10.1007/978-0-8176-4946-3}

\bibitem{okazaki1979TAIJa}
Okazaki, S.: A study of pedestrian movement in architectural space, part 1:
  {P}edestrian movement by the application of magnetic model.
\newblock Trans. of A.I.J. \textbf{283}, 111--119 (1979)

\bibitem{okazaki1979TAIJb}
Okazaki, S.: A study of pedestrian movement in architectural space, part 2:
  {C}oncentrated pedestrian movement.
\newblock Trans. of A.I.J. \textbf{284}, 101--110 (1979)

\bibitem{okazaki1979TAIJc}
Okazaki, S.: A study of pedestrian movement in architectural space, part 3:
  {A}long the shortest path, taking fire, congestion and unrecognized space
  into account.
\newblock Trans. of A.I.J. \textbf{285}, 137--147 (1979)

\bibitem{PT:13}
Pareschi, L., Toscani, G.: Interacting multi-agent systems. {K}inetic equations
  \& {M}onte {C}arlo methods.
\newblock Oxford University Press, USA (2013)

\bibitem{parisi2005PhysA}
Parisi, D.R., Dorso, C.O.: Microscopic dynamics of pedestrian evacuation.
\newblock Physica A \textbf{354}, 606--618 (2005).
\newblock \doi{http://dx.doi.org/10.1016/j.physa.2005.02.040}

\bibitem{peri2019CAIE}
Peri, D.: Hybridization of the imperialist competitive algorithm and local
  search with application to ship design optimization.
\newblock Computers \& Industrial Engineering \textbf{137}, 1--30 (2019).
\newblock \doi{10.1016/j.cie.2019.106069}

\bibitem{piccoli2009CMT}
Piccoli, B., Tosin, A.: Pedestrian flows in bounded domains with obstacles.
\newblock Contin. Mech. Thermodyn. \textbf{21}, 85--107 (2009)

\bibitem{shukla2009}
Shukla, P.K.: Genetically optimized architectural designs for control of
  pedestrian crowds.
\newblock In: K.~Korb, M.~Randall, T.~Hendtlass (eds.) Artificial life:
  Borrowing from biology, \emph{LNCS}, vol. 5865, pp. 22--31. Springer-Verlag
  Berlin Heidelberg (2009)

\bibitem{toscanibellomoenskog}
Toscani, G., Bellomo, N.: The {E}nskog-{B}oltzmann equation in the whole space
  {$\mathbb R^3$}: some global existence, uniqueness and stability results.
\newblock Comput. Math. Applic. \textbf{13}(9-11), 851--859 (1987)

\bibitem{twarogowska2014AMM}
Twarogowska, M., Goatin, P., Duvigneau, R.: Macroscopic modeling and
  simulations of room evacuation.
\newblock Appl. Math. Model. \textbf{38}(24), 5781--5795 (2014).
\newblock \doi{10.1016/j.apm.2014.03.027}

\bibitem{Vil02}
Villani, C.: Handbook of Mathematical Fluid Dynamics, vol.~1, chap. A review of
  mathematical topics in collisional kinetic theory.
\newblock Elsevier (2002)

\bibitem{wang2015PhysA}
Wang, J., Zhang, L., Shi, Q., Yang, P., Hu, X.: Modeling and simulating for
  congestion pedestrian evacuation with panic.
\newblock Physica A \textbf{428}, 396--409 (2015).
\newblock \doi{10.1016/j.physa.2015.01.057}

\bibitem{zhao2017PhysA}
Zhao, Y., Li, M., Lu, X., Tian, L., Yu, Z., Huang, K., Wang, Y., Li, T.:
  Optimal layout design of obstacles for panic evacuation using differential
  evolution.
\newblock Physica A \textbf{465}, 175--194 (2017).
\newblock \doi{10.1016/j.physa.2016.08.021}

\end{thebibliography}
\end{document}